\documentclass[journal,onecolumn,draftcls,12pt]{IEEEtranv17}

\pdfoutput=1

\newlength{\figwidth}

\usepackage[nosort]{cite}        
\usepackage{url} 
\usepackage[intlimits]{amsmath}
\usepackage{bbm}
\usepackage{graphicx}
\usepackage{paralist}
\usepackage{fancyref}

\usepackage[stretch=16,shrink=16,step=4]{microtype}

\makeatletter
\def\@IEEEinterspaceratioM{0.265}
\def\@IEEEinterspaceMINratioM{0.1651}
\def\@IEEEinterspaceMAXratioM{0.38}

\def\@IEEEinterspaceratioB{0.31}
\def\@IEEEinterspaceMINratioB{0.19}
\def\@IEEEinterspaceMAXratioB{0.38}
\@IEEEtunefonts
\makeatother

\usepackage{amssymb}
\usepackage{amsfonts}
\usepackage{mathrsfs}
\usepackage{xspace}
\usepackage{bm}
\usepackage{upgreek}

\newcommand{\safemath}[2]{\newcommand{#1}{\ensuremath{#2}\xspace}}

\safemath{\bma}{\mathbf{a}}
\safemath{\bmb}{\mathbf{b}}
\safemath{\bmc}{\mathbf{c}}
\safemath{\bmd}{\mathbf{d}}
\safemath{\bme}{\mathbf{e}}
\safemath{\bmf}{\mathbf{f}}
\safemath{\bmg}{\mathbf{g}}
\safemath{\bmh}{\mathbf{h}}
\safemath{\bmi}{\mathbf{i}}
\safemath{\bmj}{\mathbf{j}}
\safemath{\bmk}{\mathbf{k}}
\safemath{\bml}{\mathbf{l}}
\safemath{\bmm}{\mathbf{m}}
\safemath{\bmn}{\mathbf{n}}
\safemath{\bmo}{\mathbf{o}}
\safemath{\bmp}{\mathbf{p}}
\safemath{\bmq}{\mathbf{q}}
\safemath{\bmr}{\mathbf{r}}
\safemath{\bms}{\mathbf{s}}
\safemath{\bmt}{\mathbf{t}}
\safemath{\bmu}{\mathbf{u}}
\safemath{\bmv}{\mathbf{v}}
\safemath{\bmw}{\mathbf{w}}
\safemath{\bmx}{\mathbf{x}}
\safemath{\bmy}{\mathbf{y}}
\safemath{\bmz}{\mathbf{z}}
\safemath{\bmzero}{\mathbf{0}}
\safemath{\bmone}{\mathbf{1}}

\bmdefine{\biad}{a}
\bmdefine{\bibd}{b}
\bmdefine{\bicd}{c}
\bmdefine{\bidd}{d}
\bmdefine{\bied}{e}
\bmdefine{\bifd}{f}
\bmdefine{\bigd}{g}
\bmdefine{\bihd}{h}
\bmdefine{\biid}{i}
\bmdefine{\bijd}{j}
\bmdefine{\bikd}{k}
\bmdefine{\bild}{l}
\bmdefine{\bimd}{m}
\bmdefine{\bind}{n}
\bmdefine{\biod}{o}
\bmdefine{\bipd}{p}
\bmdefine{\biqd}{q}
\bmdefine{\bird}{r}
\bmdefine{\bisd}{s}
\bmdefine{\bitd}{t}
\bmdefine{\biud}{u}
\bmdefine{\bivd}{v}
\bmdefine{\biwd}{w}
\bmdefine{\bixd}{x}
\bmdefine{\biyd}{y}
\bmdefine{\bizd}{z}

\bmdefine{\bixid}{\xi}
\bmdefine{\bilambdad}{\lambda}
\bmdefine{\bimud}{\mu}
\bmdefine{\bithetad}{\theta}
\bmdefine{\biphid}{\phi}

\safemath{\bmia}{\biad}
\safemath{\bmib}{\bibd}
\safemath{\bmic}{\bicd}
\safemath{\bmid}{\bidd}
\safemath{\bmie}{\bied}
\safemath{\bmif}{\bifd}
\safemath{\bmig}{\bigd}
\safemath{\bmih}{\bihd}
\safemath{\bmii}{\biid}
\safemath{\bmij}{\bijd}
\safemath{\bmik}{\bikd}
\safemath{\bmil}{\bild}
\safemath{\bmim}{\bimd}
\safemath{\bmin}{\bind}
\safemath{\bmio}{\biod}
\safemath{\bmip}{\bipd}
\safemath{\bmiq}{\biqd}
\safemath{\bmir}{\bird}
\safemath{\bmis}{\bisd}
\safemath{\bmit}{\bitd}
\safemath{\bmiu}{\biud}
\safemath{\bmiv}{\bivd}
\safemath{\bmiw}{\biwd}
\safemath{\bmix}{\bixd}
\safemath{\bmiy}{\biyd}
\safemath{\bmiz}{\bizd}

\safemath{\bmxi}{\bixid}
\safemath{\bmlambda}{\bilambdad}
\safemath{\bmmu}{\bimud}
\safemath{\bmtheta}{\bithetad}
\safemath{\bmphi}{\biphid}

\safemath{\bA}{\mathbf{A}}
\safemath{\bB}{\mathbf{B}}
\safemath{\bC}{\mathbf{C}}
\safemath{\bD}{\mathbf{D}}
\safemath{\bE}{\mathbf{E}}
\safemath{\bF}{\mathbf{F}}
\safemath{\bG}{\mathbf{G}}
\safemath{\bH}{\mathbf{H}}
\safemath{\bI}{\mathbf{I}}
\safemath{\bJ}{\mathbf{J}}
\safemath{\bK}{\mathbf{K}}
\safemath{\bL}{\mathbf{L}}
\safemath{\bM}{\mathbf{M}}
\safemath{\bN}{\mathbf{N}}
\safemath{\bO}{\mathbf{O}}
\safemath{\bP}{\mathbf{P}}
\safemath{\bQ}{\mathbf{Q}}
\safemath{\bR}{\mathbf{R}}
\safemath{\bS}{\mathbf{S}}
\safemath{\bT}{\mathbf{T}}
\safemath{\bU}{\mathbf{U}}
\safemath{\bV}{\mathbf{V}}
\safemath{\bW}{\mathbf{W}}
\safemath{\bX}{\mathbf{X}}
\safemath{\bY}{\mathbf{Y}}
\safemath{\bZ}{\mathbf{Z}}

\safemath{\bZero}{\mathbf{0}}
\safemath{\bOne}{\mathbf{1}}
\safemath{\bDelta}{\mathbf{\Delta}}
\safemath{\bLambda}{\mathbf{\UpLambda}}
\safemath{\bPhi}{\mathbf{\Upphi}}
\safemath{\bSigma}{\mathbf{\Upsigma}}
\safemath{\bOmega}{\mathbf{\Upomega}}
\safemath{\bTheta}{\mathbf{\Uptheta}}

\bmdefine{\biAd}{A}
\bmdefine{\biBd}{B}
\bmdefine{\biCd}{C}
\bmdefine{\biDd}{D}
\bmdefine{\biEd}{E}
\bmdefine{\biFd}{F}
\bmdefine{\biGd}{G}
\bmdefine{\biHd}{H}
\bmdefine{\biId}{I}
\bmdefine{\biJd}{J}
\bmdefine{\biKd}{K}
\bmdefine{\biLd}{L}
\bmdefine{\biMd}{M}
\bmdefine{\biOd}{N}
\bmdefine{\biPd}{O}
\bmdefine{\biQd}{P}
\bmdefine{\biRd}{R}
\bmdefine{\biSd}{S}
\bmdefine{\biTd}{T}
\bmdefine{\biUd}{U}
\bmdefine{\biVd}{V}
\bmdefine{\biWd}{W}
\bmdefine{\biXd}{X}
\bmdefine{\biYd}{Y}
\bmdefine{\biZd}{Z}

\bmdefine{\biDelta}{\Delta}
\bmdefine{\biLambda}{\Lambda}
\bmdefine{\biPhi}{\Phi}
\bmdefine{\biSigma}{\Sigma}
\bmdefine{\biOmega}{\Omega}
\bmdefine{\biTheta}{\Theta}

\safemath{\bimA}{\biAd}
\safemath{\bimB}{\biBd}
\safemath{\bimC}{\biCd}
\safemath{\bimD}{\biDd}
\safemath{\bimE}{\biEd}
\safemath{\bimF}{\biFd}
\safemath{\bimG}{\biGd}
\safemath{\bimH}{\biHd}
\safemath{\bimI}{\biId}
\safemath{\bimJ}{\biJd}
\safemath{\bimK}{\biKd}
\safemath{\bimL}{\biLd}
\safemath{\bimM}{\biMd}
\safemath{\bimN}{\biNd}
\safemath{\bimO}{\biOd}
\safemath{\bimP}{\biPd}
\safemath{\bimQ}{\biQd}
\safemath{\bimR}{\biRd}
\safemath{\bimS}{\biSd}
\safemath{\bimT}{\biTd}
\safemath{\bimU}{\biUd}
\safemath{\bimV}{\biVd}
\safemath{\bimW}{\biWd}
\safemath{\bimX}{\biXd}
\safemath{\bimY}{\biYd}
\safemath{\bimZ}{\biZd}

\safemath{\bimDelta}{\biDelta}
\safemath{\bimLambda}{\biLambda}
\safemath{\bimPhi}{\biPhi}
\safemath{\bimSigma}{\biSigma}
\safemath{\bimOmega}{\biOmega}
\safemath{\bimTheta}{\biTheta}

\safemath{\setA}{\mathcal{A}}
\safemath{\setB}{\mathcal{B}}
\safemath{\setC}{\mathcal{C}}
\safemath{\setD}{\mathcal{D}}
\safemath{\setE}{\mathcal{E}}
\safemath{\setF}{\mathcal{F}}
\safemath{\setG}{\mathcal{G}}
\safemath{\setH}{\mathcal{H}}
\safemath{\setI}{\mathcal{I}}
\safemath{\setJ}{\mathcal{J}}
\safemath{\setK}{\mathcal{K}}
\safemath{\setL}{\mathcal{L}}
\safemath{\setM}{\mathcal{M}}
\safemath{\setN}{\mathcal{N}}
\safemath{\setO}{\mathcal{O}}
\safemath{\setP}{\mathcal{P}}
\safemath{\setQ}{\mathcal{Q}}
\safemath{\setR}{\mathcal{R}}
\safemath{\setS}{\mathcal{S}}
\safemath{\setT}{\mathcal{T}}
\safemath{\setU}{\mathcal{U}}
\safemath{\setV}{\mathcal{V}}
\safemath{\setW}{\mathcal{W}}
\safemath{\setX}{\mathcal{X}}
\safemath{\setY}{\mathcal{Y}}
\safemath{\setZ}{\mathcal{Z}}
\safemath{\emptySet}{\varnothing}

\safemath{\colA}{\mathscr{A}}
\safemath{\colB}{\mathscr{B}}
\safemath{\colC}{\mathscr{C}}
\safemath{\colD}{\mathscr{D}}
\safemath{\colE}{\mathscr{E}}
\safemath{\colF}{\mathscr{F}}
\safemath{\colG}{\mathscr{G}}
\safemath{\colH}{\mathscr{H}}
\safemath{\colI}{\mathscr{I}}
\safemath{\colJ}{\mathscr{J}}
\safemath{\colK}{\mathscr{K}}
\safemath{\colL}{\mathscr{L}}
\safemath{\colM}{\mathscr{M}}
\safemath{\colN}{\mathscr{N}}
\safemath{\colO}{\mathscr{O}}
\safemath{\colP}{\mathscr{P}}
\safemath{\colQ}{\mathscr{Q}}
\safemath{\colR}{\mathscr{R}}
\safemath{\colS}{\mathscr{S}}
\safemath{\colT}{\mathscr{T}}
\safemath{\colU}{\mathscr{U}}
\safemath{\colV}{\mathscr{V}}
\safemath{\colW}{\mathscr{W}}
\safemath{\colX}{\mathscr{X}}
\safemath{\colY}{\mathscr{Y}}
\safemath{\colZ}{\mathscr{Z}}

\safemath{\opA}{\mathbb{A}}
\safemath{\opB}{\mathbb{B}}
\safemath{\opC}{\mathbb{C}}
\safemath{\opD}{\mathbb{D}}
\safemath{\opE}{\mathbb{E}}
\safemath{\opF}{\mathbb{F}}
\safemath{\opG}{\mathbb{G}}
\safemath{\opH}{\mathbb{H}}
\safemath{\opI}{\mathbb{I}}
\safemath{\opJ}{\mathbb{J}}
\safemath{\opK}{\mathbb{K}}
\safemath{\opL}{\mathbb{L}}
\safemath{\opM}{\mathbb{M}}
\safemath{\opN}{\mathbb{N}}
\safemath{\opO}{\mathbb{O}}
\safemath{\opP}{\mathbb{P}}
\safemath{\opQ}{\mathbb{Q}}
\safemath{\opR}{\mathbb{R}}
\safemath{\opS}{\mathbb{S}}
\safemath{\opT}{\mathbb{T}}
\safemath{\opU}{\mathbb{U}}
\safemath{\opV}{\mathbb{V}}
\safemath{\opW}{\mathbb{W}}
\safemath{\opX}{\mathbb{X}}
\safemath{\opY}{\mathbb{Y}}
\safemath{\opZ}{\mathbb{Z}}
\safemath{\opZero}{\mathbb{O}}
\safemath{\identityop}{\opI}

\safemath{\veca}{\bma}
\safemath{\vecb}{\bmb}
\safemath{\vecc}{\bmc}
\safemath{\vecd}{\bmd}
\safemath{\vece}{\bme}
\safemath{\vecf}{\bmf}
\safemath{\vecg}{\bmg}
\safemath{\vech}{\bmh}
\safemath{\veci}{\bmi}
\safemath{\vecj}{\bmj}
\safemath{\veck}{\bmk}
\safemath{\vecl}{\bml}
\safemath{\vecm}{\bmm}
\safemath{\vecn}{\bmn}
\safemath{\veco}{\bmo}
\safemath{\vecp}{\bmmp}
\safemath{\vecq}{\bmq}
\safemath{\vecr}{\bmr}
\safemath{\vecs}{\bms}
\safemath{\vect}{\bmt}
\safemath{\vecu}{\bmu}
\safemath{\vecv}{\bmv}
\safemath{\vecw}{\bmw}
\safemath{\vecx}{\bmx}
\safemath{\vecy}{\bmy}
\safemath{\vecz}{\bmz}

\safemath{\veczero}{\bmzero}
\safemath{\vecone}{\bmone}
\safemath{\vecxi}{\bmxi}
\safemath{\veclambda}{\bmlambda}
\safemath{\vecmu}{\bmmu}
\safemath{\vectheta}{\bmtheta}
\safemath{\vecphi}{\bmphi}

\safemath{\matA}{\bA}
\safemath{\matB}{\bB}
\safemath{\matC}{\bC}
\safemath{\matD}{\bD}
\safemath{\matE}{\bE}
\safemath{\matF}{\bF}
\safemath{\matG}{\bG}
\safemath{\matH}{\bH}
\safemath{\matI}{\bI}
\safemath{\matJ}{\bJ}
\safemath{\matK}{\bK}
\safemath{\matL}{\bL}
\safemath{\matM}{\bM}
\safemath{\matN}{\bN}
\safemath{\matO}{\bO}
\safemath{\matP}{\bP}
\safemath{\matQ}{\bQ}
\safemath{\matR}{\bR}
\safemath{\matS}{\bS}
\safemath{\matT}{\bT}
\safemath{\matU}{\bU}
\safemath{\matV}{\bV}
\safemath{\matW}{\bW}
\safemath{\matX}{\bX}
\safemath{\matY}{\bY}
\safemath{\matZ}{\bZ}
\safemath{\matzero}{\bmzero}

\safemath{\matDelta}{\bDelta}
\safemath{\matLambda}{\bLambda}
\safemath{\matPhi}{\bPhi}
\safemath{\matSigma}{\bSigma}
\safemath{\matOmega}{\bOmega}
\safemath{\matTheta}{\bTheta}

\safemath{\matidentity}{\matI}
\safemath{\matone}{\matO}

\safemath{\rnda}{A}
\safemath{\rndb}{B}
\safemath{\rndc}{C}
\safemath{\rndd}{D}
\safemath{\rnde}{E}
\safemath{\rndf}{F}
\safemath{\rndg}{G}
\safemath{\rndh}{H}
\safemath{\rndi}{I}
\safemath{\rndj}{J}
\safemath{\rndk}{K}
\safemath{\rndl}{L}
\safemath{\rndm}{M}
\safemath{\rndn}{N}
\safemath{\rndo}{O}
\safemath{\rndp}{P}
\safemath{\rndq}{Q}
\safemath{\rndr}{R}
\safemath{\rnds}{S}
\safemath{\rndt}{T}
\safemath{\rndu}{U}
\safemath{\rndv}{V}
\safemath{\rndw}{W}
\safemath{\rndx}{X}
\safemath{\rndy}{Y}
\safemath{\rndz}{Z}

\safemath{\rveca}{\bimA}
\safemath{\rvecb}{\bimB}
\safemath{\rvecc}{\bimC}
\safemath{\rvecd}{\bimD}
\safemath{\rvece}{\bimE}
\safemath{\rvecf}{\bimF}
\safemath{\rvecg}{\bimG}
\safemath{\rvech}{\bimH}
\safemath{\rveci}{\bimI}
\safemath{\rvecj}{\bimJ}
\safemath{\rveck}{\bimK}
\safemath{\rvecl}{\bimL}
\safemath{\rvecm}{\bimM}
\safemath{\rvecn}{\bimN}
\safemath{\rveco}{\bomO}
\safemath{\rvecp}{\bimP}
\safemath{\rvecq}{\bimQ}
\safemath{\rvecr}{\bimR}
\safemath{\rvecs}{\bimS}
\safemath{\rvect}{\bimT}
\safemath{\rvecu}{\bimU}
\safemath{\rvecv}{\bimV}
\safemath{\rvecw}{\bimW}
\safemath{\rvecx}{\bimX}
\safemath{\rvecy}{\bimY}
\safemath{\rvecz}{\bimZ}

\safemath{\rvecxi}{\bmxi}
\safemath{\rveclambda}{\bmlambda}
\safemath{\rvecmu}{\bmmu}
\safemath{\rvectheta}{\bmtheta}
\safemath{\rvecphi}{\bmphi}

\safemath{\rmatA}{\bimA}
\safemath{\rmatB}{\bimB}
\safemath{\rmatC}{\bimC}
\safemath{\rmatD}{\bimD}
\safemath{\rmatE}{\bimE}
\safemath{\rmatF}{\bimF}
\safemath{\rmatG}{\bimG}
\safemath{\rmatH}{\bimH}
\safemath{\rmatI}{\bimI}
\safemath{\rmatJ}{\bimJ}
\safemath{\rmatK}{\bimK}
\safemath{\rmatL}{\bimL}
\safemath{\rmatM}{\bimM}
\safemath{\rmatN}{\bimN}
\safemath{\rmatO}{\bimO}
\safemath{\rmatP}{\bimP}
\safemath{\rmatQ}{\bimQ}
\safemath{\rmatR}{\bimR}
\safemath{\rmatS}{\bimS}
\safemath{\rmatT}{\bimT}
\safemath{\rmatU}{\bimU}
\safemath{\rmatV}{\bimV}
\safemath{\rmatW}{\bimW}
\safemath{\rmatX}{\bimX}
\safemath{\rmatY}{\bimY}
\safemath{\rmatZ}{\bimZ}

\safemath{\rmatDelta}{\bimDelta}
\safemath{\rmatLambda}{\bimLambda}
\safemath{\rmatPhi}{\bimPhi}
\safemath{\rmatSigma}{\bimSigma}
\safemath{\rmatOmega}{\bimOmega}
\safemath{\rmatTheta}{\bimTheta}

\usepackage{amssymb}
\usepackage{amsfonts}
\usepackage{mathrsfs}
\usepackage{xspace}
\usepackage{bm}
\usepackage{fancyref}
\usepackage{textcomp}

\newenvironment{textbmatrix}{	\setlength{\arraycolsep}{2.5pt}%
								\big[\begin{matrix}}{\end{matrix}\big]%
								\raisebox{0.08ex}{\vphantom{M}}}

\def\be{\begin{equation}}
\def\ee{\end{equation}}
\def\een{\nonumber \end{equation}}
\def\mat{\begin{bmatrix}}
\def\emat{\end{bmatrix}}
\def\btm{\begin{textbmatrix}}
\def\etm{\end{textbmatrix}}

\def\ba#1\ea{\begin{align}#1\end{align}}
\def\bas#1\eas{\begin{align*}#1\end{align*}}
\def\bs#1\es{\begin{split}#1\end{split}} 
\def\bg#1\eg{\begin{gather}#1\end{gather}}
\def\bml#1\eml{\begin{multline}#1\end{multline}}
\def\bi#1\ei{\begin{itemize}#1\end{itemize}}

\newcommand{\subtext}[1]{\text{\fontfamily{cmr}\fontshape{n}\fontseries{m}\selectfont{}#1}}
\newcommand{\lefto}{\mathopen{}\left}

\newcommand{\sub}[1]{\ensuremath{_{\subtext{#1}}}} 

\DeclareMathOperator{\tr}{tr}				
\DeclareMathOperator{\rank}{rank}			
\DeclareMathOperator{\had}{\odot}			
\DeclareMathOperator{\Exop}{\opE}			
\DeclareMathOperator{\esssup}{esssup}			

\DeclareMathOperator{\landauo}{\mathit{o}}

\newcommand{\nullspace}{\setN}	 			
\newcommand{\Ex}[2]{\ensuremath{\Exop_{#1}\lefto[#2\right]}} 	
\newcommand{\abs}[1]{\left\lvert#1\right\rvert}		

\newcommand{\vecnorm}[1]{\lVert#1\rVert}		
\newcommand{\conj}[1]{\ensuremath{#1^{*}}} 	
\newcommand{\tp}[1]{\ensuremath{#1^{T}}} 		
\newcommand{\herm}[1]{\ensuremath{#1^{H}}} 	
\newcommand{\inv}[1]{\ensuremath{#1^{-1}}} 	

\safemath{\dirac}{\delta}					
\safemath{\krond}{\dirac}					
\newcommand{\allz}[2]{\ensuremath{#1=0,1,\ldots,#2-1}}
\newcommand{\sumz}[2]{\ensuremath{\sum_{#1=0}^{#2-1}}}
\newcommand{\sumo}[2]{\ensuremath{\sum_{#1=1}^{#2}}}
\newcommand{\suminf}[1]{\ensuremath{\sum_{#1=-\infty}^{\infty}}}
\newcommand{\logdet}[1]{\log \det\lefto(#1\right)} 
\safemath{\upto}{\uparrow}
\safemath{\downto}{\downarrow}
\safemath{\iu}{j}							
\safemath{\ev}{\lambda}						
\safemath{\hilseqspace}{l^{2}}				
\newcommand{\banachfunspace}[1]{\setL^{#1}}	
\safemath{\hilfunspace}{\banachfunspace{2}}	
\newcommand{\floor}[1]{\lfloor #1 \rfloor}

\safemath{\SNR}{\text{\sc snr}} 				
\safemath{\No}{N_0}							
\safemath{\Es}{E_s}							
\safemath{\Eb}{E_b}							
\safemath{\EbNo}{\frac{\Eb}{\No}}
\safemath{\EsNo}{\frac{\Es}{\No}}

\DeclareMathOperator{\CHop}{\ensuremath{\opH}} 
\safemath{\tvir}{\rndh_{\CHop}}				
\safemath{\tvtf}{\rndl_{\CHop}}				
\safemath{\spf}{\rnds_{\CHop}}				
\safemath{\bff}{H_{\CHop}}					

\safemath{\ircf}{r_{h}}						
\safemath{\tftvcf}{r_{s}}					
\safemath{\tfcf}{r_{l}}						
\safemath{\bfcf}{r_{H}}						

\safemath{\tcorr}{c_h}						
\safemath{\scf}{c_{s}}						
\safemath{\tfcorr}{c_{l}}					
\safemath{\fcorr}{c_{H}}						

\safemath{\mi}{I}							
\safemath{\capacity}{C}						

\newcommand{\iid}{i.i.d.\@\xspace}

\newcommand{\pdf}[1]{q_{#1}}				
\newcommand{\cdf}[1]{Q_{#1}} 			
\safemath{\normal}{\mathcal{N}}			
\safemath{\jpg}{\mathcal{CN}}			
\safemath{\mchain}{\leftrightarrow}		
\newcommand{\given}{\,\vert\,}				

\safemath{\dB}{\,\mathrm{dB}}
\safemath{\dBm}{\,\mathrm{dBm}}
\safemath{\Hz}{\,\mathrm{Hz}}
\safemath{\kHz}{\,\mathrm{kHz}}
\safemath{\MHz}{\,\mathrm{MHz}}
\safemath{\GHz}{\,\mathrm{GHz}}
\safemath{\s}{\,\mathrm{s}}
\safemath{\ms}{\,\mathrm{ms}}
\safemath{\mus}{\,\mathrm{\text{\textmu}s}}
\safemath{\ns}{\,\mathrm{ns}}
\safemath{\ps}{\,\mathrm{ps}}
\safemath{\meter}{\,\mathrm{m}}
\safemath{\mm}{\,\mathrm{mm}}
\safemath{\cm}{\,\mathrm{cm}}
\safemath{\m}{\,\mathrm{m}}
\safemath{\W}{\,\mathrm{W}}
\safemath{\mW}{\, \mathrm{mW}}
\safemath{\J}{\,\mathrm{J}}
\safemath{\K}{\,\mathrm{K}}
\safemath{\bit}{\,\mathrm{bit}}
\safemath{\nat}{\,\mathrm{nat}}

\safemath{\define}{=}			

\newcommand{\sothat}{\,:\,}				
\providecommand{\inner}[2]{\ensuremath{\langle#1,#2\rangle}}
\safemath{\equivalent}{\sim}
\safemath{\distas}{\sim}					
\safemath{\sdiff}{\Delta}				

\safemath{\reals}{\mathbb{R}}
\safemath{\positivereals}{\reals_{+}}
\safemath{\integers}{\mathbb{Z}}
\safemath{\posint}{\integers_{+}}
\safemath{\naturals}{\mathbb{N}}
\safemath{\posnaturals}{\naturals_{+}}
\safemath{\complexset}{\mathbb{C}}
\safemath{\rationals}{\mathbb{Q}}

\newcommand*{\fancyrefapplabelprefix}{app}		
\newcommand*{\fancyrefthmlabelprefix}{thm}		
\newcommand*{\fancyreflemlabelprefix}{lem}		
\newcommand*{\fancyrefcorlabelprefix}{cor}		
\newcommand*{\fancyrefdeflabelprefix}{def}		
\newcommand*{\fancyrefproplabelprefix}{prop}		
\frefformat{vario}{\fancyrefseclabelprefix}{Section~#1}
\frefformat{vario}{\fancyrefthmlabelprefix}{Theorem~#1}
\frefformat{vario}{\fancyreflemlabelprefix}{Lemma~#1}
\frefformat{vario}{\fancyrefcorlabelprefix}{Corollary~#1}
\frefformat{vario}{\fancyrefdeflabelprefix}{Definition~#1}
\frefformat{vario}{\fancyreffiglabelprefix}{Fig.~#1}
\frefformat{vario}{\fancyrefapplabelprefix}{Appendix~#1}
\frefformat{vario}{\fancyrefeqlabelprefix}{(#1)}
\frefformat{vario}{\fancyrefproplabelprefix}{Property~#1}

\newcommand{\covmat}[1]{\matR_{#1}}	
\safemath{\rndproc}{h}				
\safemath{\covmatalt}{\matA}			
\safemath{\rndvec}{\vech}			
\newcommand{\corrfun}[1]{r_{#1}}		
\safemath{\rndproccorr}{\corrfun{\rndproc}}	
\newcommand{\specdensity}[1]{c_{#1}}			
\newcommand{\mvspecdensity}[1]{\matC_{#1}}	
\safemath{\rndprocspec}{\specdensity{\rndproc}}	
\safemath{\rndveccovmat}{\covmat{\rndvec}}
\safemath{\measure}{\mu}				
\safemath{\rndev}{\lambda}			
\safemath{\rndsv}{\sigma}			
\safemath{\wpone}{\text{w.p.1}}		

\let\time\undefined
\safemath{\bandwidth}{W}			
\safemath{\effdou}{\tilde{\bandwidth}}	
\safemath{\time}{t}				
\safemath{\freq}{f}				
\safemath{\dtdf}{\dtime,\dfreq}  
\safemath{\Dtime}{\Delta\time}	
\safemath{\Dfreq}{\Delta\freq}	
\safemath{\dtime}{k}				
\safemath{\dfreq}{n}				
\safemath{\Ddtime}{\dtime}		
\safemath{\Ddfreq}{\dfreq}		
\safemath{\delay}{\tau}			
\safemath{\doppler}{\nu}			
\safemath{\maxDoppler}{\doppler_{0}}		
\safemath{\maxDelay}{\delay_{0}}			
\safemath{\spread}{\Delta_{\CHop}}		
\safemath{\tstep}{T}				
\safemath{\fstep}{F}				
\safemath{\tfstep}{\tstep\fstep}	
\safemath{\fslots}{N}			
\safemath{\tslots}{K}			
\safemath{\fslotsb}{\fslots^{'}}	
\safemath{\tslotsb}{\tslots^{'}}	
\safemath{\tfslots}{\tslots\fslots}	
\safemath{\tfslotsb}{\tslotsb\fslotsb}	
\safemath{\tfsamples}{\dtime\tstep,\dfreq\fstep}	
\safemath{\tfslotsbv}{\tslotsb\fslots} 

\safemath{\CHopapprox}{\widetilde{\CHop}}
\safemath{\kernel}{k_{\CHop}}			
\safemath{\kernelapprox}{k_{\CHopapprox}}
\safemath{\kernelp}{\kernel(\time,\time')	}	
\safemath{\kernelpapprox}{\kernelapprox(\time,\time')	}	
\let\tvir\undefined
\safemath{\tvir}{h_{\CHop}}				
\safemath{\tvirp}{\tvir(\time,\delay)}	
\let\tvtf\undefined
\safemath{\tvtf}{L_{\CHop}}				
\safemath{\tvtfp}{\tvtf(\time,\freq)}		
\let\spf\undefined
\safemath{\spf}{S_{\CHop}}				
\safemath{\spfp}{\spf(\doppler,\delay)}	
\safemath{\scafun}{C_{\CHop}}			
\safemath{\scafunp}{\scafun(\doppler,\delay)}	
\safemath{\pdep}{p_{\CHop}}				
\safemath{\pdepp}{\pdep(\delay)}			
\safemath{\pDop}{q_{\CHop}}				
\safemath{\pDopp}{\pDop(\doppler)}		
\safemath{\scafunpsq}{\scafun^{2}(\doppler,\delay)}	
\safemath{\scafunn}{\tilde{C}_{\CHop}}	
\safemath{\scafunnp}{\scafunn(\tilde{\doppler},\tilde{\delay})} 
\safemath{\pathloss}{\sigma_{\CHop}^{2}}	
\safemath{\ch}{h}						
\safemath{\wgn}{w}						
\safemath{\wgnvec}{\vecw}				
\safemath{\mvwgn}{\vecw}					
\safemath{\mvswgn}{\vecw}				
\safemath{\mvswgnb}{\widetilde{\mvswgn}}	
\safemath{\mvch}{\vech}					
\safemath{\mvsch}{\vech}					
\safemath{\mvschb}{\widetilde{\mvsch}}	
\safemath{\mvschcovmat}{\covmat{\mvch}}	
\safemath{\mvchcovmat}{\covmat{\mvsch}}	
\safemath{\mvchcovmatb}{\covmat{\mvschb}}		
\safemath{\chcovmatb}{\covmat{\mvschb}[0]}	
\safemath{\chcovmat}{\covmat{\mvsch}[0]}		
\safemath{\specparam}{\theta}			
\safemath{\altspecparam}{\varphi}		
\safemath{\chcorr}{R_{\CHop}}			
\safemath{\chcorrp}{\chcorr(\time,\freq)}    
\safemath{\chspecfun}{\specdensity{}}		
\safemath{\chspecfunoned}{\specdensity{}}	
\safemath{\chspecfunp}{\specdensity{}(\specparam,\altspecparam)}
\safemath{\mvchspecfun}{\mvspecdensity{}}	
\safemath{\mvchspecfunp}{\mvchspecfun(\specparam)}	
\safemath{\mvchspecfunentry}{c}			
\safemath{\mvchspecfunentryp}{\mvchspecfunentry_{\dfreq}(\specparam)}
\safemath{\peakiness}{\kappa_{\CHop}}		
\safemath{\leftsf}{u}					
\safemath{\rightsf}{v}					
\safemath{\ef}{z}						
\safemath{\logon}{g}						
\safemath{\eftime}{T_{0}}	
\safemath{\efband}{F_{0}} 
\safemath{\logonf}{G} 
\safemath{\logonfp}{\logonf(\freq)}
\safemath{\logonalt}{f}
\safemath{\logonp}{\logon(\time)}
\safemath{\logonaltp}{\logonalt(\time)}
\safemath{\slogonct}{\logon_{(\alpha,\beta)}}
\safemath{\slogonctalt}{\logon_{(\alpha',\beta')}}
\safemath{\slogon}{\logon_{\dtime,\dfreq}}	
\safemath{\af}{A}						
\safemath{\afp}{\af_{\logon}(\doppler,\delay)}	
\safemath{\rngspace}{\setR_{\CHop}}		
\let\nullspace\undefined
\safemath{\nullspace}{\setN_{\CHop}}		

\safemath{\inp}{x}				
\safemath{\altinp}{s}				
\safemath{\inpb}{\widetilde{\inp}}	
\safemath{\inpd}{x}				
\safemath{\inpmagd}{q}			
\safemath{\outp}{y}				
\safemath{\outpd}{y}				
\safemath{\nfoutp}{r}			
\safemath{\mvinp}{\vecx}			
\safemath{\mvoutp}{\vecy}		
\safemath{\mvsinp}{\vecx}		
\safemath{\mvsinpd}{\vecx}		
\safemath{\mvsinpmagd}{\vecq}	
\safemath{\cmiidinp}{u}			
\safemath{\mvscmiidinp}{\vecu}	
\safemath{\mvsoutp}{\vecy}		
\safemath{\mvsjinp}{\vecs}		
\safemath{\mvscminpb}{\vecd}		
\safemath{\mvsinpb}{\widetilde{\mvsinp}}		
\safemath{\mvsoutpb}{\widetilde{\mvsoutp}}	
\safemath{\ampfactor}{b}			
\safemath{\dutycycle}{\zeta}		
\safemath{\setxTpeak}{\setX}
\safemath{\altinpvec}{\vecs}

\safemath{\Pave}{P}				
\safemath{\Ppeak}{\papr\Pave}	
\safemath{\papr}{\beta}			
\safemath{\avP}{\Ex{}{\vecnorm{\mvsinp}^{2}}\le\tslots\Pave\,\tstep}
\safemath{\avPnorm}{(1/\tstep)\Ex{}{\vecnorm{\mvsinp}^{2}}\le\tslots\Pave}
\safemath{\pparam}{\alpha}		
\safemath{\avPeq}{\Ex{}{\vecnorm{\mvsinp}^{2}}=\pparam\tslots\Pave\,\tstep}
\safemath{\distset}{\setQ}		
\safemath{\dsetpapr}{\distset}
\safemath{\dsetpeaktime}{\setS}
\safemath{\dsetres}{\distset\rvert_{\pparam}}	
\safemath{\dsetpeaktimeres}{\dsetpeaktime\rvert_{\pparam}}
\safemath{\snrmax}{\overline{\SNR}}			
\safemath{\snrmaxp}{\snrmax(\spread,\papr)} 	

\safemath{\capacityp}{\capacity(\bandwidth)}	
\safemath{\infcapacity}{\capacity_{\infty}}	
\safemath{\ubcoh}{\mathrm{U}\sub{c}}			
\safemath{\ubcohp}{\ubcoh(\bandwidth)}		
\safemath{\ubTFpeak}{\mathrm{U}\sub{1}}		
\safemath{\ubTFpeakp}{\ubTFpeak(\bandwidth)}	
\safemath{\lbTFpeak}{\mathrm{L}_{1}}			
\safemath{\lbTFpeakp}{\lbTFpeak(\bandwidth)} 
\safemath{\lbapprox}{\mathrm{L}\sub{a}}		 
\safemath{\lbapproxp}{\lbapprox(\bandwidth)} 
\safemath{\lbaapprox}{\mathrm{L}\sub{aa}}
\safemath{\lbaapproxp}{\lbaapprox(\bandwidth)} 
\safemath{\lbTFpeakunopt}{\lbTFpeak_{\param}}	
\safemath{\lbTFpeakcf}{\mathrm{L}_{2}}		
\safemath{\lbTFpeakcfp}{\lbTFpeakcf(\bandwidth)}
\let\SNR\undefined
\safemath{\SNR}{\rho}

\safemath{\avpopt}{\alpha}						
\safemath{\avpoptp}{\alpha(\bandwidth)}			
\safemath{\ubpenalty}{A}							
\safemath{\ubpenaltyp}{\ubpenalty(\bandwidth)}	
\safemath{\approxTFpeak}{\tilde{\capacity}}		
\safemath{\taylorzero}{c_{0}}					
\safemath{\taylorone}{c}							
\safemath{\tayloronelb}{\underline{c}}   			
\safemath{\lbpenalty}{B}		      				
\safemath{\lbpenaltyp}{\lbpenalty(\bandwidth)}	
\safemath{\lbpenaltylb}{\underline{\lbpenalty}}	
\safemath{\lbpenaltylbp}{\lbpenaltylb(\bandwidth)}	
\safemath{\lbpenaltyub}{\overline{\lbpenalty}}		
\safemath{\lbpenaltyubp}{\lbpenaltyub(\bandwidth)}	
\safemath{\lbTpeak}{\mathrm{L}_{\infty}}			
\safemath{\lbTpeakp}{\lbTpeak}					

\safemath{\ubTpeak}{\mathrm{U}_{\infty}}			
\safemath{\ubTpeakp}{\ubTpeak}					

\renewcommand{\setminus}{-}						
\safemath{\limintime}{\lim_{\tslots\to\infty}}	
\safemath{\liminfreq}{\lim_{\fslots\to\infty}}	
\safemath{\liminbw}{\lim_{\bandwidth\to\infty}}	
\safemath{\imat}{\matidentity}
\newcommand{\diag}[1]{\mathrm{diag}\lefto(#1\right)}	
\newcommand{\spreadint}[1]{\iint_{\doppler\:\delay} #1 d\delay d\doppler}
\newcommand{\spreadintlog}[1]{\spreadint{\log\lefto( #1 \right)}}
\newcommand{\dopplerint}[1]{\int_{\doppler} #1 d\doppler}
\newcommand{\delayint}[1]{\int_{\delay} #1 d\delay}
\newcommand{\dopplerintlog}[1]{\dopplerint{\log\lefto( #1 \right) } }
\newcommand{\iset}{\setM}						
\newcommand{\cex}[1]{e^{\iu2\pi #1}}				
\newcommand{\cexn}[1]{e^{-\iu2\pi #1}}			
\safemath{\supavp}{\sup_{0\le\pparam\le 1}}		
\safemath{\param}{\gamma}						
\newcommand{\osmall}[1]{\landauo\lefto(#1\right)}	
\safemath{\indvar}{i}							
\safemath{\altindvar}{j}							
\safemath{\aerr}{\epsilon}						
\safemath{\diagelcirc}{d}						
\safemath{\diagelcirci}{\diagelcirc_{\indvar}}
\safemath{\diagelcircp}{\diagelcirci(\specparam)}
\safemath{\toep}{t}
\safemath{\spec}{s}
\safemath{\specp}{\spec{(\altspecparam})}
\safemath{\taylexp}{L}
\safemath{\taylexpindex}{l}

\let\esssup\undefined
\safemath{\esssup}{p}
\safemath{\genfun}{f}
\safemath{\genfunp}{\genfun(1/\fslots)}
\safemath{\modtime}{l}
\safemath{\modfreq}{m}
\safemath{\modtimeindex}{p}
\safemath{\modfreqindex}{q}

\safemath{\numnz}{M}					
\safemath{\dftmat}{\matF}			
\safemath{\dftcol}{\vecf}			
\safemath{\subcindex}{i}				
\newcommand{\kulleib}[2]{D\lefto(\cdf{#1}\|\cdf{#2} \right) }
\newcommand{\kulleibprob}[2]{D\lefto(#1\lefto\|\right.#2\right) }
\safemath{\nepero}{e}

\newtheorem{thm}{Theorem}
\newtheorem{lem}[thm]{Lemma}
\newtheorem{rem}{Remark}
\newtheorem{cor}[thm]{Corollary}
\newtheorem{dfn}[thm]{Definition}
\newtheorem{prop}{Property}

\safemath{\Sfun}{S_{\fslots}} 
\safemath{\Sfunplus}{S_{\fslots+1}}
\safemath{\Sfunp}{\Sfun(\mvswgn)}
\safemath{\sfun}{s_{\subcindex}}
\safemath{\sfunp}{\sfun(\mvswgn)}
\safemath{\Vfun}{V_{\fslots}}
\safemath{\Vfunp}{\Vfun(\mvswgn)}
\safemath{\Vfunonep}{V_{1}(\mvswgn)}
\safemath{\sigmaSet}{\setG_{\fslots}}
\safemath{\sfunzero}{s_{0}}
\safemath{\sfunzerop}{\sfunzero(\mvswgn)}
\newcommand{\fun}[1]{r\lefto(#1\right)}
\safemath{\Vrv}{v}

\newcommand{\pulsersii}{{{\sc Pulsers} Phase~II}\xspace}

\begin{document}

\IEEEoverridecommandlockouts
\pdfinfo{
	/Title		(Noncoherent Capacity of Underspread Fading Channels)
	/Author		(Giuseppe Durisi, Ulrich Schuster, Helmut Boelcskei,	Shlomo
				Shamai (Shitz))
	/Subject		(SVN revision , 2371 \today)
	/Keywords	(test)
}

\title{Noncoherent Capacity of Underspread Fading Channels}
\author{Giuseppe~Durisi,~\IEEEmembership{Member,~IEEE,}
             Ulrich~G.~Schuster,~\IEEEmembership{Student~Member,~IEEE,}
			Helmut~B\"{o}lcskei,~\IEEEmembership{Senior~Member,~IEEE,}
			Shlomo~Shamai~(Shitz),~\IEEEmembership{Fellow,~IEEE}%
\thanks{This work was supported in part by the Swiss
{\em Kommission f\"ur Technologie und Innovation (KTI)} under grant
6715.2~ENS-ES, and by the European Commission as part of the Integrated Project
\pulsersii under contract~FP6-027142, and as part of the FP6 Network of
Excellence~NEWCOM.}%
\thanks{G. Durisi and H. B\"{o}lcskei are with the
Communication  Technology Laboratory, ETH Zurich, 8092 Zurich, Switzerland
(e-mail: \{gdurisi, boelcskei\}@nari.ee.ethz.ch).}%
\thanks{U. G. Schuster  was with the Communication  Technology Laboratory, ETH Zurich, and is now with Celestrius AG, Zurich, Switzerland.}
\thanks{S. Shamai (Shitz) is with Technion, Israel Institute of Technology,
32000 Haifa, Israel (e-mail: sshlomo@ee.technion.ac.il).}
\thanks{
This paper was presented in part at the IEEE International Symposium on Information Theory, Seattle, WA, U.S.A., July 2006, and at the IEEE International Symposium on Information Theory, Nice, France, June 2007.}}%
\maketitle

\begin{abstract}
We derive bounds on the noncoherent capacity of wide-sense stationary uncorrelated scattering~(WSSUS) 
channels that are selective both in time and  frequency, and are underspread, i.e., the product of the 
channel's delay spread and Doppler spread is small. 
For input signals that are peak constrained in time and frequency,
we obtain upper and lower bounds on capacity that are explicit in the channel's scattering function, are accurate for
a large range of bandwidth and allow to coarsely identify the capacity-optimal bandwidth
as a function of the peak power and the channel's scattering function. We also obtain
a closed-form expression for the first-order Taylor series expansion of capacity in the limit of
large bandwidth, and show that our bounds are tight in the wideband regime. 
For input signals that are peak constrained in time only (and, hence, allowed to be
peaky in frequency), we provide
upper and lower bounds on the infinite-bandwidth capacity and find cases when the bounds
coincide and the infinite-bandwidth capacity is characterized exactly. Our lower bound is closely related to a result by Viterbi~(1967).

The analysis in this paper is based on a discrete-time discrete-frequency approximation of 
WSSUS  time- and frequency-selective channels. This discretization explicitly takes
into account the underspread property, which is satisfied by virtually all wireless communication channels.
\end{abstract}

\section{Introduction and Outline}
\label{sec:introduction}
\subsubsection{Models for fading channels}
Channel capacity is a benchmark for the design of any communication system. 
The techniques used to compute, or at least to bound, channel capacity often
provide guidelines for the design of practical systems, e.g., how to best
utilize the resources bandwidth and power, and how to design efficient
modulation and coding schemes~\cite[Sec.~III.3]{biglieri98-10a}. 
Our goal in this paper is to analyze the capacity of wireless communication
channels that are of direct practical importance. We believe that an accurate
stochastic model for such channels should take the following aspects into
account:
\begin{itemize}
\item The channel is selective in time and frequency, i.e., it exhibits memory in frequency and in time, respectively.
\item Neither the transmitter nor the receiver knows the instantaneous
	realization of the channel.
\item The peak power of the input signal is limited.
\end{itemize}
These aspects are important because they arise from practical limitations
of real-world communication systems: temporal variations of the environment and
multipath propagation are responsible for channel selectivity in time and
frequency, respectively~\cite{vaughan03a,tse05a}; perfect channel knowledge at
the receiver is impossible to obtain because channel state information needs to be extracted from the received signal; finally, realizable transmitters are always
limited in their peak output power~\cite{gray01a}.
The above aspects are also  fundamental as they significantly impact
the behavior of channel capacity: for example, the capacity of a block-fading
channel behaves differently from the capacity of a channel that is stationary in
time~\cite{lapidoth05-07a}; channel capacity with perfect channel knowledge
at the receiver is always larger than the capacity without channel
knowledge~\cite{medard00-05a}, and the signaling schemes necessary to achieve
capacity are also very different in the two
cases~\cite{biglieri98-10a}; finally, a peak constraint on the transmit signal can
lead to vanishing capacity in the large-bandwidth limit~\cite{telatar00-07a,medard02-04a,subramanian02-04a}, while without a peak constraint the infinite-bandwidth AWGN capacity can be attained
asymptotically~\cite{telatar00-07a,jacobs63-01a,pierce66-01a,kennedy69,gallager68a,verdu02-06a,durisi06-07a}. 

Small scale fading of wireless channels can be sensibly modeled as a stochastic Gaussian
linear time-varying~(LTV) system~\cite{vaughan03a};  in
particular, we base our developments on the widely used wide-sense stationary
uncorrelated scattering~(WSSUS) model for random
LTV~channels~\cite{bello63-12a,kennedy69}. Like most models for real-world
channels, the WSSUS~model is time continuous; however, almost all tools for
information-theoretic analysis of noisy channels require a discretized
representation of the channel's input-output relation. Several approaches to
discretize random LTV~channels are proposed in the literature, e.g.,
sampling~\cite{medard02-04a,bello63-12a,sayeed99-01a} or  basis 
expansion~\cite{ma03-07a,zemen05-09a}; all these discretized models incur an
approximation error with respect to the continuous-time WSSUS~model that is
often difficult to quantify. As virtually all wireless channels of practical interest are {\em
underspread}, i.e., the product of maximum delay and maximum Doppler shift is
small, we build our information-theoretic analysis upon a discretization of
LTV~channels, proposed by Kozek~\cite{kozek97a}, that explicitly takes into
account the underspread property to minimize the approximation error in the 
mean-square sense.

\subsubsection{Capacity of noncoherent WSSUS channels}
Throughout  the paper, we assume that both the transmitter and 
receiver know the channel law\footnote{This implies that the codebook and the
decoding strategy can be optimized accordingly~\cite{lapidoth98-10a}.} but both
are ignorant of the channel realization, a setting often called {\em noncoherent}. In the following, we refer to channel capacity in the noncoherent setting simply as ``capacity''. In contrast, in the coherent setting the receiver is also assumed to know the channel realization perfectly; the corresponding capacity is termed coherent capacity.

A general closed-form expression for the capacity of Rayleigh-fading channels is
not known, even if the channel is memoryless~\cite{abou-faycal01-05a}. However,
several asymptotic results are available. If only a constraint on the average
transmitted power is imposed, the  AWGN capacity can be
achieved in the infinite-bandwidth limit also in the presence of fading. This result is quite robust, as it
holds for a wide variety of channel
models~\cite{telatar00-07a,jacobs63-01a,pierce66-01a,kennedy69,gallager68a,verdu02-06a,durisi06-07a}.
Verd\'u showed that {\em flash signaling}, which implies unbounded peak power of the
input signal, is necessary and sufficient to achieve the infinite-bandwidth
AWGN~capacity on block-memoryless fading channels~\cite{verdu02-06a}; a form of flash
signaling is also infinite-bandwidth optimal for the more general time- and
frequency-selective channel model used in the present
paper~\cite{durisi06-07a}. In contrast, if the peakiness of the input signal is restricted,
the infinite-bandwidth capacity behavior of most fading channels changes drastically,
and the limit depends on the type
of peak constraint imposed~\cite{telatar00-07a,medard02-04a,subramanian02-04a,gallager68a,viterbi67-07a}. In this paper,
we shall distinguish between a peak constraint in time and a peak constraint in time
and frequency.

\paragraph{Peak constraint in time}
No closed-form capacity expression, not even in the infinite-bandwidth limit,
seems to exist to date for time- and frequency-selective WSSUS channels.
Viterbi's analysis~\cite{viterbi67-07a} provides a result that can be {\em interpreted} as a lower bound on the
infinite-bandwidth capacity of time- and frequency-selective channels. This lower bound is in the
form of the infinite-bandwidth AWGN~capacity minus a penalty term that
depends on the channel's power-Doppler profile~\cite{bello63-12a}.
For channels that are time selective but frequency flat, structurally similar
expressions were found for the infinite-bandwidth
capacity~\cite{sethuraman06-07a,zhang07-01a} and for the capacity per unit
energy~\cite{sethuraman05-09a}.

\paragraph{Peak constraint in time and frequency}
Although a closed-form capacity expression valid for all bandwidths is not
available, it is known that the infinite-bandwidth capacity is zero for various
channel models~\cite{telatar00-07a,medard02-04a,subramanian02-04a}.
This asymptotic capacity behavior implies that signaling schemes that
spread the transmit energy uniformly across time and frequency perform poorly
in the  large-bandwidth regime. Even more useful for performance assessment would be capacity bounds for finite bandwidth. For frequency-flat
time-selective channels, such bounds can be found in~\cite{sethuraman05-09b,sethuraman08a},
while for the more general time- and frequency-selective case treated in the
present paper, upper bounds seem to exist only on the rates achievable with
particular signaling schemes, namely for  orthogonal
frequency-division multiplexing~(OFDM) with constant-modulus symbols~\cite{schafhuber04-06a}, and for 
multiple-input multiple-output~(MIMO) OFDM with unitary space-frequency codes over
frequency-selective block-fading channels~\cite{borgmann05-09a}.

\subsubsection{Contributions}

We use the discrete-time discrete-frequency approximation of
continuous-time underspread WSSUS channels proposed in~\cite{kozek97a}, to obtain the
following results:
\begin{itemize}
\item We derive upper and lower bounds on capacity under a constraint on the
	average power and under a peak constraint in both time and frequency. These
	bounds are valid for any bandwidth, are explicit in the channel's
	scattering function, and generalize the results on achievable rates
	in~\cite{schafhuber04-06a}. In particular, our bounds allow to coarsely
	identify the capacity-optimal bandwidth for a given peak constraint and a
	given scattering function.
\item Under the same peak constraint in time and frequency, we find the
	first-order Taylor series expansion of channel capacity in the limit of
	infinite bandwidth. This result extends the asymptotic capacity analysis
	for frequency-flat time-selective channels
	in~\cite{sethuraman08a} to channels that are selective in
	both time and frequency.
\item In the infinite-bandwidth limit and for transmit signals that are
	peak-constrained in time only, we recover Viterbi's capacity lower
	bound~\cite{viterbi67-07a}. In addition, we derive an upper bound that
	is shown to coincide with the lower bound for a specific class of channels; hence,
	the infinite-bandwidth capacity for this class of channels is established.
\end{itemize}

The results in this paper rely on several flavors of
Szeg\"o's theorem on the asymptotic eigenvalue distribution of Toeplitz
matrices~\cite{grenander84a,gray05a}; in particular, we use various extensions
of Szeg\"o's theorem to {\em two-level Toeplitz} matrices, i.e., block-Toeplitz
matrices that have Toeplitz
blocks~\cite{voois96a,miranda00-02a}.
Another key ingredient for several of our proofs is the relation between mutual
information and minimum mean-square error~(MMSE) discovered recently by Guo
{\it et al.}~\cite{guo05-04a}. Furthermore, we use a property of the information
divergence of orthogonal signaling schemes derived by Butman and
Klass~\cite{butman73-09a}.

\subsubsection{Notation}

Uppercase boldface letters denote matrices and lowercase boldface letters designate vectors. 
The superscripts~$\tp{}$, $\conj{}$, and~$\herm{}$ stand for transposition,
element-wise conjugation, and Hermitian transposition, respectively.
For two matrices~\matA and~\matB of appropriate dimensions, 
the {\em Hadamard product} is denoted as~$\matA\had\matB$.
We designate the identity matrix of dimension~$\fslots\times\fslots$
as~$\imat_{\fslots}$ and the all-zero vector of appropriate dimension as~\veczero.
We let~$\diag{\vecx}$ denote a diagonal square matrix whose main diagonal contains
the elements of the
vector~\vecx. The determinant, trace, and rank of the
matrix~\matX are denoted as~$\det(\matX)$, $\tr(\matX)$, and~$\rank(\matX)$,
respectively, and~$\ev_{\indvar}(\matX)$ is the $i$th eigenvalue of a square
matrix~\matX. The function~$\dirac(x)$ is the Dirac distribution,
and~$\krond[n]$ is defined as~$\dirac[0]=1$ and~$\dirac[n]=0$ for all~$n\neq 0$.
All logarithms are to the base~\nepero. The real part of the complex number~$z$
is denoted~$\Re\{z\}$. We write~$\setA\setminus\setB$ for the set difference
between the sets~\setA and~\setB. For two functions~$f(x)$ and~$g(x)$, the
notation~$f(x) = \osmall{g(x)}$ for~$x\to0$ means that~$\lim_{x \to 0}f(x)/
g(x)= 0$. With~$\floor{x}$ we denote the largest integer smaller or equal to~$x \in \reals$. A {\em signal} is an element of the Hilbert space~\hilfunspace of
square integrable functions. The inner product between two signals~$f(x)$
and~$g(x)$ is denoted as~$\inner{f}{g}=\int_{-\infty}^{\infty}f(x)\conj{g}(x)
dx$. For a random variable~(RV)~$x$ with distribution~$\cdf{x}$, we
write~$x\distas\cdf{x}$. We denote expectation by~$\Ex{}{\cdot}$, and use the
notation~$\Ex{x}{\cdot}$ to stress that the expectation is taken with
respect to the RV~$x$. We write~$\kulleib{x}{y}$ for the
Kullback-Leibler~(KL) divergence between the two distributions~$\cdf{x}$
and~$\cdf{y}$. Finally, $\jpg(\bmm,\covmat{})$ stands for the distribution of a
jointly proper Gaussian (JPG) random vector with mean~\bmm and covariance
matrix~$\covmat{}$.

\section{Channel and System Model}
\label{sec:model}

A channel model needs to strike a balance between generality, accuracy,
engineering relevance, and mathematical tractability. In the following, we start
from the classical WSSUS~model for LTV~channels~\cite{bello63-12a,kennedy69}
because it is a fairly general, yet accurate and mathematically tractable model that is widely
used. This model has a continuous-time input-output relation, which is difficult
to use as a basis for information-theoretic studies. However, if the channel is
{\em underspread} it is possible to closely approximate the original~WSSUS
input-output relation by a discretized input-output relation that is especially suited for the
derivation of capacity bounds. In particular, the bounds we
derive in this paper can be
directly related to the underlying continuous-time WSSUS~channel as they are
explicit in its scattering function.

\subsection{Time- and Frequency-Selective Underspread Fading Channels}
\label{sec:wssus-model}
\subsubsection{The channel operator}
A wireless channel can be described as a linear 
operator~$\CHop :\hilfunspace\to\rngspace$ that maps an input signal~$\inp(\time)$ into an output signal~$\nfoutp(\time)\in\rngspace$, where~$\rngspace\subset \hilfunspace$ denotes the range space of~$\CHop$~\cite{naylor82a}.
The corresponding noise-free input-output relation is then~$\nfoutp(\time)=
(\CHop\!\inp)(\time)$.

It is sensible to model wireless channels as random, for one because a
deterministic description of the physical propagation environment is too complex
in most cases of practical interest, and second because a stochastic description
is much more robust, in the sense that systems designed on the basis of a
stochastic channel model can be expected to work in a variety of different
propagation environments~\cite{tse05a}. Consequently, we assume that~$\CHop$ is
a random operator.

\subsubsection{System functions}
Because communication takes
place over a finite bandwidth and a finite time duration, we can assume that
each realization of~$\CHop$ is a Hilbert-Schmidt
operator~\cite{dunford63a,lax02a}. Hence, the noise-free input-output relation
of the LTV~channel can be written as\footnote{All integrals are from~$-\infty$ to~$\infty$ unless stated otherwise.}\cite[p.~1083]{dunford63a}
\ba
	\nfoutp(\time) = \bigl(\CHop\!\inp\bigr)(\time)=\int_{\time'}\kernelp
		\inp(\time')d\time'
\label{eq:ltv-kernel-io}
\ea
where the {\em kernel}~\kernelp can be interpreted as the channel response at
time~\time to a Dirac impulse at time~$\time'$. Instead of two variables that
denote absolute time, it is common in the engineering literature to use absolute
time~\time and delay~\delay. This leads to the {\em time-varying impulse
response}~$\tvirp\define\kernel(\time,\time-\delay)$ and the corresponding
noise-free input-output relation~\cite{bello63-12a}
\ba
	\nfoutp(\time) = \int_{\delay}\tvirp\inp(\time-\delay)d\delay.
\label{eq:ltv-io}
\ea
Two more {\em system functions} that will be important in the following
developments are the {\em time-varying transfer function}\footnote{
As~$\CHop$ is of Hilbert-Schmidt type, the time-varying impulse response~\tvirp is square integrable, and the Fourier transforms in~\eqref{eq:transfer function}
and~\eqref{eq:spreading function} are well defined.}
\ba\label{eq:transfer function}
	\tvtfp &\define\int_{\delay}\tvirp\cexn{\freq\delay}d\delay
\ea
and the  {\em spreading function}
\ba\label{eq:spreading function}
	\spfp &\define\int_{\time}\tvirp\cexn{\doppler\time}d\time 
		= \iint_{\time\:\freq}\tvtfp\cexn{(\doppler\time-\delay\!\freq)} 
		d\time d\freq.
\ea
In particular, if we rewrite the input-output relation~\fref{eq:ltv-io} in
terms of the spreading function~\spfp as
\ba
	\nfoutp(\time)=\spreadint{\spfp\inp(\time-\delay)\cex{\time\doppler}}
\label{eq:ltv-altio}
\ea
we obtain an intuitive physical interpretation: the output
signal~$\nfoutp(\time)$ is a weighted superposition of copies of the input
signal~$\inp(\time)$ that are shifted in time by the delay~\delay and in
frequency by the {\em Doppler shift}~\doppler.

\subsubsection{Stochastic characterization and WSSUS assumption}
For mathematical tractability, we need to make additional assumptions
on the system functions. First,  we assume that~\tvtfp is a zero-mean JPG random
process in~\time and~\freq. Indeed, the Gaussian distribution is
empirically supported for narrowband channels~\cite{vaughan03a}, and even
ultrawideband (UWB) channels with bandwidth up to several gigahertz can be modeled as
Gaussian distributed~\cite{schuster07-07a}.
By virtue of the Gaussian assumption,~\tvtfp is completely characterized by its
correlation function. Yet, this correlation function is four-dimensional in general and
thus difficult to work with. A further simplification is possible if we assume
that the channel process is {\em wide-sense stationary} in time~\time and {\em
uncorrelated} in delay~\delay, the so-called WSSUS
assumption~\cite{bello63-12a}. As a consequence, \tvtfp is wide-sense
stationary both in time~\time and frequency~\freq, or, equivalently, \spfp is uncorrelated in
Doppler~\doppler and delay~\delay~\cite{bello63-12a}:
\bas
	\Ex{}{\tvtfp\conj{\tvtf}(\time',\freq')}&=\chcorr(\time-
		\time',\freq-\freq')\\
	\Ex{}{\spfp\conj{\spf}(\doppler',\delay')}&=\scafunp\dirac(\doppler-
		\doppler')\dirac(\delay-\delay').
\eas
The function~\chcorrp is called the channel's (time-frequency)
correlation function, and~\scafunp is called the {\em scattering function} of the
channel~$\CHop$. The two functions are related by a two-dimensional Fourier
transform,
\ba
\label{eq:scafun-chcorr}
	\scafunp =\iint_{\time\:\freq}\chcorr(\time,\freq)\cexn{(\doppler \time -
		\delay\!\freq)}d\time d\freq.
\ea
As~$\chcorr(\time,\freq)$ is stationary in~$\time$ and~$\freq$, \scafunp is nonnegative and real-valued for all~\doppler
and~\delay, and can be interpreted as the spectrum of the
channel process.  The {\em power-delay profile} of~$\CHop$ is defined as
\bas
	\pdepp&\define\int_{\doppler}\scafunp d\doppler
\intertext{and the {\em power-Doppler profile} as}
	\pDopp&\define\int_{\delay}\scafunp d\delay.
\eas
The WSSUS assumption is widely used in wireless channel
modeling~\cite{bello63-12a,kennedy69,vaughan03a,biglieri98-10a,proakis01,matz03a}. It is in
good agreement with measurements of tropospheric scattering channels~\cite{kennedy69}, and provides a reasonable model for many types of mobile radio
channels~\cite{cox73-04a,cox73-11a,jakes74a}, at least over a limited time
duration and bandwidth~\cite{bello63-12a}. Furthermore, the scattering
function can be directly estimated from measured
data~\cite{gaarder68-09a,artes04-05a}, so that capacity expressions and bounds
that explicitly depend on the channel's scattering function can be evaluated for
many channels of practical interest.

Formally, the WSSUS assumption is mathematically incompatible with the requirement that~$\CHop$
is of Hilbert-Schmidt type, or, equivalently, that
the system functions are square integrable, because stationarity in time~\time
and frequency\freq of~\tvtfp implies that~\tvtfp cannot decay to zero
for~$\time\to\infty$ and~$\freq\to\infty$. Similarly to the engineering model of
white noise, this incompatibility is a mathematical artifact and not a problem
of real-world wireless channels: in fact, every communication system transmits over a
finite time duration and over a finite bandwidth.\footnote{A more detailed account
on solutions to overcome the mathematical incompatibility between stationary and finite-energy models
can be found in~\cite[Sec. 7.5]{gallager08a}.} We believe that the
simplification the WSSUS assumption entails justifies this mathematical
inconsistency.

\subsection{The Underspread Assumption and its Consequences}
\label{sec:underspread}

Because the velocity of the transmitter, of the receiver, and of the objects in
the propagation environment is limited, so is the maximum Doppler shift~\maxDoppler
experienced by the transmitted signal. We also assume that the maximum delay is
strictly smaller than~$2\maxDelay$. For simplicity and without loss of
generality, throughout this paper, we consider scattering functions that are
centered at~$\delay=0$ and~$\doppler=0$, i.e., we remove any overall fixed delay
and Doppler shift. The assumptions of limited Doppler shift and delay then imply
that the scattering function is supported on a rectangle of
{\em spread}~$\spread\define 4\maxDoppler\maxDelay$,
\ba\label{eq:support condition}
	\scafunp=0\quad\text{for }(\doppler,\delay)\notin[-\maxDoppler,\maxDoppler]
		\times[-\maxDelay,\maxDelay].
\ea
Condition~\eqref{eq:support condition} in turn implies that the spreading function~\spfp is also
supported on the same rectangle with probability~1~(\wpone). If~$\spread<1$, the
channel is said to be {\em underspread}~\cite{bello63-12a,kennedy69,kozek97a}.
Virtually all channels in wireless communication are highly underspread,
with~$\spread\approx 10^{-3}$ for typical land-mobile channels and as low
as $10^{-7}$ for some indoor channels with restricted mobility of
the terminals~\cite{hashemi93-07a,parsons00a,rappaport02}. The underspread
property of typical wireless channels is very important, first because only
(deterministic) underspread channels can be completely identified from
measurements~\cite{kailath63-10a,pfander06-11a}, and second because underspread
channels have a well-structured set of approximate eigenfunctions that can be
used to discretize the channel operator, as described next.

\subsubsection{Approximate diagonalization of underspread channels}
\label{sec:diagonalization}

As~$\CHop$ is a Hilbert-Schmidt operator, its kernel can be expressed in terms
of its positive singular values~$\{\rndsv_{\indvar}\}$, its left singular
functions~$\{\leftsf_{\indvar}(\time)\}$, and its right singular
functions~$\{\rightsf_{\indvar}(\time)\}$~\cite[Th. 6.14.1]{naylor82a}, according to
\ba\label{eq:svd}
	\kernel(\time,\time')=\sum_{\indvar=-\infty}^{\infty}\rndsv_{\indvar}
		\leftsf_{\indvar}(\time)\conj{\rightsf}_{\indvar}(\time').
\ea
We denote by~\nullspace the null space of~$\CHop$, i.e., the space of input
signals that the channel maps onto~$0$. The set~$\{\rightsf_{\indvar}(\time)
\}$ is an orthonormal basis for the linear span of~$\hilfunspace\setminus\nullspace$, and~$\{\leftsf_{\indvar}(\time)\}$ is an orthonormal basis for the
range space~\rngspace.
Any input signal  in~\nullspace is of no utility for communication purposes;
the remaining input signals in the linear span of~$\hilfunspace  \setminus \nullspace$, which we denote
in the remainder of the paper as {\em input space}, can be
completely characterized by their projections onto the set~$\{\rightsf_{\indvar}(\time)
\}$. Similarly, the output signal~$\nfoutp(\time)=(\CHop\inp)(\time)$ is completely
described by its projections onto the set~$\{\leftsf_{\indvar}(\time)\}$.
These projections together with the kernel decomposition~\fref{eq:svd} yield a
countable set of scalar input-output relations, which we refer to as the {\em diagonalization} of~$\CHop$.

Because the right and left singular functions depend on the realization
of~$\CHop$, diagonalization requires perfect channel knowledge. But
this knowledge is not available in the noncoherent setting. In contrast, if the
singular functions of the random channel~$\CHop$ did not depend on its
particular realization, we could diagonalize~$\CHop$ without
knowledge of the channel realization. This is the case, for example, for random
linear time-invariant (LTI) channels, where complex sinusoids are always
eigenfunctions, independently of the realization of the channel's impulse response. Fortunately, the singular
functions of underspread random LTV~channels can be well approximated by
deterministic functions. More precisely, an underspread channel~$\CHop$ has the
following properties~\cite{kozek97a}:
\begin{enumerate}
\item All realizations of the underspread channel~$\CHop$ are approximately
	{\em normal}, so that the singular value decomposition~\fref{eq:svd} can be
	replaced by an eigenvalue decomposition.
\item Any deterministic unit-energy signal~\logonp that is well
	localized\footnote{We measure the joint time-frequency localization of a
	signal~\logonp by the product between its  {\em effective duration}
	and its {\em effective bandwidth}, defined in~\eqref{eq:efftime and bandwidth}.} in time and frequency is an
	{\em approximate eigenfunction} of~$\CHop$ in the mean-square sense, i.e.,
	the mean-square error $\Ex{}{\vecnorm{\inner{\CHop\logon}
	{\logon}	\logon-\CHop	\logon}^{2}}$ is small if~$\CHop$ is underspread.
	This error can be further reduced by an appropriate choice of~\logonp, where
	the choice depends on the scattering function~\scafunp.
\item If~\logonp is an approximate eigenfunction as defined in the previous point, then so
	is~$\slogonct(\time)\define\logon(\time-\alpha)\cex{\beta\time}$ for any
	time shift~$\alpha\in\reals$ and any frequency shift~$\beta\in\reals$.
\item For any~$(\alpha,\beta)$, the time-varying transfer function
	$\tvtf(\alpha,\beta)$ is an {\em approximate eigenvalue} of~$\CHop$
	corresponding to the approximate eigenfunction~$\slogonct(\time)$, in the
	sense that the mean-square error~$\Exop\bigl[\abs{\inner{\CHop\slogonct}
	{\slogonct}-\tvtf(\alpha,\beta)}^{2}\bigr]$ is small.
\end{enumerate}
We use these properties of underspread operators to construct an approximation~\CHopapprox
of the random channel~$\CHop$ that has a well-structured set of deterministic
eigenfunctions. The errors incurred by this approximation are discussed in detail in \fref{app:ch-approx-error}. We then diagonalize this approximating
operator and exclusively consider the corresponding discretized input-output
relation in the reminder of the paper. Property~1, the approximate normality
of~$\CHop$, together with Property~2 implies that the kernel of the
approximating operator~\CHopapprox can be synthesized as
$
	\sum_{\indvar=-\infty}^{\infty}\rndev_{\indvar}\ef_{\indvar}(\time)
		\conj{\ef}_{\indvar}	(\time'),
$
where, differently from~\fref{eq:svd}, the~$\rndev_{\indvar}$ are now random
eigenvalues instead of random singular values, and the~$\ef_{\indvar}(\time)$
constitute a set of deterministic orthonormal eigenfunctions instead of random
singular functions. Property~2 means that we are at liberty to choose the approximate eigenfunctions~$\ef_{\indvar}(\time)$ among all signals that are well localized
in time and frequency. In particular, we would like the resulting approximating kernel to be convenient to work with
and the approximate eigenfunctions~$\ef_{\indvar}(\time)$ easy to implement, as discussed in~\fref{sec:ofdm-interpretation}; therefore, we choose the set of
approximate eigenfunctions to be highly structured. By Property~3, it is
possible to use time- and frequency-shifted versions of a single well-localized
prototype function~\logonp as eigenfunctions. Furthermore, because the support
of~\spfp is strictly limited in Doppler~\doppler and delay~\delay, it follows
from the sampling theorem and the Fourier transform relation~\fref{eq:spreading function} that the
samples~$\tvtf(\tfsamples)$, taken on a rectangular grid with~$\tstep\le1/(2
\maxDoppler)$ and~$\fstep\le1/(2\maxDelay)$, are sufficient to characterize
$\tvtfp$ exactly. Hence, we take as our set of approximate eigenfunctions the so-called
{\em Weyl-Heisenberg set}~$\{\slogon(\time)\}$, where~$\slogon(\time)\define
\logon(t-\dtime\tstep)\cex{\dfreq\fstep\time}$ are orthonormal signals. The
requirement that the~$\slogon(\time)$ are orthonormal and at the same time well
localized in time and frequency implies~$\tfstep>1$~\cite{christensen03a}, as a
consequence of the Balian-Low theorem~\cite[Ch.~8]{groechenig01a}. 
Large values of the product~\tfstep
allow for better time-frequency localization of~\logonp, but result in a loss of
dimensions in signal space compared with the critically sampled case~$\tfstep=
1$. The Nyquist condition~$\tstep\le1/(2\maxDoppler)$ and~$\fstep\le1/(2
\maxDelay)$  can be readily satisfied for all underspread channels.

The samples~$\tvtf(\tfsamples)$ are approximate eigenvalues of~$\CHop$ by
Property~4; hence, our choice of approximate eigenfunctions results in the
following approximating eigenvalue decomposition for~\kernelp
\ba
	\kernelp\approx\kernelpapprox=\sum_{\dtime=-\infty}^{\infty}\sum_{\dfreq=
		-\infty}^{\infty}\tvtf(\tfsamples)\slogon(\time)\conj{
		\slogon}(\time')
\label{eq:approx-kernel}
\ea
where~\kernelpapprox denotes the kernel of the approximating operator~\CHopapprox.
For~$\tfstep>1$, the Weyl-Heisenberg set~$\{\slogon(\time)\}$ is not complete in~\hilfunspace~\cite[Th.~8.3.1]{christensen03a}. Therefore, the null space of~\CHopapprox is nonempty. As~\kernelpapprox
 is only an approximation of~\kernelp, this null space
might differ from~\nullspace. Similarly, the range space of~\CHopapprox might differ from~\rngspace. The
characterization of the difference between these spaces is an important open
problem.

\subsubsection{Canonical characterization of signaling schemes}
The approximating random channel operator~\CHopapprox
has a highly structured set of deterministic orthonormal eigenfunctions.
We can, therefore, diagonalize the input-output relation of the approximating channel without the need for channel knowledge at both transmitter and
receiver. Any input signal~$\inp(\time)$ that lies in the input space of the
approximating operator is uniquely characterized by its projections onto
the set~$\{\slogon(\time)\}$. 
All physically realizable transmit signals are {\em effectively band limited}.
As the prototype function~\logonp is well concentrated in frequency by
construction, we can model the effective band limitation of~$\inp(\time)$ by
using only a finite number of slots~\fslots in frequency. The resulting
transmitted signal 
\ba
	\inp(\time)=\sum_{\dtime=-\infty}^{\infty}\sum_{\dfreq=0}^{\fslots-1}
		\underbrace{\inner{\inp}{\slogon}}_{\define\inp[\dtdf]} \slogon(\time)
\label{eq:canonical-input}
\ea
then has effective bandwidth~$\bandwidth=\fslots\fstep$.
We call the coefficient~$\inp[\dtdf]$ the {\em transmit symbol} in the {\em
time-frequency} slot~$(\dtdf)$. The received signal can be expanded in the
same basis. To compute the resulting projections, we substitute~\kernelpapprox and the canonical input
signal~\fref{eq:canonical-input} into the integral input-output
relation~\fref{eq:ltv-kernel-io}, add white Gaussian noise~$\wgn(\time)$, and
project the resulting noisy received signal $\outp(\time)=(\CHopapprox \inp)(\time) +\wgn(\time)$ onto the functions~$\{\slogon(\time)\}$, i.e.,
\be
\bs
	\outp[\dtdf] &= \inner{\outp}{\slogon} =\inner{\CHopapprox\inp}{\slogon} + \underbrace{\inner{\wgn}{\slogon}}_{\wgn[\dtdf]}\\
	&= \sum_{\dtime',\dfreq'}\inp[\dtime',\dfreq']\inner{\CHopapprox\logon_{\dtime',\dfreq'}}{\slogon}
	+\wgn[\dtdf] \\
	&=\underbrace{\tvtf(\tfsamples)}_{\ch[\dtdf]}\inp[\dtdf]+\wgn[\dtdf]
\es
\label{eq:scalar-io}
\ee
for all time-frequency~slots~$(\dtdf)$. The last step in~\eqref{eq:scalar-io} follows from the orthonormality of the set~$\{\slogon(\time)\}$. Orthonormality also
implies that the discretized noise signal~$\wgn[\dtdf]$ is JPG, independent and identically distributed~(\iid) over time~\dtime and
frequency~\dfreq; for convenience, we normalize the noise variance so
that~$\wgn[\dtdf]\distas\jpg(0,1)$ for all~\dtime and~\dfreq. The diagonalized input-output relation~\fref{eq:scalar-io} is completely generic, i.e., it is not limited to a specific signaling scheme.

\subsubsection{OFDM interpretation of the approximating channel model}
\label{sec:ofdm-interpretation}

The canonical signaling scheme~\fref{eq:canonical-input} and the corresponding
discretized input-output relation~\fref{eq:scalar-io}, are not just
tools to analyze channel capacity, but also lead to a practical transmission
system. The decomposition of the channel input signal~\fref{eq:canonical-input}
can be interpreted as pulse-shaped~(PS)~OFDM~\cite{kozek98-10a}, where discrete
data symbols~$\inp[\dtdf]$ are modulated onto a set of orthogonal signals,
indexed by~\dtime and~\dfreq.
In addition, this perspective leads to an operational interpretation of
the error incurred when approximating~$\kernelp$ as in~\fref{eq:approx-kernel}.
The time- and frequency-dispersive nature of LTV~channels leads to intersymbol
interference~(ISI) and intercarrier interference~(ICI) in the received PS-OFDM
signal. This is apparent if we project~$\nfoutp(\time)$ onto the 
function~$\slogon(\time)$:
\begin{multline}
	\inner{\nfoutp}{\slogon}=\inner{\CHop\!\inp}{\slogon}=	\suminf{\dtime'}
		\sumz{\dfreq'}{\fslots}\inp[\dtime',\dfreq']\inner{\CHop\!
		\logon_{\dtime',\dfreq'}}{\slogon}\\
	= \inner{\CHop\!\slogon}{\slogon}\inp[\dtdf] + \mathop{
		\suminf{\dtime'}\sumz{\dfreq'}{\fslots}}_{(\dtime',\dfreq') \neq
		(\dtdf)} \inp[\dtime',\dfreq']\inner{\CHop\!\logon_{\dtime',\dfreq'}}
		{\slogon}.
\label{eq:psofdm-rx}
\end{multline}
The second term on the right-hand side (RHS) of~\fref{eq:psofdm-rx} corresponds to~ISI and~ICI,
while the first term is the desired signal; we can approximate the first term as~$\tvtf(\tfsamples)\inp[\dtdf]$ by Property~4. Comparison of~\fref{eq:scalar-io}
and~\fref{eq:psofdm-rx} then shows that the input-output
relation~\fref{eq:scalar-io}, which results from the
approximation~\fref{eq:approx-kernel}, can be interpreted as PS-OFDM
transmission over the original channel~$\CHop$ if all~ISI and~ICI
terms are neglected.

With proper design of the prototype  signal~\logonp and choice of the grid
parameters~\tstep and~\fstep, both ISI and ICI can be reduced~\cite{kozek98-10a,liu04-11a,matz07-05a}. The larger the product~\tfstep, the more
effective the reduction in~ISI and~ICI, as discussed in~\fref{app:ch-approx-error}. Heuristically, a good compromise
between loss of dimensions in signal space and reduction of the interference terms
seems to result for~$\tfstep\simeq 1.2$~\cite{kozek98-10a,matz07-05a}. The cyclic prefix~(CP) in a conventional CP-OFDM system incurs a similar dimension loss.

In~\eqref{eq:isi-ici-error}, we provide an upper bound on mean-square energy
of the interference term in~\fref{eq:psofdm-rx}, and show how this upper bound
can be minimized by a careful choice of the signal~\logonp and of the grid
parameters~\tstep and~\fstep~\cite{kozek97a,sayeed99-01a,matz07-05a}. For general
scattering functions, the optimization of the triple~$\{\logonp,\tstep,\fstep\}$
needs to be performed numerically; a general guideline is to choose~\tstep and~\fstep such that
(see~\fref{app:ch-approx-error})
\ba
\label{eq:grid-matching}
	\frac{\tstep}{\fstep}=\frac{\maxDelay}{\maxDoppler}.
\ea 

To summarize, in this section we constructed an approximation~\CHopapprox of the random linear operator~$\CHop$ on the basis of the underspread property. The kernel of the approximating operator is synthesized from the Weyl-Heisenberg set~$\{\slogon(\time)\}$ as in~\eqref{eq:approx-kernel}, so that~$\{\slogon(\time)\}$ is an orthonormal basis for the input space and the range space of~\CHopapprox. The decomposition of the input signal~\eqref{eq:canonical-input} can be interpreted as PS-OFDM: this interpretation sheds light on one of the errors resulting from the approximation~\eqref{eq:approx-kernel}. Finally, an important open problem is the characterization of the difference between the input spaces of~$\CHop$ and~\CHopapprox, and between the range spaces of~$\CHop$ and~\CHopapprox.

\subsection{Linear Time-Invariant and Linear Frequency-Invariant Channels}
\label{sec:LTI and LFI}
The properties of LTV underspread channels we listed in~\fref{sec:underspread} are similar to the properties of LTI and linear frequency-invariant (LFI) channels: both LTI and LFI channel operators are normal and have a well-structured set of deterministic eigenfunctions (sinusoids parametrized by frequency for LTI channels,  and Dirac functions parametrized by time for LFI channels), with corresponding eigenvalues equal to the samples of a channel system function (e.g., the transfer function in the LTI case).  Intuitively, LTI and LFI channels are limiting cases within the class of LTV channels analyzed in this section; in fact, an LTV channel reduces to an LTI channel when~$\maxDoppler=0$, and to an LFI channel when~$\maxDelay=0$. Both LTI and LFI channels are then underspread, according to our definition. Yet,  since LTI and LFI channel operators are not of  Hilbert-Schmidt type~\cite[App. A]{matz00-11a}, the kernel diagonalization  presented in~\fref{sec:underspread} does not apply to these two classes of channels; consequently, the capacity bounds we derive in Sections~\ref{sec:capbounds} and~\ref{sec: peak in time} do not reduce to capacity bounds for the LTI or the LFI case when~$\maxDoppler=0$ or~$\maxDelay=0$, respectively.\footnote{For deterministic LTI channels, a channel discretization that is useful for
information-theoretic analysis is discussed in~\cite[Sec. 8.5]{gallager68a}.} 

Quasi-LTI channels, i.e., channels that are slowly time varying (\maxDoppler small but positive), and quasi-LFI channels, i.e., channels that are slowly frequency varying (\maxDelay small but positive), can instead be approximately diagonalized as described in~\fref{sec:underspread}, as long as they are underspread.

\subsection{Discrete-Time Discrete-Frequency Input-Output Relation}
\label{sec:dtdf-io}

The discrete-time discrete-frequency channel coefficients~$\{\ch[\dtdf]\}$
constitute a two-dimensional discrete-parameter stationary random process that
is JPG with zero mean and correlation function 
\be
\label{eq:chcorr}
	\chcorr[\Ddtime,\Ddfreq]=\Ex{}{\ch[\dtime'+\Ddtime,\dfreq'+\Ddfreq]
		\conj{\ch}[\dtime',\dfreq']}=
	\Ex{}{\tvtf\bigl((\dtime'+\Ddtime)\tstep,(\dfreq'+\Ddfreq)\fstep
		\bigr)\conj{\tvtf}(\dtime'\tstep,\dfreq'\fstep)}.
\ee
The two-dimensional power spectral density of~$\{\ch[\dtdf]\}$ is defined as
\ba
\label{eq:chspecfun}
	\chspecfunp\define\sum_{\Ddtime=-\infty}^{\infty}
		\sum_{\Ddfreq=-\infty}^{\infty}\chcorr[\Ddtime,\Ddfreq]\cexn{(\Ddtime
		\specparam-\Ddfreq\altspecparam)},\quad\abs{\specparam},
		\abs{\altspecparam}\le 1/2.
\ea
We shall often need the following expression for~\chspecfunp in terms of the scattering
function~\scafunp:
\be
\bs
	\chspecfunp &\stackrel{(a)}{=} \sum_{\Ddtime=-\infty}^{\infty}
		\sum_{\Ddfreq=-\infty}^{\infty} \cexn{(\Ddtime\specparam - \Ddfreq
		\altspecparam)} \spreadint{\scafunp \cex{(\Ddtime\tstep\doppler -\Ddfreq
		\fstep\delay)}}\\
	&= \spreadint{\scafunp \sum_{\Ddtime=-\infty}^{\infty} \cex{\Ddtime\tstep
		\left(\doppler-\frac{\specparam}{\tstep}\right)} 
		\sum_{\Ddfreq=-\infty}^{\infty} \cexn{\Ddfreq\fstep\left(\delay -
		\frac{\altspecparam}{\fstep}\right)}}\\
	&\stackrel{(b)}{=} \frac{1}{\tfstep} \spreadint{\scafunp
		\sum_{\Ddtime=-\infty}^{\infty}
		\dirac\lefto(\doppler - \frac{\specparam-\Ddtime}{\tstep}\right)
		\sum_{\Ddfreq=-\infty}^{\infty} \dirac\lefto(\delay - 
		\frac{\altspecparam-\Ddfreq}{\fstep}\right)}\\
	&=\frac{1}{\tfstep}\sum_{\Ddtime=-\infty}^{\infty}
		\sum_{\Ddfreq=-\infty}^{\infty} \scafun\lefto(\frac{\specparam -
		\Ddtime}{\tstep},\frac{\altspecparam - \Ddfreq }{\fstep}\right)
\es	 \label{eq:specfun-scafun}
\ee
where~(a) follows from the Fourier transform relation~\fref{eq:scafun-chcorr}, and~(b)
results from Poisson's summation formula. 
The variance of each channel
coefficient is given by
\be
\bs
	\pathloss&\define\int_{-1/2}^{1/2} \int_{-1/2}^{1/2}\chspecfunp
		d\specparam d\altspecparam \\
	&\stackrel{(a)}{=}\frac{1}{\tfstep}\sum_{\Ddtime=-\infty}^{\infty}
		\suminf{\Ddfreq}\int_{-1/2}^{1/2}\int_{-1/2}^{1/2}\scafun\lefto(
		\frac{\specparam - \Ddtime}{\tstep},\frac{\altspecparam - \Ddfreq }
		{\fstep}\right)d \specparam d\altspecparam\\
	&\stackrel{(b)}{=} \frac{1}{\tfstep}\int_{-1/2}^{1/2} \int_{-1/2}^{1/2}
		\scafun\lefto(\frac{\specparam} {\tstep},\frac{\altspecparam  }{\fstep}
		\right)d \specparam d\altspecparam\\
	&\stackrel{(c)}{=}\spreadint{\scafunp} 
\es
\label{eq:pathloss}
\ee
where~(a) follows from~\fref{eq:specfun-scafun}, and~(b) results because we
chose the grid parameters to satisfy the Nyquist conditions~$\tstep\le1/(2
\maxDoppler)$ and~$\fstep\le1/(2\maxDelay)$, so that periodic repetitions of the
compactly supported scattering function lie outside the integration region.
Finally,~(c) follows from the change of variables~$\doppler=\specparam/\tstep$
and~$\delay=\altspecparam/\fstep$. For ease of notation, we
normalize~$\pathloss=1$ throughout the paper.

For each time slot~\dtime, we arrange the discretized input signal~$\inp[\dtdf]
$, the discretized output signal~$\outp[\dtdf]$, the channel
coefficients~$\ch[\dtdf]$, and the noise samples~$\wgn[\dtdf]$ in corresponding
vectors. For example, the \fslots-dimensional vector that contains the
input symbols in the $\dtime$th time slot is defined as
\bas
	\mvsinp[\dtime]\define\tp{\mat\inp[\dtime,0]\; \inp[\dtime,1]\; \cdots\;
		\inp[\dtime,\fslots-1]\emat}.
\eas
The output vector~$\mvoutp[\dtime]$, the channel vector~$\mvch[\dtime]$, and
the noise vector~$\mvwgn[\dtime]$ are defined analogously.
This notation allows us to rewrite the input-output
relation~\fref{eq:scalar-io} as
\ba
\label{eq:iorel-mv}
	\mvoutp[\dtime] = \mvch[\dtime]\had\mvinp[\dtime]+\mvwgn[\dtime]
\ea
for all~\dtime. In this formulation, the channel is a multivariate stationary
process~$\{\mvch[\dtime]\}$ with matrix-valued correlation function
\be
\label{eq:mv-covmat}
	\mvchcovmat[\dtime]\define\Ex{}{\mvsch[\dtime'+\dtime]
		\herm{\mvsch}[\dtime']} 
	=\mat
		\chcorr[\dtime,0] & \conj{\chcorr}[\dtime,1] & \hdots & 
			\conj{\chcorr}[\dtime,\fslots-1] \\
		\chcorr[\dtime,1] & \chcorr[\dtime,0] & \hdots &\conj{\chcorr}[\dtime,
			\fslots-2]\\
		\vdots & \vdots & \ddots & \vdots \\
		\chcorr[\dtime, \fslots-1] & \chcorr[\dtime, \fslots-2]  & \hdots
			& \chcorr[\dtime,0]
	\emat.
\ee 

In most of the following analyses, we initially consider a finite number~\tslots
of time slots and then take the limit~$\tslots\to\infty$. To obtain a compact
notation, we stack~\tslots contiguous elements of the multivariate input,
channel, and output processes just defined. For the channel input, this
results in the $\tslots\fslots$-dimensional vector
\ba
	\mvsinp\define\tp{\mat\tp{\mvsinp}[0]\; \tp{\mvsinp}[1]\; \cdots\;
		\tp{\mvsinp}[\tslots-1]\emat}.
\label{eq:stacked-input}
\ea
Again, the stacked vectors~\mvsoutp, $\mvsch$, and~$\mvswgn$ are defined
analogously. With these definitions, we can now compactly express the
input-output relation~\fref{eq:scalar-io} as
\ba
	\mvsoutp=\mvsinp\had\mvsch+\mvswgn.
\label{eq:vec-io}
\ea
We denote the correlation matrix of the stacked channel vector~\mvsch
by~$\mvschcovmat\define\Ex{}{\mvsch\herm{\mvsch}}$. Because the channel
process~$\{\ch[\dtdf]\}$ is stationary in time and in frequency, \mvschcovmat is a
two-level Hermitian Toeplitz matrix, given by
\ba
	\mvschcovmat=\mat
		\mvchcovmat[0] & \herm{\mvchcovmat}[1] & \hdots &
			\herm{\mvchcovmat}[\tslots-1]\\
		\mvchcovmat[1] & \mvchcovmat[0] & \hdots &
			\herm{\mvchcovmat}[\tslots-2]\\
		\vdots & \vdots & \ddots & \vdots\\
		\mvchcovmat[\tslots-1] &\mvchcovmat[\tslots-2] & \hdots & \mvchcovmat[0] 
	\emat.
\label{eq:two-level-covmat}
\ea
\subsection{Power Constraints}\label{sec:power constraints}

Throughout the paper, we assume that the average power of the transmitted signal
is constrained as~\avPnorm. In addition, we limit the peak power to be no larger
than \papr~times the average power, where~$\papr\ge1$ is the {\em nominal
peak-  to average-power ratio}~(PAPR).

The multivariate input-output relation~\fref{eq:vec-io} allows to constrain the
peak power in several different ways. We analyze the following two cases:
\begin{enumerate}
\item {\em Peak constraint in time:} The power of the transmitted signal in
	each time slot~\dtime is limited as
	\ba\label{eq:peak-per-tslot}
 		\frac{1}{\tstep}\sum_{\dfreq=0}^{\fslots-1}\abs{\inp[\dtdf]}^2
		\leq \papr\Pave\qquad \wpone.
	\ea
	This constraint models the fact that physically realizable power amplifiers
	can only provide limited output power~\cite{gray01a}.
\item {\em Peak constraint in time and frequency:} Regulatory bodies sometimes
	limit the peak power in certain frequency bands, e.g., for UWB systems. We
	model this type of constraint by imposing a limit on the squared amplitude
	of the transmitted symbols~$\inp[\dtdf]$ in each time-frequency
	slot~$(\dtdf)$ according to
	\ba
	\label{eq:peak-per-tfslot}
		(1/\tstep)\abs{\inp[\dtdf]}^2 \leq {\papr\Pave}/{\fslots}\qquad \wpone.
	\ea
	This type of constraint is more stringent than the  peak constraint in time
	given in~\fref{eq:peak-per-tslot}.
\end{enumerate} 
Both peak constraints above are imposed on the input symbols~$\inp[\dtdf]$,
i.e., in the eigenspace of the approximating channel operator. This limitation
is mathematically convenient; however, the peak value of the corresponding
transmitted  continuous-time signal~$\inp(\time)$ in~\fref{eq:canonical-input}
also depends on the prototype signal~\logonp, so that a limit on~$\inp[\dtdf]$
does not generally imply that~$\inp(\time)$ is  peak limited.

\section{Capacity Bounds under a Peak Constraint in Time and Frequency}
\label{sec:capbounds}

In the present section, we analyze the capacity of the discretized channel in~\fref{eq:scalar-io}
subject to the peak constraint in time and frequency  specified by~\fref{eq:peak-per-tfslot}.
The link between the discretized channel~\fref{eq:scalar-io} and the continuous-time channel model established in \fref{sec:model} then allows us to express the resulting bounds in terms of the scattering
function~\scafunp of the underspread WSSUS channel~$\CHop$. 

As we assumed that the channel process~$\{\ch[\dtdf]\}$ has a spectral density [given in~\eqref{eq:specfun-scafun}],
the vector process~$\{\mvch[\dtime]\}$ is ergodic~\cite{maruyama49a} and the capacity of the discretized underspread channel~\eqref{eq:vec-io} is given by~\cite[Ch.~12]{gray07b}
\ba
\label{eq:capacityPeakTF} 
 	\capacity(\bandwidth)=\limintime \frac{1}{\tslots\tstep} \sup_{\dsetpapr}
		\mi(\mvsoutp;\mvsinp)\qquad\text{[nat/s]}
\ea
for a given bandwidth~$\bandwidth=\fslots\fstep$. Here, the supremum is taken over the set~\dsetpapr of all input distributions
that satisfy the peak constraint~\fref{eq:peak-per-tfslot} and the average-power
constraint~$\Ex{}{\vecnorm{\mvsinp}^{2}}\le\tslots\Pave\tstep$.

The capacity of fading channels with finite bandwidth has so far
resisted all attempts at closed-form solutions~\cite{taricco97-07a,abou-faycal01-05a,lapidoth03-10a},
even for the memoryless case; thus, we resort to bounds to characterize
the capacity~\fref{eq:capacityPeakTF}. In particular, we present the
following bounds:
\begin{itemize}
\item An upper bound~\ubcohp, which we refer to as coherent upper bound, that is based on the
assumption that the receiver has perfect knowledge of the channel realizations. 
This bound is standard; it turns out to be useful for small bandwidth.
\item An upper bound~\ubTFpeakp that is useful for medium to large bandwidth.
	This bound is explicit in the channel's scattering function and extends the
	upper bound~\cite[Prop.~2.2]{sethuraman08a} on the capacity of
	frequency-flat time-selective channels to general underspread channels that are
	selective in time and frequency.
\item A lower bound~\lbTFpeakp that extends the lower
	bound~\cite[Prop.~2.2]{sethuraman05-09b} to general underspread
	channels	 that are selective in time and frequency. This bound is explicit
	in the channel's scattering function only for large bandwidth.
\end{itemize}

\subsection{Coherent Upper Bound}

The assumption that the receiver perfectly knows the instantaneous channel realizations furnishes 
the following capacity upper bound:
\be
\label{eq:coh-ub-deriv}
\bs
	\frac{1}{\tslots\tstep}\sup_{\dsetpapr}\mi(\mvsoutp;\mvsinp)
	&\stackrel{(a)}{\le}\frac{1}{\tslots\tstep}\sup_{\dsetpapr}\mi(\mvsoutp;
		\mvsinp\given\mvsch)\\
	&\stackrel{(b)}{\le}\frac{1}{\tslots\tstep}\sup_{\avP}\mi(\mvsoutp;\mvsinp
		\given\mvsch) \\
	&\stackrel{(c)}{=}\frac{1}{\tslots\tstep}\sup_{\covmat{\mvsinp}}
		\Ex{\mvsch}{\logdet{\imat_{\tslots\fslots}+(\mvsch\herm{\mvsch})
		\had\covmat{\mvsinp}}}\\
	&\stackrel{(d)}{\le}\frac{\fslots}{\tstep}\Ex{\ch}{\log
		\lefto(1+\frac{\Pave\tstep}{\fslots}\abs{\ch}^{2}\right)}.
\es
\ee
Here, (a) holds because the coherent mutual information,~$\mi(\mvsoutp;\mvsinp\given\mvsch)$, is an upper bound on the corresponding mutual information in the noncoherent setting.
Inequality~(b) follows as we drop the peak constraint and thus enlarge the
set of admissible input distributions. The supremum of~$\mi(\mvsoutp;\mvsinp
\given\mvsch)$ over the resulting relaxed input constraint is achieved by a
zero-mean JPG input vector~\mvsinp with covariance
matrix~$\covmat{\mvsinp}=\Ex{}{\mvsinp\herm{\mvsinp}}$ that
satisfies~$\tr(\covmat{\mvsinp})\le\tslots\Pave\,\tstep$~\cite{tse05a}. To
obtain~(c), we use that, conditioned on~\mvsch, the output vector~\mvsoutp is
JPG and its covariance matrix can be expressed as
\bas
	\Ex{}{\mvsoutp\herm{\mvsoutp}\given \mvsch} = \imat_{\tslots\fslots} +
		\Ex{\mvsinp}{(\mvsinp \had \mvsch) \herm{(\mvsinp \had
	\mvsch)}} = \imat_{\tslots\fslots} + (\mvsch \herm{\mvsch})\had
		\covmat{\mvsinp}
\eas
where the last equality results from the following elementary relation between
Hadamard products and outer products:
\ba
	(\mvsinp\had\mvsch) \herm{(\mvsinp\had\mvsch)} = \mvsinp\herm{\mvsinp}\had
		\mvsch\herm{\mvsch}.
\label{eq:had-gramian}
\ea
Finally, (d)~follows from Hadamard's inequality, from the fact that by Jensen's inequality
the supremum is achieved by $\covmat{\mvsinp}=(\Pave\tstep/
\fslots)\imat_{\tslots\fslots}$, and because the channel coefficients all have the same distribution~$\ch[\dtdf]\distas\ch\distas~\jpg(0,1)$.  As the upper bound~\fref{eq:coh-ub-deriv} does not depend on~\tslots, we obtain an upper 
bound~\ubcohp on capacity~\fref{eq:capacityPeakTF} as a function of bandwidth~\bandwidth if we set~$\bandwidth=\fslots\fstep$:
\ba
\label{eq:ub-coh}
 	\capacityp\leq \ubcohp\define\frac{\bandwidth}{\tfstep}
		\Ex{\ch}{\log\lefto(1 + \frac{\Pave\tfstep}{\bandwidth}\abs{\ch}^2
		\right)}.
\ea
For a discretization of the WSSUS channel~$\CHop$ different from the one in
\fref{sec:underspread}, M\'edard and Gallager~\cite{medard02-04a} showed that
the corresponding
capacity vanishes with increasing bandwidth if the peakiness of the input signal
is constrained in a way that includes our peak
constraint~\fref{eq:peak-per-tfslot}. As the upper bound~\ubcohp 
monotonically increases in~\bandwidth, it is sensible to conclude that~\ubcohp does not accurately reflect the capacity
behavior for large bandwidth. However, we
demonstrate in \fref{sec:num-eval} by means of a numerical example that~\ubcohp
can be quite useful for small and medium bandwidth.

\subsection{An Upper Bound for Large but Finite Bandwidth}
\label{sec:ub-noncoh}

To better understand the capacity behavior at large bandwidth, we derive an upper bound~\ubTFpeakp that captures the effect of diminishing capacity in the large-bandwidth regime. The upper bound \ubTFpeakp is explicit in the channel's scattering function~\scafunp.

\subsubsection{The upper bound}
\begin{thm}
\label{thm:ubTFpeak}
Consider an underspread Rayleigh-fading channel with scattering
function~$\scafunp$; assume that the channel input~\mvsinp satisfies the
average-power constraint~\avP and the peak 
constraint~$\abs{\inp[\dtdf]}^{2}\le {\papr\Pave\tstep}/{\fslots}$~\wpone.
The capacity of this channel is upper-bounded as~$\capacityp
\leq\ubTFpeakp$, where
\begin{subequations}
\ba
	\ubTFpeakp&=\frac{\bandwidth}{\tfstep} \log\lefto(1+\avpoptp\Pave
		\frac{\tfstep}{\bandwidth}\right)
		-\avpoptp \ubpenaltyp \label{eq:ubTFpeak-core}\\
\intertext{with}
	\avpoptp&\define\min\lefto\{1,\, \frac{\bandwidth}{\tfstep}\left(\frac{1}
		{\ubpenaltyp}-\frac{1}{\Pave}\right)\right\}
		\label{eq:ubTFpeak-avpopt}\\
\intertext{and}
 	\ubpenaltyp&\define\frac{\bandwidth}{\papr}\spreadint{\log\lefto(1+
		\frac{\papr\Pave}{\bandwidth}\scafunp\right)}.
		\label{eq:ubTFpeak-penalty}
\ea\label{eq:ubTFpeak}
\end{subequations}
\end{thm}

\begin{IEEEproof}
To bound $\sup_{\dsetpapr} \mi(\mvsoutp;\mvsinp)$, we first use the chain rule for mutual
information, $\mi(\mvsoutp;\mvsinp)=\mi(\mvsoutp;\mvsinp,\mvsch)-\mi(\mvsoutp;
\mvsch\given\mvsinp)$. Next, we split the supremum over~\dsetpapr into two
parts, similarly as in the proof of~\cite[Prop.~2.2]{sethuraman08a}:
one supremum over a restricted set of input distributions~\dsetres that satisfy
the peak constraint~\eqref{eq:peak-per-tfslot} and have a prescribed average power, i.e.,~\avPeq for some
fixed parameter~$\pparam\in[0,1]$, and another supremum over the
parameter~\pparam. Both steps together yield the upper bound
\be
\bs
	\sup_{\dsetpapr}\mi(\mvsoutp;\mvsinp) &= \sup_{\dsetpapr}\{\mi(\mvsoutp;
		\mvsinp,\mvsch) - \mi(\mvsoutp;\mvsch\given\mvsinp)\}\\
	&= \supavp\sup_{\dsetres}\{\mi(\mvsoutp;\mvsinp,\mvsch) -
		\mi(\mvsoutp;\mvsch\given\mvsinp)\}\\
	&\le\supavp\Biggl\{\sup_{\dsetres}\mi(\mvsoutp;\mvsinp,\mvsch) -
		\inf_{\dsetres}\mi(\mvsoutp;\mvsch\given\mvsinp)\Biggr\}.
\es\label{eq:ubTFpeak-step1}
\ee
Next, we bound the two terms inside the braces individually. While standard steps suffice for the bound on the first term, the second term requires some more effort; we relegate some of the more technical steps to \fref{app:mmse-mi}.

\paragraph{Upper bound on the first term}
The output vector~\mvsoutp depends on the input vector~\mvsinp only
through~$\mvsjinp\define\mvsinp\had\mvsch$, so that~$\mi(\mvsoutp;\mvsinp,
\mvsch)=\mi(\mvsoutp; \mvsjinp)$. To upper-bound the mutual
information~$\mi(\mvsoutp;\mvsjinp)$, we take~\mvsjinp as JPG with zero mean and covariance
matrix~$\Ex{}{\mvsjinp\herm{\mvsjinp}}=\Ex{}{\mvsinp\herm{\mvsinp}
}\had\mvschcovmat$. An upper bound on the first term inside the braces in~\fref{eq:ubTFpeak-step1} now results if we drop the peak constraint on~\mvsjinp. Then,
\be
\bs
	\sup_{\dsetres}\mi(\mvsoutp;\mvsinp,\mvsch) &\leq\sup_{\avPeq}\logdet{
		\imat_{\tfslots}+ \Ex{}{\mvsinp\herm{\mvsinp}}\had\mvschcovmat}\\
    &\stackrel{(a)}{\leq} \sup_{\avPeq}\sum_{\dtime=0}^{\tslots-1}
    		\sum_{\dfreq=0}^{\fslots-1}\log\lefto(1 + \Ex{}{\abs{\inp[\dtdf]}^{2}}
		\right)\\
	& \stackrel{(b)}{\leq}\tfslots \log\lefto(1 + \frac{\pparam\Pave\tstep}
	{\fslots}\right)
\es
\label{eq:ubTFpeak-term1-ub}
\ee
where~(a) follows from Hadamard's inequality and~(b) from Jensen's inequality.

\paragraph{Lower bound on the second term}

We use the fact that the channel~\mvsch is JPG, so
that~$\mi(\mvsoutp;\mvsch\given\mvsinp)=\Ex{\mvsinp}{\logdet{\imat_{\tfslots}+
(\mvsinp\herm{\mvsinp})\had\mvschcovmat}}$. Next, we expand the expectation
operator as follows:
\be
\bs
	\inf_{\dsetres}\mi(\mvsoutp;\mvsch\given\mvsinp)&=\inf_{\dsetres}
		\Ex{\mvsinp}{\logdet{\imat_{\tfslots} + (\mvsinp\herm{\mvsinp})\had
		\mvschcovmat	}}\\
	&=\inf_{\cdf{}\in\dsetres}\int_{\mvsinpd \in \setxTpeak}\Biggl(\frac{\logdet{
		\imat_{\tfslots} + (\mvsinpd\herm{\mvsinpd})\had\mvschcovmat}}
		{\vecnorm{\mvsinpd}^{2}}\Biggr)\vecnorm{\mvsinpd}^{2}d\cdf{}
\es\label{eq:ubTFpeak-term2-expectation}
\ee
where~$\setxTpeak=\{\mvsinpd \in \complexset^{\tslots\fslots} \sothat \abs{\inp[\dtdf]}^{2}\leq\papr \Pave\tstep/\fslots, \forall \dtime, \dfreq\}$ is the integration domain because the input distribution~$\cdf{}$ satisfies the peak constraint~\eqref{eq:peak-per-tfslot}.
Both factors under the integral are nonnegative; hence, we obtain a lower bound on the expectation
if we replace the first factor by its infimum over~\setxTpeak.
\be
\bs
	\inf_{\dsetres}\mi(\mvsoutp;\mvsch\given\mvsinp)
	&\ge\inf_{\cdf{}\in\dsetres}\int_{\tilde{\mvsinpd} \in \setxTpeak}\Biggl(
		\inf_{\mvsinpd \in \setxTpeak}\frac{\logdet{\imat_{\tfslots} + (\mvsinpd
		\herm{\mvsinpd})\had\mvschcovmat}}{\vecnorm{\mvsinpd}^{2}}\Biggr)
		\left(\vecnorm{\tilde{\mvsinpd}}^{2}\right)d\cdf{}\\
	&=\inf_{\mvsinpd \in \setxTpeak}\frac{\logdet{\imat_{\tfslots} + (\mvsinpd\herm{\mvsinpd})
		\had\mvschcovmat}}{\vecnorm{\mvsinpd}^{2}}
		\underbrace{\left(\inf_{\cdf{}\in\dsetres}\int\vecnorm{\mvsinpd
		}^{2}d\cdf{}\right)}_{\inf_{
		\dsetres}\avPeq}\\
	&=\pparam\tslots\Pave\,\tstep\inf_{\mvsinpd \in \setxTpeak}\frac{\logdet{
		\imat_{\tfslots} + (\mvsinpd\herm{\mvsinpd})\had\mvschcovmat}}
		{\vecnorm{\mvsinpd}^{2}}.
	\es\label{eq:ubTFpeak-term2-inf}      
\ee 
As the matrix~\mvschcovmat is positive semidefinite, the above infimum
is achieved on the boundary of the admissible set~\cite[Sec.~VI.A]{sethuraman05-09a},
i.e., by a vector~\mvsinpd whose entries satisfy~$\abs{\inp[\dtdf]}^{2}\in\{0,\Ppeak\tstep/\fslots\}$. We use this fact and the relation between mutual information and MMSE, recently discovered by Guo {\it et al.} \cite{guo05-04a}, to further lower-bound the infimum on the~RHS in~\fref{eq:ubTFpeak-term2-inf}. The corresponding derivation is detailed in \fref{app:mmse-mi}; it results in
\be
\label{eq:ubTFpeak-term2-immse-lb}
	\inf_{\mvsinpd \in \setxTpeak}\frac{\logdet{
		\imat_{\tfslots} + (\mvsinpd\herm{\mvsinpd})\had\mvschcovmat}}
		{\vecnorm{\mvsinpd}^{2}}
	\ge \frac{\fslots}{\papr\Pave\tstep}\int_{-1/2}^{1/2}\int_{-1/2}^{1/2} \log\lefto(1 + \frac{\Ppeak\tstep}
		{\fslots} \chspecfunp\right)d\specparam 
		d\altspecparam
\ee
where~\chspecfunp, defined in~\fref{eq:chspecfun}, is the two-dimensional power
spectral density of the channel process~$\{\ch[\dtdf]\}$.
Finally, we use the bound~\fref{eq:ubTFpeak-term2-immse-lb}  in~\fref{eq:ubTFpeak-term2-inf}, relate~\chspecfunp to the scattering function~\scafunp by means of~\fref{eq:specfun-scafun} and get
\be
\bs
	\inf_{\dsetres}\mi(\mvsoutp;\mvsch\given\mvsinp)&\ge
		\frac{\pparam \tfslots}{\papr} \int_{-1/2}^{1/2} \int_{-1/2}^{1/2}
		\log\lefto(1 + \frac{\Ppeak}{\fslots\fstep} \sum_{\Ddtime=
		-\infty}^{\infty} \sum_{\Ddfreq=-\infty}^{\infty}\scafun\lefto(
		\frac{\specparam - \Ddtime}{\tstep},\frac{\altspecparam - \Ddfreq}
		{\fstep}\right)\right)d\specparam d\altspecparam \\
	&= \frac{\pparam \tfslots}{\papr}\int_{-1/2}^{1/2}
		\int_{-1/2}^{1/2}\log\lefto(1 +\frac{\Ppeak}{\fslots\fstep} \scafun
		\lefto(\frac{\specparam}{\tstep},\frac{\altspecparam}{\fstep}\right)
		\right)d\specparam d\altspecparam \\
	&=\frac{\pparam \tfslots\tstep\fstep}{\papr}\spreadint{
		\log\lefto(1 +\frac{\Ppeak}{\fslots\fstep}\scafunp\right)}
\es
\label{eq:ubTFpeak-term2-lb}
\ee
where the last two equalities result from steps similar to the ones used in~\fref{eq:pathloss}.

\paragraph{Completing the proof}
We insert~\fref{eq:ubTFpeak-term1-ub} and~\fref{eq:ubTFpeak-term2-lb}
in~\fref{eq:ubTFpeak-step1}, divide by~$\tslots\tstep$, and set~$\bandwidth=
\fslots\fstep$ to obtain the following upper bound on
capacity~\fref{eq:capacityPeakTF}
\be
\bs
	\capacityp\le\supavp\left\{\frac{\bandwidth}{\tfstep}
		\log\lefto(1 +\frac{\pparam\Pave\tfstep}{\bandwidth}\right)
		- \frac{\pparam\bandwidth}{\papr}
		\spreadintlog{1 +\frac{\papr \Pave}{\bandwidth}\scafunp}\right\}.
\es\label{eq:ubTFpeak-to-maximize}
\ee
As the function to maximize in~\fref{eq:ubTFpeak-to-maximize} is concave
in~\pparam, the maximizing value is unique. To conclude the proof and obtain the
bound~\fref{eq:ubTFpeak}, we perform an elementary optimization over~\pparam to
find the maximizing~\avpoptp given in~\fref{eq:ubTFpeak-avpopt}.
\end{IEEEproof}

The upper bound in \fref{thm:ubTFpeak} generalizes the upper
bound~\cite[Eq.~(2)]{schafhuber04-06a}, which holds only for constant
modulus signals, i.e., for signals whose magnitude~$\abs{\inp[\dtdf]}$ is the same for all~\dtime and~\dfreq. The bounds~\fref{eq:ubTFpeak-core}
and~\cite[Eq.~(2)]{schafhuber04-06a} are both explicit in the channel's
scattering function, have similar structure, and coincide for~$\papr=1$
when~$\avpoptp=1$ in~\fref{eq:ubTFpeak-avpopt}.

\subsubsection{Conditions for~$\avpoptp=1$}
\label{sec:avpopt-cond}
If~$\avpoptp=1$ independently of~\bandwidth, the first term of the upper bound~\ubTFpeakp
in~\fref{eq:ubTFpeak-core} can be interpreted as the capacity of an effective
AWGN~channel with receive power~\Pave and~$\bandwidth/(\tfstep)$ degrees of
freedom, while the second term can be seen as a penalty term that characterizes
the capacity loss because of channel uncertainty. We highlight the relation between this penalty term and the error in predicting the channel from its noisy past and future in \fref{app:mmse-mi}. For~$\avpoptp<1$, the upper
bound~\fref{eq:ubTFpeak-core} has a more complicated structure, which is
difficult to interpret. We show in
\fref{app:avpopt-proof} that a sufficient condition for~$\avpoptp=1$ is\footnote{More precisely, in~\fref{app:avpopt-proof} we derive a sufficient condition for~$\avpoptp=1$ that implies~\eqref{eq:avpopt-conditions}.} 
\begin{subequations}
\label{eq:avpopt-conditions}
\ba
\label{eq:avpopt-spread}
	\spread&\le{\papr}/{(3\tfstep)}
\intertext{and}
\label{eq:avpopt-snr}
	0\le\frac{\Pave}{\bandwidth}&<\frac{\spread}{\papr}\left[\exp\lefto(
		\frac{\papr}{2\tfstep\spread}\right)-1\right].
\ea
\end{subequations}
As virtually all wireless channels are highly underspread, as~$\papr\geq1$, and as, typically,~$\tfstep\approx 1.25$,
condition~\eqref{eq:avpopt-spread} is satisfied in all cases of practical interest, so that the only relevant condition is~\eqref{eq:avpopt-snr}; but even for large channel spread~\spread, this condition holds for all SNR values\footnote{Recall
that we normalized~$\No=1$.}$\Pave/\bandwidth$ of practical interest. As an example,
consider a system with~$\papr=1$ and spread~$\spread=10^{-2}$; for this choice,~\fref{eq:avpopt-snr} is satisfied for
all SNR values less than~$153\dB$. As this value is far in excess of the receive SNR encountered in practical systems, we can safely claim
that a capacity upper bound of practical interest results if we
substitute~$\avpoptp=1$ in~\fref{eq:ubTFpeak-core}.

\subsubsection{Impact of channel characteristics}
\label{sec:ubTFpeak-characterisitcs}
The spread~\spread and the shape of the scattering
function~\scafunp are important characteristics of wireless channels. As the
upper bound~\fref{eq:ubTFpeak} is explicit in the scattering function, we can
analyze its behavior as a function of~\spread and~\scafunp. We restrict our discussion to the practically relevant case~$\avpoptp=1$.

\paragraph{Channel spread}
For fixed shape of the scattering function, the upper bound~\ubTFpeakp
decreases for increasing spread~\spread. To see this, we define a normalized scattering function~\scafunnp with unit spread,\footnote{Recall that we normalized~$\pathloss=1$ in~\eqref{eq:pathloss}.} so that~$\scafunp=
\scafunn\bigl(\doppler/(2\maxDoppler),\delay/(2\maxDelay)\bigr)/\spread$. By a change of variables, the penalty term can now be written as
\be
\bs
 	\ubpenaltyp&=\frac{\bandwidth}{\papr}\spreadint{\log\lefto(1+
		\frac{\papr\Pave}{\bandwidth}\scafunp\right)} \\
	&= \frac{\bandwidth\spread}{\papr}\int_{-1/2}^{1/2}
		\int_{-1/2}^{1/2}\log\lefto(1+ \frac{\papr\Pave}
		{\bandwidth\spread}\scafunnp\right)d\tilde{\delay}d\tilde{\doppler}.
\es
\label{eq:ubTFpeak-penalty-dou}
\ee
Because~$\spread\log(1+\rho/\spread)$ is monotonically increasing in~\spread for
any positive constant~$\rho>0$, the penalty term~\ubpenaltyp increases with
increasing spread~\spread. As the first term in~\fref{eq:ubTFpeak-core} does not
depend on~\spread, the upper bound~\ubTFpeakp decreases with increasing spread.

\paragraph{Shape of the scattering function}\label{sec:shape scattering function}
For fixed spread~\spread, the scattering function that results in the lowest
upper bound~\ubTFpeakp is the ``brick-shaped'' scattering function:~$\scafunp=
1/\spread$ for~$(\doppler,\delay) \in [-\maxDoppler,\maxDoppler]\times
[-\maxDelay,\maxDelay]$. We prove this claim in two steps. First, we apply
Jensen's inequality to the penalty term in~\fref{eq:ubTFpeak-penalty}:
\be
\bs
	\spreadintlog{1 +\frac{\papr \Pave}{\bandwidth}\scafunp}
	&\leq \spread \log\lefto( 1 + \frac{\papr\Pave}{\bandwidth\spread}
		\spreadint{\scafunp} \right) \\
	&= \spread \log\lefto( 1 + \frac{\papr\Pave}{\spread\bandwidth}\right).
\es
\label{eq:ubTFpeak-penalty-ub}
\ee
Second, we note that a brick-shaped scattering function achieves this upper
bound.

The observation that a brick-shaped scattering function minimizes the upper
bound~\ubTFpeakp sheds some light on the common practice to use~\maxDoppler and~\maxDelay, rather than~\scafunp in the design of a communication system. A design on the basis of~\maxDoppler and~\maxDelay is implicitly targeted at a channel with brick-shaped scattering function, i.e., at the worst-case channel.

\subsection{Lower Bound}
\label{sec:lbTFpeak}

\subsubsection{A lower bound in terms of the multivariate spectrum
of~$\{\mvch[\dtime]\}$}

To state our lower bound on the capacity~\fref{eq:capacityPeakTF}, we require the following definitions.
\begin{itemize}
\item Let~$\mvchspecfunp$ denote the matrix-valued power spectral density of the
	multivariate channel process~$\{\mvch[\dtime]\}$, i.e.,
	\ba
		\mvchspecfunp\define\sum_{\Ddtime=-\infty}^{\infty}
			\mvchcovmat[\Ddtime]\cexn{\Ddtime\specparam},\quad\abs{\specparam}
			\le\frac{1}{2}.
	\label{eq:mvspec}
	\ea
\item Let~$\mi(\outp;\inp\given\ch)$ denote the coherent mutual
	information of a scalar,	memoryless Rayleigh-fading channel~$\outp=\ch\inp+
	\wgn$ with~$\ch\distas\jpg(0,1)$, additive noise~$\wgn\distas
	\jpg(0,1)$, and zero-mean constant-modulus input signal, i.e.,
	$\abs{\inp}^{2}=\param\Pave\tstep/{\fslots}$~\wpone.
\end{itemize}

\begin{thm}
\label{thm:lbTFpeak} 

Consider an underspread Rayleigh-fading channel with scattering
function~$\scafunp$. Assume that the channel input~\mvsinp satisfies the
average-power constraint~\avP and the peak 
constraint~$\abs{\inp[\dtdf]}^{2}\le\Ppeak\tstep/\fslots$~\wpone.
The capacity of this channel is lower-bounded
as~$\capacityp\ge\lbTFpeakp$, where
\ba
	\lbTFpeakp=\max_{1\le\param\le\papr}\Biggl\{\frac{\bandwidth}{
		\param\tfstep}\mi(\outp;\inp\given\ch)-\frac{1}{\param\tstep}
		\int_{-1/2}^{1/2}\logdet{\imat_{\fslots}+\frac{
		\param\Pave\tfstep}{\bandwidth}\mvchspecfunp}
		d\specparam\Biggr\}.
\label{eq:lbTFpeak}
\ea
\end{thm}
\begin{IEEEproof}
We obtain a lower bound on capacity by computing the mutual information for a
specific input distribution. A  simple scheme is to send symbols that have
zero mean, are~\iid over
time and frequency slots and have constant magnitude, i.e., $\abs{
\inp[\dtdf]}^{2}=\Pave\tstep/\fslots$ for~$\allz{\dtime}{\tslots}$
and~$\allz{\dfreq}{\fslots}$. The average power constraint is then satisfied
with equality. We denote a \tfslots-dimensional input vector that follows
this distribution by~\mvscmiidinp; this vector has entries~$\cmiidinp[\dtdf]$
that are first stacked in frequency and then in time, analogously to the definitions of~\mvsinp and~\mvsoutp in \fref{sec:dtdf-io}.

We use the chain rule for mutual information and the fact that mutual information is nonnegative to obtain the following bound:
\be
\bs
	\mi(\mvsoutp;\mvscmiidinp) &= \mi(\mvsoutp;\mvscmiidinp,\mvsch) 
		- \mi(\mvsoutp;\mvsch\given\mvscmiidinp)\\
	& = \mi(\mvsoutp;\mvsch)+\mi(\mvsoutp;\mvscmiidinp\given\mvsch) -
		\mi(\mvsoutp;\mvsch\given\mvscmiidinp)\\
	&\ge \mi(\mvsoutp;\mvscmiidinp\given\mvsch) -
		\mi(\mvsoutp;\mvsch\given\mvscmiidinp).
\es
\label{eq:lbTFpeak-mi-decomp}
\ee
Next, we evaluate the two terms on the RHS of the above inequality separately.
The first term satisfies
\ba
	\mi(\mvsoutp;\mvscmiidinp\given\mvsch)=\tfslots\: \mi(\outp;\cmiidinp
		\given\ch)
\label{eq:lbTFpeak-term1} 
\ea
where we set~$\ch=\ch[\dtdf]$ and~$\cmiidinp=\cmiidinp[\dtdf]$ for arbitrary~\dtime and~\dfreq because
\begin{inparaenum}[(i)]
\item the input vector~\mvscmiidinp has~\iid entries, and
\item all channel coefficients have the same distribution.
\end{inparaenum}
The second term equals 
\be
\bs
	\mi(\mvsoutp;\mvsch\given\mvscmiidinp)&=\Ex{\mvscmiidinp}{\logdet{
		\imat_{\tfslots}+\left(\mvscmiidinp\herm{\mvscmiidinp}\right)
		\had\mvschcovmat } }\\
	&=\Ex{\mvscmiidinp}{\logdet{\imat_{\tfslots}+\diag{\mvscmiidinp}
		\mvschcovmat\herm{\diag{\mvscmiidinp}}}}\\
	&\stackrel{(a)}{=}\Ex{\mvscmiidinp}{\logdet{\imat_{\tfslots}+\herm{
		\diag{\mvscmiidinp}}\diag{\mvscmiidinp}\mvschcovmat}}\\
	&\stackrel{(b)}{=}\logdet{\imat_{\tfslots} + \frac{\Pave\tstep}{\fslots}
	\mvschcovmat}
\es
\label{eq:lbTFpeak-term2}
\ee
where~(a) follows from the identity~$\det\left(\imat + \matA
\herm{\matB}\right) = \det\left(\imat +\herm{\matB}\matA\right)$ 
for any~$\matA$ and~$\matB$ of appropriate
dimension~\cite[Th.~1.3.20]{horn85a}, and~(b) follows from the constant
modulus
assumption. We now combine the two terms~\fref{eq:lbTFpeak-term1}
and~\fref{eq:lbTFpeak-term2}, set~$\bandwidth =\fslots\fstep$, divide
by~$\tslots\tstep$, and take the limit~$\tslots\to\infty$ to obtain the following lower bound:
\be
\bs
	\capacityp&\ge\limintime\frac{1}{\tslots\tstep}
		\mi(\mvsoutp;\mvscmiidinp) \\
	&\ge \frac{\bandwidth}{\tfstep}\mi(\outp;\cmiidinp\given\ch)-\limintime
		\frac{1}	{\tslots\tstep}\logdet{\imat_{\tfslots} + \frac{\Pave\tfstep}
		{\bandwidth}	\mvschcovmat}.
\es
\label{eq:lbTFpeak-non-timeshare}
\ee
The correlation matrix~\mvschcovmat is two-level Toeplitz, with blocks that
are~$\fslots\times\fslots$ correlation matrices~$\mvchcovmat[\Ddtime]$, as shown in~\fref{eq:two-level-covmat} and~\fref{eq:mv-covmat}, respectively.
Hence, we can explicitly evaluate the limit on the RHS
of~\fref{eq:lbTFpeak-non-timeshare} and express it in terms of an integral over
the matrix-valued power spectral density~$\mvchspecfunp$ of the multivariate
channel process~$\{\mvch[\dtime]\}$. By direct application
of~\cite[Th.~3.4]{miranda00-02a}, an extension of Szeg\"o's theorem 
(on the asymptotic eigenvalue distribution of Toeplitz matrices) to two-level Toeplitz matrices, we obtain
\be
	\limintime\frac{1}{\tslots\tstep}\logdet{\imat_{\tfslots}
		+\frac{\Pave\tfstep}{\bandwidth}\mvschcovmat}
	=\frac{1}{\tstep}\int_{-1/2}^{1/2}\logdet{\imat_{\fslots}
		+\frac{\Pave\tfstep}{\bandwidth}\mvchspecfunp}d\specparam.
\label{eq:lbTFpeak-block-szegoe}
\ee
The lower bound that results upon substitution of~\fref{eq:lbTFpeak-block-szegoe} into~\fref{eq:lbTFpeak-non-timeshare} can be tightened by time-sharing~\cite[Cor.~2.1]{sethuraman05-09b}:
we allow the input signal to have squared magnitude~$\param\Pave\tfstep/
\bandwidth$ during a fraction~$1/\param$ of the total transmission time,
where~$1\le\param\le\papr$; that is, we set~$\inp=\sqrt{\param}\cmiidinp$ during
this time; for the remaining transmission time, the transmitter is silent, so
that the constraint on the average power is satisfied.
\end{IEEEproof}

 The evaluation of~\lbTFpeakp in~\fref{eq:lbTFpeak} is complicated by two facts:
\begin{inparaenum}[(i)]
\item the mutual information~$\mi(\outp;\inp\given\ch)$ in the first term
	on the RHS of~\fref{eq:lbTFpeak} needs to be evaluated for a 
	constant-modulus input;
\item the eigenvalues of~\mvchspecfunp in the second term (the penalty term) can
in general not be
	derived in closed form.
\end{inparaenum}
While efficient numerical algorithms exist to evaluate the coherent mutual information~$\mi(\outp;\inp\given\ch)$ for constant-modulus inputs~\cite{he05-05a}, numerically computing the eigenvalues of
the~$\fslots \times \fslots$ matrix~\mvchspecfunp is challenging for channels of very wide bandwidth because the matrix~\mvchspecfunp will be large.
In the following lemma, we present two bounds on the second term of~\lbTFpeakp that are easy to compute.

\begin{lem}
\label{lem:lbTFpeak-penalty-bound}
Let
\ba
\label{eq:lb-diag-elem}
	\diagelcirci=\Re\lefto\{\frac{2}{\fslots}\sumz{\dfreq}{\fslots}
		(\fslots-\dfreq)\chcorr[0,\dfreq]\cexn{\frac{\indvar\dfreq}{\fslots}}
		\right\}	-1.
\ea
Then, the penalty term in~\fref{eq:lbTFpeak} (for the case $\param=1$) can be bounded as follows:
\begin{multline}
	2\maxDoppler \sumz{\indvar}{\fslots}\log\lefto(1 + \frac{\Pave\fstep}
		{2\maxDoppler\bandwidth}\diagelcirci\right)\\
	\ge	 \frac{1}{\tstep}\!\int_{-1/2}^{1/2}\logdet{\imat_{\fslots}
		+\frac{\Pave\tfstep}{\bandwidth}\mvchspecfunp}d\specparam \\
	\ge \bandwidth\!\spreadintlog{ 1 + \frac{\Pave}{\bandwidth}\scafunp}.
\label{eq:lbTFpeak-penalty-bound}	
\end{multline}
Furthermore, the following asymptotic results hold:
\begin{itemize}
\item The penalty term and its lower bound
	in~\fref{eq:lbTFpeak-penalty-bound} have the same Taylor series expansion
	around the point~$1/\bandwidth=0$ up to any order.
\item For scattering functions that are flat in the Doppler domain, i.e., that
	satisfy\footnote{The multiplication by~$1/(2\maxDoppler)$ in~\eqref{eq:doppler-flat-scafun} follows from the normalization~$\pathloss=1$.}
	\ba
	\label{eq:doppler-flat-scafun}
		\scafunp&=\frac{1}{2\maxDoppler}\pdepp, \qquad (\doppler, \delay)
		\in [-\maxDoppler,\maxDoppler]\times [-\maxDelay,\maxDelay],
	\ea
	the upper bound and the lower
	bound in~\fref{eq:lbTFpeak-penalty-bound} have the same Taylor series expansion 
	around the point~$1/\bandwidth=0$ up to any order.
	\end{itemize} 
\end{lem}
\begin{IEEEproof}
See \fref{app:penalty term}.
\end{IEEEproof}

The bounds~\fref{eq:lbTFpeak-penalty-bound} on the penalty term allow us to further bound~\lbTFpeakp.
If we replace the penalty term in~\fref{eq:lbTFpeak} with its upper bound
in~\fref{eq:lbTFpeak-penalty-bound}, we obtain the following lower bound on~\lbTFpeakp
and, hence, on capacity
\ba
\label{eq: second lower bound}
	\lbTFpeakp \geq \lbTFpeakcfp = \max_{1\le\param\le\papr}\Biggl\{\frac{\bandwidth}{
		\param\tfstep}\mi(\outp;\inp\given\ch)- \frac{2\maxDoppler}{\param} 
		\sumz{\indvar}{\fslots} \log\lefto(1 + \frac{\param\Pave \fstep}
		{2\maxDoppler \bandwidth}\diagelcirci \right) \Biggl\}.
\ea
The lower bound~\lbTFpeakcfp can be evaluated
numerically in a much more efficient way than~\lbTFpeakp because the
coefficients~$\{\diagelcirci\}$ can be computed from the samples $\{(\fslots-\dfreq)\chcorr[0,n]\}$ through the discrete Fourier Transform~(DFT).
If, instead, we replace the penalty term in~\fref{eq:lbTFpeak} with its lower bound
in~\fref{eq:lbTFpeak-penalty-bound} we obtain
\ba\label{eq:lbapprox}
	\lbTFpeakp \leq\lbapproxp\define \max_{1\le\param\le\papr}\Biggl\{
		\frac{\bandwidth}{\param\tfstep}\mi(\outp;\inp\given\ch)-
		\frac{\bandwidth}{\param}\spreadintlog{1+\frac{\param\Pave}
		{\bandwidth}\scafunp}\Biggl\}.
\ea
Furthermore, for large bandwidth we can replace the coherent mutual information~$\mi(\outp;\inp\given\ch)$
in~\fref{eq:lbapprox} with its second-order Taylor series
expansion~\cite[Th.~14]{verdu02-06a} to obtain the approximation
\ba
\label{eq:lbcoarserapprox}
	\lbapproxp\approx\lbaapproxp\define \max_{1\le\param\le\papr}\Biggl\{\Pave
		-\frac{\param\Pave^2\tfstep}{\bandwidth}- \frac{\bandwidth}{\param}
		\spreadintlog{1+\frac{\gamma\Pave}{\bandwidth}\scafunp}\Biggr\}.
\ea
It follows from~\fref{lem:lbTFpeak-penalty-bound} that~$\lbTFpeakp$ and $\lbapproxp$
have the same Taylor series expansion around~$1/\bandwidth=0$ up to any order, so
that~$\lbTFpeakp \approx \lbapproxp\approx\lbaapproxp$ for large enough $\bandwidth$. Furthermore,
for scattering functions that satisfy~\fref{eq:doppler-flat-scafun} (e.g., a brick-shaped 
scattering function), also~$\lbTFpeakp$ and $\lbTFpeakcfp$ have the same Taylor series expansion around~$1/\bandwidth=0$ up to any order. Hence,~$\lbTFpeakcfp \approx \lbTFpeakp \approx \lbapproxp$ for large enough~$\bandwidth$, for scattering functions that satisfy~\fref{eq:doppler-flat-scafun}.

\subsection{Numerical Example}
\label{sec:num-eval}

\begin{figure}
\centering
	\includegraphics[width=\figwidth]{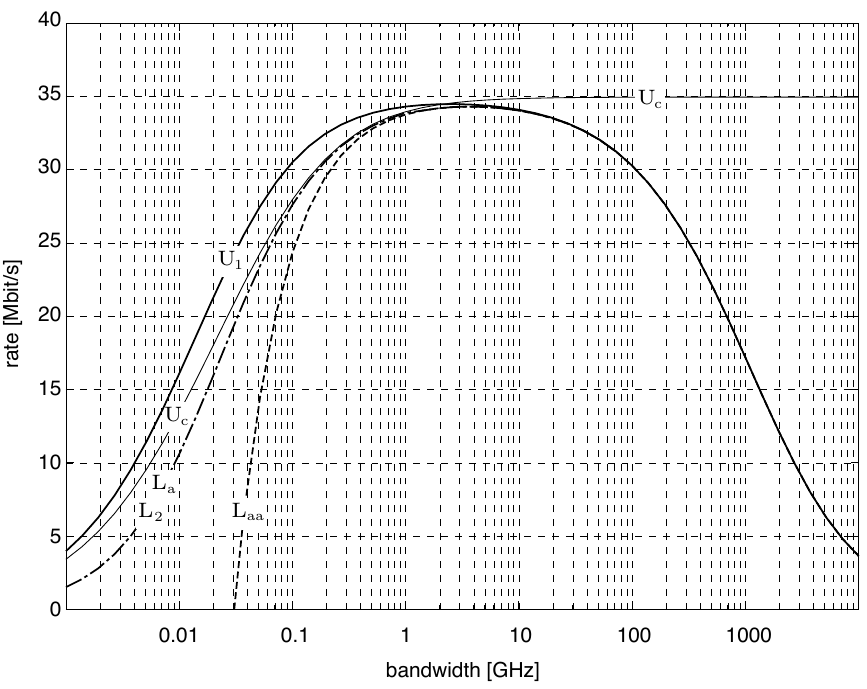}
  	\caption{The upper bounds~\ubcohp in~\fref{eq:ub-coh} and~\ubTFpeakp
		in~\fref{eq:ubTFpeak}, as well as the
		lower bound~\lbTFpeakcfp in~\fref{eq: second lower bound}, and the 
		large-bandwidth approximations of~\lbTFpeakp in~\fref{eq:lbapprox}
		and~\fref{eq:lbcoarserapprox}  for~$\papr=1$ and a brick-shaped
		scattering function with spread~$\spread=10^{-5}$.}
\label{fig:bounds-TFpeak}
\end{figure}

We next evaluate the bounds found in the previous section for the following set of
practically relevant system parameters:
\begin{itemize}
\item Brick-shaped scattering function with maximum delay~$\maxDelay=0.5\mus$,
	maximum Doppler shift~$\maxDoppler=5\Hz$, and corresponding
	spread~$\spread=4\maxDelay\maxDoppler=10^{-5}$.
\item Grid parameters~$\tstep=0.35\ms$ and~$\fstep=3.53\kHz$, so that~$\tfstep
	\approx1.25$ and~$\tstep/\fstep=\maxDelay/\maxDoppler$, as suggested by the
	design rule~\fref{eq:grid-matching}.
\item Receive power normalized with respect to the noise spectral density 
	\bas
		\frac{\Pave}{1 \W/\Hz}=2.42\cdot10^{7}\inv{\sec}.
	\eas
\end{itemize} 
These parameter values are representative for several different types of systems. For example:
\begin{enumerate}[(a)]
\item An IEEE~802.11a system with transmit power of~200\,mW, pathloss
	of~118\dB, and receiver noise figure~\cite{razavi98a} of~5\dB; the
	pathloss is rather pessimistic for typical indoor link distances and
	includes the attenuation	of the signal, e.g.,  by a concrete wall.
\item A UWB system with transmit power of~0.5\,mW, pathloss of~77\dB,
	and receiver noise figure of~20\dB.
\end{enumerate}

\fref{fig:bounds-TFpeak} shows the upper bounds~\ubcohp in~\fref{eq:ub-coh}
and~\ubTFpeakp in~\fref{eq:ubTFpeak}, as well as the lower bound~\lbTFpeakcfp
in~\fref{eq: second lower bound}, and the large-bandwidth approximations~\lbapproxp in~\fref{eq:lbapprox} and~\lbaapproxp in~\fref{eq:lbcoarserapprox}, all
for~$\papr=1$. 

As brick-shaped scattering functions are flat in the Doppler domain, i.e., they
satisfy the condition in~\fref{eq:doppler-flat-scafun}, it follows from~\fref{lem:lbTFpeak-penalty-bound} that the difference
between~\lbapproxp and the lower bound~\lbTFpeakcfp
in~\fref{eq: second lower bound} vanishes as~$\bandwidth\to\infty$. For our
choice of parameters, this difference is so small even for finite bandwidth
that the curves for~\lbapproxp and the lower
bound~\lbTFpeakcfp  cannot be distinguished in
\fref{fig:bounds-TFpeak}.
As~$\lbTFpeakcfp \leq\lbTFpeakp \leq \lbapproxp$, the lower bound~\lbTFpeakp is fully characterized as well.

The upper bound~\ubTFpeakp and the lower bound~\lbTFpeakp take on their maximum at
a large but finite bandwidth; beyond this {\em critical} bandwidth, additional
bandwidth is detrimental and the capacity approaches zero as bandwidth
increases further.
In particular, we can see from \fref{fig:bounds-TFpeak} that many current
wireless systems operate well below the critical bandwidth. It can furthermore
be verified numerically that  the critical bandwidth
increases with decreasing spread, consistent with our analysis in
\fref{sec:ubTFpeak-characterisitcs}. We also observed that the gap between upper and lower bounds increases with increasing~\papr.

For bandwidths smaller than the critical bandwidth,~\lbTFpeakp comes quite close to the coherent upper bound~\ubcohp; this seems to validate, at least for the setting considered, the standard receiver design principle to first estimate the channel, and then use the resulting estimates as if they were perfect.

The approximate lower bound~\lbaapproxp in~\fref{eq:lbcoarserapprox} is accurate for
bandwidths above the critical bandwidth and very loose otherwise. Furthermore,
\ubTFpeakp and~\lbaapproxp seem to
fully characterize~\capacityp in the large-bandwidth regime. We will
make this statement precise in the next section, where we relate~\ubTFpeakp
and~\lbTFpeakp to the first-order Taylor series expansion of~\capacityp around the
point~$1/\bandwidth=0$.

\subsection{Capacity in the Infinite-Bandwidth Limit}
\label{sec:TFpeak-cap-infbw}

The plots in \fref{fig:bounds-TFpeak} of the upper bound~\ubTFpeakp and the lower bound~\lbTFpeakp seem to coincide for large bandwidth, yet it is not clear a priori if the two bounds allow to characterize capacity in the limit for~$\bandwidth\to\infty$. To address this question, we next investigate if both bounds have the same first-order Taylor series expansion in~$1/\bandwidth$ around the point~$1/\bandwidth=0$. 

Because the upper bound~\ubTFpeakp in~\fref{eq:ubTFpeak} takes on two different forms,
depending on the value of the parameter~\avpoptp in~\fref{eq:ubTFpeak-avpopt},
its first-order Taylor series is somewhat tedious to derive. We state the result
in the following lemma and provide the derivation in
\fref{app:ubTFpeak-taylor-proof}.

\begin{lem}
\label{lem:ubTFpeak-taylor}
Let
\ba
	\peakiness &\define \spreadint{\scafunpsq}.
\label{eq:peakiness}
\ea
Then, the upper bound~\fref{eq:ubTFpeak} in \fref{thm:ubTFpeak} admits
the following first-order Taylor series expansion around the point~$1/
\bandwidth=0$:
\begin{subequations}
\ba
	\ubTFpeakp &= \frac{\taylorone}{\bandwidth}+\osmall{\frac{1}
		{\bandwidth}}\\
\intertext{where}
	\taylorone = \liminbw\bandwidth\ubTFpeakp &= 
	\begin{cases}
		\dfrac{\Pave^{2} }{2}\left(\papr\peakiness-\tfstep\right), 
		&\quad\text{if } \papr > \dfrac{2\tfstep}{\peakiness}\\[0.3cm]
		\dfrac{(\papr\Pave\peakiness)^{2}}{8\tfstep},
			&\quad\text{if }\papr\leq \dfrac{2\tfstep}
			{\peakiness}.
	\end{cases}
\label{eq:TFpeak-taylorone-coeff}
\ea
\label{eq:TFpeak-taylor}
\end{subequations}
\end{lem}

We show in \fref{app:lbTFpeak-taylor-proof} that the corresponding Taylor series
expansion of the lower bound~\lbTFpeakp in~\fref{eq:lbTFpeak} does not have the
same first-order term~\taylorone. This result is formalized in the following
lemma.

\begin{lem}
\label{lem:lbTFpeak-taylor}
The lower bound~\fref{eq:lbTFpeak} in \fref{thm:lbTFpeak}
admits the following first-order Taylor series expansion around the point~$1/
\bandwidth=0$:
\begin{subequations}
\ba
	\lbTFpeakp &= \frac{\tayloronelb}{\bandwidth}+\osmall{\frac{1}
		{\bandwidth}} \\
\intertext{where}
	\tayloronelb = \liminbw\bandwidth\lbTFpeakp&= \papr \Pave^{2}
		\left(\frac{\peakiness}{2} - \tfstep\right).
		\label{eq:lbTFpeak-taylor}
\ea	
\end{subequations}
\end{lem}

As~\taylorone in~\fref{eq:TFpeak-taylorone-coeff} and~\tayloronelb
in~\fref{eq:lbTFpeak-taylor} are different, the two bounds~\ubTFpeakp and~\lbTFpeakp do not fully
characterize~\capacityp in the wideband limit. In the next theorem, we show, however, that the first-order Taylor series of~\ubTFpeakp in \fref{lem:ubTFpeak-taylor} indeed correctly characterizes~\capacityp for~$\bandwidth\to\infty$.

\begin{thm}
\label{thm:TFpeak-taylor}
Consider an underspread Rayleigh-fading channel with scattering
function~$\scafunp$. Assume that the channel input~\mvsinp satisfies the
average-power constraint~\avP and the peak 
constraint~$\abs{\inp[\dtdf]}^{2}\le\Ppeak\tstep/\fslots$~\wpone.
The capacity~\capacityp of this channel has a first-order Taylor series
expansion  around the point~$1/\bandwidth=0$ equal
to the first-order Taylor series expansion in~\fref{eq:TFpeak-taylor}.
\end{thm}

\begin{IEEEproof}
We need a capacity lower bound different from~\lbTFpeakp with the same asymptotic behavior for~$\bandwidth\to\infty$ as the upper bound~\ubTFpeakp. The key element in the
derivation of this new lower bound is an extension of the block-constant
signaling scheme used in~\cite{sethuraman08a} to prove
asymptotic capacity results for frequency-flat time-selective channels.
In particular, we use input signals with uniformly distributed phase whose
magnitude is toggled on and off at random with a prescribed probability; hence, information is encoded jointly in the amplitude and in the phase. In
comparison, the signaling scheme used to obtain~\lbTFpeakp transmits a signal of
constant amplitude in all time-frequency slots. We present the details of the
proof in \fref{app:altlbTFpeak-taylor}.
\end{IEEEproof}

Similar to the capacity behavior of a discrete-time frequency-flat time-selective channel for vanishing SNR~\cite{sethuraman08a}, the first-order Taylor series coefficient in~\fref{eq:TFpeak-taylorone-coeff} can take on two different forms as a function of the channel parameters. However, the link in~\fref{eq:specfun-scafun} between the discretized channel and the WSSUS channel~$\CHop$ allows us to conclude that~$\papr>2\tfstep/\peakiness$ and thus~$\taylorone=\Pave^{2}(\papr\peakiness-\tfstep)/2$ for virtually all channels of practical interest.
In fact, by Jensen's inequality, $\peakiness\ge\inv{\spread}$ (with equality for brick-shaped scattering functions), so that~$2\tfstep\spread\ge2\tfstep/\peakiness$, and a sufficient condition for~$\papr>2\tfstep/\peakiness$ is~$\papr>2\tfstep\spread$. For typical values of~\tfstep (e.g.,~$\tfstep\approx 1.25$) and typical values of~\spread (e.g.,~$\spread<10^{-2}$), this latter condition is satisfied for any admissible~\papr.

We state in \fref{lem:lbTFpeak-taylor} that the first-order term~\tayloronelb in the Taylor series expansion of the lower bound~\lbTFpeakp does not match the corresponding term~\taylorone of the Taylor series expansion of capacity, not even for realistic channel parameters as just discussed. Yet, the plots of the upper bound~\ubTFpeakp and the lower bound~\lbTFpeakp in \fref{fig:bounds-TFpeak} seem to coincide at large bandwidth. This observation is not surprising as the ratio
\be
	\tayloronelb/\taylorone = \frac{\papr(\peakiness /2 -\tfstep)}{(1/2)
		(\peakiness \papr -\tfstep)}
\een
approaches~$1$ for~\papr and~\tfstep fixed as~\peakiness grows large. For example, we have~$\tayloronelb/
\taylorone=0.998$ for the same parameters we used for the numerical evaluation
in \fref{sec:num-eval}, i.e.,~$\spread=10^{-3}$, $\papr=1$, and~$\tfstep=1.25$.

\section{Infinite-Bandwidth Capacity under a Peak Constraint in Time}
\label{sec: peak in time}

So far we considered a peak constraint in time and frequency; we now analyze the
case when the input signal is subject to a peak constraint in time only,
according to~\fref{eq:peak-per-tslot}. The average-power constraint~\avP
remains in force. In addition, we focus on the infinite-bandwidth limit. By
means of a capacity lower bound that is explicit in the channel's scattering
function, we show that the phenomenon of vanishing capacity in the wideband
limit can be eliminated if we allow the transmit signal to be peaky in
frequency. Furthermore,  using the same approach as in the proof of
\fref{thm:ubTFpeak}, we obtain an upper bound on the infinite-bandwidth
capacity that, for~$\fstep=1/(2\maxDelay)$, differs from the corresponding lower
bound only by a Jensen penalty term. The two bounds
coincide for brick-shaped scattering functions when~$F=1/(2\maxDelay)$.

The infinite-bandwidth capacity of the channel~\fref{eq:scalar-io} is defined as
\be
	\infcapacity =  \liminfreq \limintime \sup_{\dsetpeaktime} \frac{1}{\tslots
		\tstep} \mi(\mvsoutp;\mvsinp),
\label{eq:infinite-bandwidth capacity}
\ee
where the supremum is taken over the set~\dsetpeaktime of all input distributions
that satisfy the peak constraint~\fref{eq:peak-per-tslot} and the
constraint~$\Ex{}{\vecnorm{\mvsinp}^{2}}\le\tslots\Pave\tstep$ on the average
power.

\subsection{Lower Bound}

We obtain a lower bound on~\infcapacity by evaluating the mutual information in~\eqref{eq:infinite-bandwidth capacity} for a specific signaling scheme. As signaling scheme, we 
consider a generalization in the channel's eigenspace of the on-off FSK scheme proposed
in~\cite{gursoy06a}. The resulting lower bound is given in the
following theorem.

\begin{thm}
\label{thm:viterbi} 
Consider an underspread Rayleigh-fading channel with scattering
function~$\scafunp$; assume that the channel input~\mvsinp satisfies the
average-power constraint~\avP and the peak 
constraint~$\sum_{\dfreq=0}^{\fslots-1}\abs{\inp[\dtdf]}^{2}\le {\papr\Pave
\tstep}$~\wpone. The infinite-bandwidth capacity of this channel is
lower-bounded as~$\infcapacity \geq \lbTpeak$, where
\ba
\label{eq:viterbi} 
	\lbTpeakp &= \Pave -\frac{1}{\papr} \dopplerintlog{1 + \papr \Pave \pDopp}
\ea
and~$\pDopp =\delayint{\scafunp}$ denotes the power-Doppler
profile of the channel.
\end{thm}
\begin{IEEEproof}
See \fref{app:viterbi-result-proof}.
\end{IEEEproof}
For~$\papr=1$, the lower bound in~\fref{eq:viterbi} coincides with Viterbi's result
on the rates achievable on an AWGN channel with complex Gaussian input
signals with spectral density~\pDopp, modulated by FSK tones~\cite[Eq.~(39)]{viterbi67-07a}. 
Viterbi's setup is relevant for our analysis, because, for a WSSUS channel with power-Doppler profile~\pDopp,
the output signal that corresponds to an FSK tone can be well-approximated by Viterbi's transmit signal whenever
the observation interval at the receiver is large and the maximum delay $\maxDelay$ of the channel is much smaller
than the observation interval~\cite[Sec.~8.6]{gallager68a}.
The proof technique used to
obtain \fref{thm:viterbi} is, however, conceptually different from that in~\cite{viterbi67-07a}.   On the basis of the interpretation of Viterbi's signaling scheme provided above, we can summarize the proof technique in~\cite{viterbi67-07a} as follows: first, a signaling scheme is chosen, namely FSK, for transmission over a WSSUS channel; then, the resulting stochastic process at the channel output is discretized by means of a Karhunen-Lo\`{e}ve decomposition; finally, the result on the achievable rates in~\cite[Eq.~(39)]{viterbi67-07a} follows from an error exponent analysis of the discretized stochastic process and from~\cite[Lemma~8.5.3]{gallager68a}---Szeg\"o's theorem on the asymptotic eigenvalue distribution of self-adjoint Toeplitz operators. 

To prove \fref{thm:viterbi}, on the other hand, we first discretize the
WSSUS underspread channel;  the rate achievable
for a specific signaling scheme, which resembles~FSK, yields then the infinite-bandwidth
capacity lower bound~\eqref{eq:viterbi}. The main tool used in the
proof of \fref{thm:viterbi} is a property of the information divergence of
FSK~constellations, first presented by Butman \& Klass~\cite{butman73-09a}.

For~$\papr\to\infty$, i.e., when the input signal is subject only to an
average-power constraint,~\lbTpeak in~\fref{eq:viterbi} approaches the
infinite-bandwidth capacity of an AWGN channel with the same receive
power, as previously demonstrated by Gallager~\cite{gallager68a}. The signaling scheme used in the proof of \fref{thm:viterbi} 
is, however, not the only scheme that approaches this limit when no peak
constraints are imposed on the input signal. In~\cite{durisi06-07a} we presented
another signaling scheme, namely, \emph{TF pulse position modulation}, which
exhibits the same behavior. The proof of~\cite[Th.~1]{durisi06-07a} is
similar to the proof of \fref{thm:viterbi} in \fref{app:viterbi-result-proof}.

\subsection{Upper Bound}

In \fref{thm:inf-bw-ub} below we present an upper
bound on~\infcapacity and identify a class of scattering functions for which
this upper bound and the lower bound~\fref{eq:viterbi} coincide if~$\fstep
= 1/(2\maxDelay)$. Differently, from the lower bound, which can be obtained
both by Viterbi's approach and through our approach, the upper bound presented below
is heavily built on the discretization of the continuous-time WSSUS underspread
channel presented in~\fref{sec:diagonalization}.

\begin{thm}
\label{thm:inf-bw-ub}
Consider an underspread Rayleigh-fading channel with scattering
function~$\scafunp$; assume that the channel input~\mvsinp satisfies the
average-power constraint~\avP and the peak 
constraint~$\sum_{\dfreq=0}^{\fslots-1}\abs{\inp[\dtdf]}^{2}\le {\papr\Pave
\tstep}$~\wpone. The infinite-bandwidth capacity of this channel is
upper-bounded as~$\infcapacity \leq \ubTpeak$, where
\ba
\label{eq:ubTpeak}
     \ubTpeakp = \Pave - \frac{\fstep}{\papr} \spreadintlog{1+ \frac{\papr\Pave}
     	{\fstep} \scafunp}.
\ea
\end{thm} 
\begin{IEEEproof}
See \fref{app:viterbi-ub-proof}.
\end{IEEEproof}

As the upper bound~\fref{eq:ubTpeak} is a decreasing function of~\fstep, and 
as~\fstep has to satisfy the Nyquist condition~$\fstep\leq 1/(2\maxDelay)$, the
upper bound is minimized when~$\fstep=1/(2\maxDelay)$. 
For this value of~\fstep, Jensen's inequality applied to the second term on the RHS of~\fref{eq:ubTpeak}
yields:
\be
\label{eq: Jensen's penalty}
\bs	
	\frac{1}{2\maxDelay \papr}\spreadintlog{1+2\maxDelay\papr\Pave\scafunp}
		&\leq \frac{1}{\papr}\dopplerintlog{1+\papr\Pave\!\delayint{\!\scafunp}}\\
	&=\frac{1}{\papr}\dopplerintlog{1+\papr\Pave \pDopp}.
\es
\ee
Hence, for~$\fstep=1/(2\maxDelay)$, the upper bound~\fref{eq:ubTpeak} and the lower
bound~\fref{eq:viterbi} differ only by a Jensen penalty term.
It is interesting to observe that the Jensen penalty
in~\fref{eq: Jensen's penalty} is zero whenever  the scattering function is flat in
the delay domain, i.e., whenever~\scafunp is of the form\footnote{The multiplication by~$1/(2\maxDelay)$ in~\eqref{eq: scattering function flat in delay domain} follows from the normalization~$\pathloss=1$.}
\ba
\label{eq: scattering function flat in delay domain}
	\scafunp = \frac{1}{2\maxDelay} \pDopp,\qquad (\doppler, \delay)
		\in [-\maxDoppler,\maxDoppler]\times [-\maxDelay,\maxDelay].
\ea
In this case, upper bound and lower bound coincide and the infinite
bandwidth capacity~\infcapacity is fully characterized by
\ba
\label{eq: infinite bandwidth capacity, parallel channels}	
	\infcapacity=\Pave -\frac{1}{\papr} \dopplerintlog{1 + \papr \Pave
	\pDopp}.
\ea

Expressions similar to~\fref{eq: infinite bandwidth capacity, parallel channels} 
were found in~\cite{sethuraman05-09a} for the capacity per unit energy of a discrete-time frequency-flat time-selective
channel, and in~\cite{sethuraman06-07a,zhang07-01a} for the
infinite-bandwidth capacity of the continuous-time counterpart of the same
channel; in all cases a peak
constraint is imposed on the input signals. However, the results
in~\cite{sethuraman06-07a,zhang07-01a,sethuraman05-09a} and our results are not directly related, as discussed next.

\subsubsection{Comparison with~\cite{sethuraman06-07a,zhang07-01a}}
The continuous-time time-selective frequency-flat channel analyzed
in~\cite{sethuraman06-07a,zhang07-01a} belongs to the class of LFI channels.
As explained in~\fref{sec:LTI and LFI}, the kernel of an LFI channel cannot be diagonalized as was done
in \fref{sec:diagonalization} because LFI channels are not of Hilbert-Schmidt type.  Hence, the
infinite-bandwidth capacity expressions found
in~\cite{sethuraman06-07a,zhang07-01a} cannot be obtained from our upper and
lower bounds simply by an appropriate choice of the scattering function~\scafunp
and of the grid parameters~\tstep and~\fstep.

\subsubsection{Comparison with~\cite{sethuraman05-09a}}
For scattering functions that are flat in the delay domain [see~\eqref{eq: scattering function flat in delay domain}], the discrete correlation function~$\chcorr[\Ddtime,\Ddfreq]$ of our channel is given by
\bas
	\chcorr[\Ddtime,\Ddfreq] &=\spreadint{\scafunp \cex{(\dtime\tstep\doppler
		-\dfreq\fstep\delay)}} \\
	&= \frac{\sin(2\pi\dfreq\fstep\maxDelay)}{2\pi\dfreq\fstep\maxDelay}
		\dopplerint{\pDopp\cex{\dtime\tstep\doppler}}.
\eas
If we replace~$\fstep$ by~$1/ (2\maxDelay)$, we obtain 
\bas
	\chcorr[\Ddtime,\Ddfreq] =\krond[\dfreq]\dopplerint{\pDopp\cex{\dtime\tstep
		\doppler}}.
\eas
Hence, for scattering functions that satisfy~\fref{eq: scattering function flat in delay domain}, and
for~$\fstep=1/ (2\maxDelay)$, the discrete channel~$\ch[\dtime,\dfreq]$
is uncorrelated in frequency~$\dfreq$.
Consequently, the input-output relation~\fref{eq:vec-io} reduces to the
input-output relation of~\fslots parallel~\iid flat fading channels that are
selective in time. However, as both the average power constraint and  the peak
constraint are imposed on the overall channel and not on each parallel channel
separately, the infinite-bandwidth capacity~\fref{eq: infinite bandwidth
capacity, parallel channels} does not follow simply from the capacity per unit
energy of one of the parallel channels obtained in~\cite{sethuraman05-09a}.

\section{Conclusions}
\label{sec:conclusions}

The underspread Gaussian WSSUS channel with a peak constraint on the input
signal is a fairly accurate and general model for wireless channels. Despite
the model's mathematical elegance and simplicity, it appears to be difficult to compute the
corresponding capacity. To nonetheless study capacity as a function of
bandwidth, we have taken a three-step approach: we first approximated the
kernel of the continuous-time WSSUS channel by a kernel that can be diagonalized,
and obtained an equivalent discretized channel;
in a second step, we derived upper and lower bounds on
the capacity of this discretized channel, and in a third step we expressed these bounds in terms of the scattering function of the original continuous-time WSSUS
channel. In \fref{sec:model} and~\fref{app:ch-approx-error}, we partially characterize the approximation error that arises when the 
original continuous-time underspread WSSUS channel operator is replaced by a normal operator whose eigenfunctions are a Weyl-Heisenberg set. A complete characterization of the approximation error would require to quantify the difference between the null spaces and between the range spaces of the original operator and its approximation. This characterization is a fundamental open problem, even for deterministic operators.

The capacity bounds derived in this paper are explicit in the channel's scattering function, a quantity that can be obtained from channel measurements. Furthermore, the capacity bounds may serve as an efficient design tool even when the scattering function is not known completely, and the channel is only characterized coarsely by its maximum delay~$\maxDelay$ and maximum Doppler shift~\maxDoppler. In particular, one can assume that the scattering function is brick-shaped within its support area~$[-\maxDoppler,\maxDoppler]\times[-\maxDelay,\maxDelay]$ and evaluate the corresponding bounds. As shown in~\fref{sec:shape scattering function} a brick-shaped scattering function results  in the lowest upper bound for given~\maxDelay and~\maxDoppler. Furthermore, the bounds are particularly easy to evaluate for brick-shaped scattering functions and result in analytical expressions explicit in the channel spread~\spread. Extensions of the capacity bounds for input signals subject to a peak constraint in time and frequency to the case of spatially correlated MIMO channels are provided in~\cite{schuster08-01a}.

The multivariate discrete-time channel model considered in this paper,~$\mvoutp[\dtime]=\mvch[\dtime]\had\mvinp[\dtime]+\mvwgn[\dtime]$, and
the corresponding capacity bounds are also of interest in their own right,
without the connection to the underlying WSSUS channel. The individual elements of
the vector~$\mvch[\dtime]$ do not necessarily need to be interpreted as discrete
frequency slots; for example, the block-fading model with correlation across
blocks in~\cite{liang04-09a} can be cast into the form of our multivariate
discrete-time model as well.

As our model is a generalization of the time-selective, frequency-flat channel model, it is not
surprising that the structure of our bounds for the case of a
peak constraint both in time and frequency, and a peak constraint in time only,
is similar to the corresponding
results in~\cite{sethuraman05-09b,sethuraman08a}
and~\cite{sethuraman06-07a,zhang07-01a,sethuraman05-09a},
respectively. The key difference between our proofs and  the proofs
in~\cite{sethuraman05-09a,sethuraman08a,sethuraman06-07a} is that 
our derivation of the upper
bounds~\fref{eq:ubTFpeak} and~\fref{eq:ubTpeak} (see
\fref{app:mmse-mi} and \fref{app:viterbi-ub-proof},
respectively) is based on the
relation between mutual information and MMSE described in~\cite{guo05-04a}.
Compared to the proof
in~\cite[Sec.~VI]{sethuraman05-09a}, our approach has the advantage that it
can easily be generalized to multiple dimensions---in our case time and frequency---and 
provides the new lower bound~\eqref{eq:logdetinf-1d}.

Numerical evaluation indicates that our bounds are surprisingly accurate over a
large range of bandwidth. For small bandwidth and hence high SNR, however, 
our bounds are no longer tight, and a refined analysis along the lines
of~\cite{lapidoth05-07a,chen07-12a} is called for. In the time-selective frequency-flat
case, it was shown in~\cite{lapidoth05-07a} that the high-SNR capacity behavior
depends heavily on the spectral density of the channel process. In particular, if the spectral
density is zero on a set of positive measure, capacity grows logarithmically in~SNR, otherwise
the growth is slower, and can even be double-logarithmic. For the more general time- and frequency-selective channel
considered in this paper, the assumption that the scattering function is compactly supported 
implies that the matrix-valued spectral density~\fref{eq:mvspec} of the multivariate discrete-time process is
zero on a set of positive measure whenever~$\tstep<1/(2\maxDoppler)$. This implies that the capacity of the approximating channel operator grows logarithmically at high SNR~\cite{chen07-12a} whenever the sampling rate in time is strictly larger than the Nyquist rate. The high-SNR behavior of the capacity of the original channel operator might be different, though. In the approximating discrete-time discrete-frequency input-output relation~\eqref{eq:scalar-io},~ISI and~ICI are neglected~[see~\eqref{eq:psofdm-rx}]. But the high-SNR behavior of a fading channel is heavily influenced by~ISI and~ICI, as recently shown in~\cite{koch08-05a}.

The approximate kernel diagonalization presented in~\fref{sec:diagonalization} can be extended to WSSUS channels with non-compactly supported scattering function, as long as the area of the {\em effective} support of the scattering function is small~\cite{matz98-08a}. The capacity bounds corresponding to a non-compactly supported scattering function are, however, more difficult to evaluate numerically, because the periodic repetitions of the scattering function in~\eqref{eq:specfun-scafun} fall inside the integration region.

A challenging open problem is to characterize the capacity behavior of
{\em overspread} channels, i.e., channels with spread~$\spread >1$. The major difficulty resides in the fact that
a set of deterministic eigenfunctions can no longer be used to diagonalize the random kernel of the channel.

\appendices

\section{}
\label{app:ch-approx-error}

\subsection{Approximate Eigenfunctions and Eigenvalues of the Channel Operator}

The construction of the approximating channel operator in
\fref{sec:diagonalization} relies on the following two properties of underspread operators:
\begin{itemize}
\item Time and frequency shifts
of a time- and frequency-localized prototype signal~\logonp  matched to the channel's
scattering function~\scafunp, are approximate eigenfunctions of~$\CHop$.
\item Samples of the time-varying transfer function~\tvtfp are the corresponding
approximate eigenvalues. 
\end{itemize}
In this appendix, we make these claims more precise and
give bounds on the mean-square approximation error---averaged with respect to the channel's realizations---for both approximate
eigenfunctions and eigenvalues. The results presented in the remainder of this appendix
are not novel, as they already appeared elsewhere, sometimes in different form~\cite{kozek97a,matz98-08a,kozek98-10a,matz03a}; the goal of this appendix is rather to provide
a self-contained exposition.
\subsubsection{Ambiguity function}
The design problem for~\logonp can be restated in terms of its {\em ambiguity
function}~\afp, which is defined
as~\cite{woodward53a}
\bas
	\afp\define\int_{\time}\logon(\time)\conj{\logon}(\time-\delay)
		\cexn{\doppler\time} d\time.
\eas
Without loss of generality, we can assume that~\logonp is normalized, so
that~$\af_{\logon}(0,0)=\vecnorm{\logon}^{2}=1$.  For two signals~\logonp and~\logonaltp,
the {\em cross-ambiguity function} is defined as
\bas
 \af_{\logon,\logonalt}(\doppler,\delay)\define\int_{\time}\logon(\time)\conj{\logonalt}(\time-\delay)
		\cexn{\doppler\time} d\time.
\eas
The following properties of  the (cross-) ambiguity function are important in our context:
%
%
\begin{prop}\label{prop:volume}
  The volume under the so-called {\em ambiguity surface}~$\abs{\afp}^{2}$ is
	constant~\cite{wilcox91a}. In particular, if~\logonp has unit energy, then
	\be
		\spreadint{\abs{\afp}^{2}}=1.
	\een
\end{prop}
%
%
\begin{prop}\label{prop:ambiguity surface}
The ambiguity surface attains its maximum magnitude at the origin:
	$\abs{\afp}^{2}\le\bigl[\af_{\logon}(0,0)\bigr]^{2}=1$, for all~\doppler and~\delay.
	This property follows from the Cauchy-Schwarz inequality, as shown in~\cite{groechenig01a}.
\end{prop}
%
\begin{prop}\label{prop:cross-ambiguity}
The cross-ambiguity function between the two time- and frequency-shifted signals 
$\slogonct(\time)\define\logon(\time-\alpha)
\cex{\beta\time}$ and $\slogonctalt(\time)\define\logon(\time-\alpha')
\cex{\beta'\time}$
is given by
\be
\label{eq:property cross-amb}
\bs
	\af_{\slogonct,\slogonctalt}(\doppler,\delay)&=\int_{\time} \logon(\time-\alpha)\cex{\beta\time}
		\conj{\logon}(\time-\alpha'-\delay)		\cexn{\beta'(\time-\delay)}\cexn{\doppler\time}d\time	\\
	&\stackrel{(a)}{=}\cex{\beta'\delay} \cexn{(\doppler+\beta'-\beta)\alpha}\int_{\time'} \logon(\time')\conj{\logon}(\time'-(\alpha'-\alpha)-\delay)
	\cexn{(\doppler+\beta'-\beta)\time'}	d\time'\\
	&=\af_{\logon}(\doppler+\beta'-\beta,\delay+\alpha'-\alpha)\cexn{(\doppler\alpha-\delay\beta')}\cexn{(\beta'-\beta)\alpha}
\es
\ee
where~(a) follows from the change of variables~$\time'=\time-\alpha$. As a direct consequence of~\eqref{eq:property cross-amb}, we have
\ba
	\af_{\slogonct}(\doppler,\delay)=\afp\cexn{(\doppler\alpha-\delay\beta)}.
\label{eq:property amb}
\ea
\end{prop}
%
\begin{prop}\label{prop:derivatives}
Let the unit-energy signal~\logonp have Fourier transform~\logonfp, and denote by~\eftime and~\efband, defined as
\ba
\label{eq:efftime and bandwidth}
\eftime^{2}=\int_{\time}\time^{2}\abs{\logonp}^{2}d\time, \quad \efband^{2}=\int_{\freq}\freq^{2}\abs{\logonfp}^{2}d\freq,
\ea
the {\em effective duration} and the {\em effective bandwidth} of~\logonp. Then~$\eftime^{2}$ and~$\efband^{2}$ are proportional to the second-order derivatives of~\afp at the point~$(\doppler,\delay)=(0,0)$~\cite{wilcox91a}
\bas
	\left.\frac{\partial^{2}\afp}{\partial\doppler^{2}}\right|_{(\doppler,\delay)=(0,0)}&=-4\pi^{2}\eftime^{2}\\
	\left.\frac{\partial^{2}\afp}{\partial\delay^{2}}\right|_{(\doppler,\delay)=(0,0)}&=-4\pi^{2}\efband^{2}.
\eas
\end{prop}
%
\begin{prop}\label{prop:channel operator}
 For the channel operator~$\CHop$ in~\fref{sec:wssus-model},
\bas
	\inner{\CHop\logon}{\logonalt}&\stackrel{(a)}{=}\iiint_{\time\,\,\delay\,\,\doppler}\spfp\logon(\time-\delay)\cex{\time\doppler}
	\conj{\logonalt}(\time)d\delay d\doppler d\time\\
	&=\spreadint{\spfp\conj{\Biggl[ \int_{\time}\logonaltp\conj{\logon}(\time-\delay)\cexn{\time\doppler} d\time\Biggr]}}\\
	&=\spreadint{\spfp\conj{\af}_{\logonalt,\logon}(\doppler,\delay)}=\inner{\spf}{\af_{\logonalt,\logon}}
\eas
where in~(a) we used~\eqref{eq:ltv-altio}.
\end{prop}

Properties~\ref{prop:volume} and~\ref{prop:ambiguity surface}, which constitute the {\em radar uncertainty principle}, imply that
it is not possible to find a signal~\logonp with a corresponding ambiguity
function~\afp that is arbitrarily well concentrated in~\doppler and~\delay~\cite{wilcox91a}. The radar uncertainty principle is a manifestation of the classical {\em Heisenberg uncertainty principle}, which states that the effective duration~\eftime and the effective bandwidth~\efband [both defined in~\eqref{eq:efftime and bandwidth}] of any signal in~\hilfunspace satisfy~$\eftime\efband\geq1/(4\pi)$~\cite[Th.~2.2.1]{groechenig01a}. In fact, when~\logonp has effective duration~\eftime, and effective bandwidth~\efband, the corresponding ambiguity function~\afp is highly concentrated on a rectangle of area~$4\eftime\efband$; but this area cannot be made arbitrarily small.
\subsubsection{Approximate Eigenfunctions}
%
\begin{lem}[\! {\cite[Ch.~4.6.1]{kozek97a}}]\label{lem:eigenfunction}
Let~$\CHop$ be a WSSUS channel with scattering
function~\scafunp. Then, for any unit-energy signal~\logonp, the mean-square approximation
error incurred by assuming that~\logonp is an eigenfunction of~$\CHop$ is given by
\ba
	\aerr_{1}=\Ex{}{\vecnorm{\inner{\CHop\logon}{\logon}\logon-\CHop\logon}^{2}}
		=\spreadint{\scafunp\left(1-\abs{\afp}^{2}\right)}.
\label{eq: eigenvalue error}
\ea
\end{lem}

\begin{IEEEproof}
We decompose~$\aerr_{1}$ as follows:
\be
\bs
\Ex{}{\vecnorm{\inner{\CHop\logon}{\logon}\logon-\CHop\logon}^{2}}&=\Ex{}{\vecnorm{\inner{\CHop\logon}{\logon}\logon}^{2}}+\Ex{}{\vecnorm{\CHop\logon}^{2}}-2\Ex{}{\abs{\inner{\CHop\logon}{\logon}}^{2}}\\
&=\Ex{}{\vecnorm{\CHop\logon}^{2}}-\Ex{}{\abs{\inner{\CHop\logon}{\logon}}^{2}}.
\es
\label{eq:decomposition proof eigenvector}
\ee
Here, the last steps follows  because~\logonp has unit energy by assumption.
We now compute the two terms in~\eqref{eq:decomposition proof eigenvector} separately. The first term is equal to
\be
\label{eq:first term eigenvector}
\bs
	\Ex{}{\vecnorm{\CHop\logon}^{2}}&\stackrel{(a)}{=}\Ex{}{\int_{\time}\abs{\spreadint{\spfp\logon(\time-\delay)\cex{\time\doppler}}}^{2}d\time}\\
	&\stackrel{(b)}{=}\spreadint{\scafunp\int_{\time}\logon(\time-\delay)\conj{\logon}(\time-\delay)d\time}\\
	&\stackrel{(c)}{=}\spreadint{\scafunp}
\es
\ee
where~(a) follows from~\eqref{eq:ltv-altio},~(b) from the WSSUS property, and~(c) from the energy normalization of~\logonp. For the second term we have
\be
\label{eq:second term eigenvector}
\bs
	\Ex{}{\abs{\inner{\CHop\logon}{\logon}}^{2}}&\stackrel{(a)}{=}\Ex{}{\abs{\inner{\spf}{\af_{\logon}}}^{2}}
	=\Ex{}{\abs{\spreadint{\spfp\conj{\af}_{\logon}(\doppler,\delay)}}^{2}}\\
	&\stackrel{(b)}{=}\spreadint{\scafunp\abs{\afp}^{2}}
\es
\ee
where~(a) follows from~\fref{prop:channel operator} and~(b) follows from the WSSUS property.
To conclude the proof, we substitute~\eqref{eq:first term eigenvector} and~\eqref{eq:second term eigenvector} in~\eqref{eq:decomposition proof eigenvector}.
\end{IEEEproof}

The error~$\aerr_{1}$ in~\eqref{eq: eigenvalue error} is minimized if~\logonp is chosen so that~$\afp\approx\af_{\logon}(0,0)=1$ over the support of the scattering function. If the channel is highly  underspread, we can replace~\afp on the RHS of~\eqref{eq: eigenvalue error} with its second-order Taylor series expansion around the point~$(\doppler,\delay)=(0,0)$; \fref{prop:derivatives} now shows that good time and frequency localization of~\logonp is {\em necessary} for~$\aerr_{1}$ to be small. If~\logonp is taken to be real and even, the second-order Taylor series expansion of~\afp around the point~$(\doppler,\delay)=(0,0)$ takes on a particularly simple form because the first-order term is zero, and we can approximate~\afp around~$(0,0)$ as follows~\cite{wilcox91a}:
\bas
	\afp\approx 1-2\pi\left[ \eftime^{2}\doppler^{2}+\efband^{2}\delay^{2}-\iu\doppler\delay/(4\pi)\right].
\eas
Hence, when~\logonp is real and even, good time and frequency localization of~\logonp is also {\em sufficient} for~$\aerr_{1}$ to be small.

\subsubsection{Approximate Eigenvalues}
\begin{lem}[\!{\cite{matz98-08a,matz03a}}]
\label{lem:ev-error-bound}
Let~$\CHop$ be a WSSUS channel with time-varying
transfer function~\tvtfp and scattering function~\scafunp. Then, for any unit-energy signal~$\slogonct(\time)\define\logon(\time-\alpha)\cex{\beta\time}$, the mean-square approximation
error incurred by assuming that~$\tvtf(\alpha,\beta)$ is an eigenvalue of~$\CHop$ associated
to~$\slogonct(\time)$ is given by
\bas
	\aerr_{2}&=\Ex{}{\abs{\inner{\CHop\slogonct}{\logon_{(\alpha,\beta)}}
		-\tvtf(\alpha,\beta)}^{2}}=\spreadint{\scafunp\abs{1-\afp}^{2}}.
\eas
\end{lem}

\begin{IEEEproof}
We use \fref{prop:channel operator} and the Fourier transform relation~\eqref{eq:spreading function} to write~$\aerr_{2}$ as
\be
\label{eq:proof eigenvalue}
\bs
\aerr_{2}&=\Ex{}{\abs{\spreadint{\spfp\left[\conj{\af}_{\slogonct}(\doppler,\delay)-\cex{(\doppler\alpha-\delay\beta)}\right]}}^{2}}\\
	&\stackrel{(a)}{=}\Ex{}{\abs{\spreadint{\spfp\cex{(\doppler\alpha-\delay\beta)}\left[\conj{\af}_{\logon}(\doppler,\delay)-1\right]}}^{2}}\\
	&\stackrel{(b)}{=}\spreadint{\scafunp\abs{1-\afp}^{2}}.
\es
\ee
Here,~(a) follows from \fref{eq:property amb} and~(b) is a consequence of the WSSUS property.
\end{IEEEproof}
Similarly to what was stated for~$\aerr_{1}$ in the previous section, also in this case good time and frequency localization of~\logonp leads to small mean-square error~$\aerr_{2}$ if the channel is underspread.
\subsection{OFDM Pulse Design for Minimum~ISI and~ICI}\label{sec:bounds on interference term}

In \fref{sec:ofdm-interpretation} we introduced the concept of a
PS-OFDM system that uses an orthonormal Weyl-Heisenberg transmission set~$\{\slogon(\time)
\}$,  where~$\slogon(\time)=\logon(\time-\dtime\tstep)\cex{\dfreq\fstep\time}$, and
provided the criterion~\eqref{eq:grid-matching} for the choice of  the grid
parameters~\tstep and~\fstep to jointly minimize~ISI and~ICI. In this section, we 
detail the derivation that leads to~\eqref{eq:grid-matching}.
Let~$\nfoutp(\time)=(\CHop\inp)(\time)$ denote the noise-free channel output
when the channel input~$\inp(\time)$ is a PS-OFDM signal given by
\be
	\inp(\time)=\sum_{\dtime=-\infty}^{\infty}\sum_{\dfreq=-\infty}^{\infty}
		\inp[\dtdf]\slogon(\time).
\een
For mathematical convenience, we consider the case of an infinite time and
frequency horizon, and assume that the input symbols~$\{\inp[\dtdf]\}$ are~\iid,
with zero mean and~$\Ex{}{\abs{\inp[\dtdf]}^{2}}\le1$, $\forall\dtdf$.

\begin{figure}
\centering
	\includegraphics[width=\figwidth]{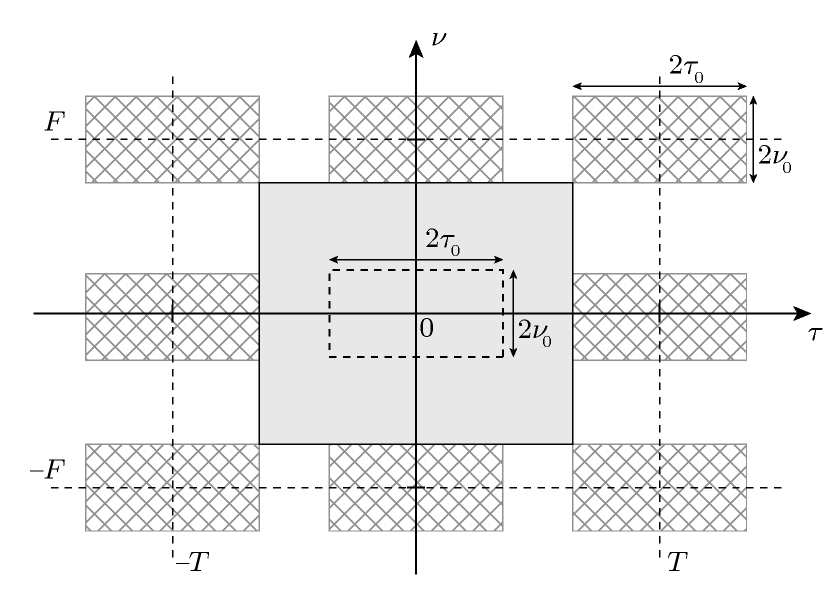}
	\caption{The support set of the periodized scattering function
		in~\fref{eq:isi-ici-error} are the rectangles with crisscross pattern,
		while the area on which the ambiguity function~\afp should be
		concentrated to minimize~$\aerr_{4}$ is shaded in grey.}
\label{fig:matchingrule}
\end{figure}

We want to quantify the mean-square error incurred by assuming that the projection of the received
signal~$\nfoutp(\time)$ onto the function~$\slogon(\time)$
equals~$\inp[\dtdf]\tvtf(\tfsamples)$, i.e., the error
\be
	\aerr_{3}\define\Ex{}{\abs{\inner{\nfoutp}{\slogon}- \inp[\dtdf]
		\tvtf(\tfsamples)}^{2}}
\een
where the expectation is over the channel realizations and the input symbols.
We bound~$\aerr_{3}$
as follows:
\be
\bs
	\aerr_{3}&=\Exop\Bigl[\bigl\lvert\inner{\nfoutp}{\slogon}-\inp[\dtdf]
		\inner{\CHop	\slogon}{\slogon} \\
		&\hphantom{\le\Exop\Bigl[\lvert\inner{\nfoutp}{\slogon}}+ \inp[\dtdf]
		\bigl(\inner{\CHop\slogon}{\slogon}- \tvtf(\tfsamples)\bigr)\bigr\rvert^{2}
		\Bigr] \\
	&\stackrel{(a)}{\leq}2\underbrace{\Ex{}{\abs{\inner{\nfoutp}{\slogon}-
		\inp[\dtdf]\inner{\CHop\slogon}{\slogon}}^{2}}}_{\aerr_{4}}\\
		&\hphantom{\le\Exop\Bigl[\lvert\inner{\nfoutp}{\slogon}}+2\Ex{}{
		\abs{\inp[\dtdf]\bigl(\inner{\CHop\slogon}{\slogon}-\tvtf(\tfsamples)
		\bigr)}^{2}}\\
	&=2\aerr_{4}+2\Ex{}{\abs{\inp[\dtdf]}^{2}}\underbrace{\Ex{}{\bigl\lvert
		\inner{\CHop	\slogon}	{\slogon}-\tvtf(\tfsamples)\bigr\rvert^{2}}}_{\aerr_{2}}\\
	&\le2\aerr_{4} +2\aerr_{2}
\es
\een
where~(a) holds because for any two complex numbers~$u$ and~$v$ we have that~$\abs{u+v}^{2} \leq 2\abs{u}^{2} +2 \abs{v}^{2}$. 
The error~$\aerr_{2}$ is the same as the one computed in
\fref{lem:ev-error-bound}. The error~$\aerr_{4}$ results from neglecting ISI and ICI and can be bounded as follows:
\be
\bs
	\aerr_{4} =& \Ex{}{\abs{\inner{\nfoutp}{\slogon}}^{2}  } + 
	\Ex{}{\abs{\inp[\dtdf]}^{2}}\Ex{}{\abs{\inner{\CHop\slogon}{\slogon}}^{2}}\\
	&-2\Re\lefto\{\Ex{}{\conj{\inp}[\dtdf]\inner{\nfoutp}{\slogon}\conj{\inner{\CHop\slogon}{\slogon}}}\right\}\\
	\stackrel{(a)}{=}&\mathop{\sum_{\dtime'=-\infty}^{\infty}\sum_{\dfreq'=-\infty}^{\infty}}_{(\dtime',\dfreq')\neq (\dtdf)}
	\Ex{}{\abs{\inp[\dtime',\dfreq']}^{2}}\Ex{}{\abs{\inner{\CHop\logon_{\dtime',\dfreq'}}{\slogon}}^{2}}\\
	\stackrel{(b)}{\leq}&\mathop{\sum_{\dtime'=-\infty}^{\infty}\sum_{\dfreq'=-\infty}^{\infty}}_{(\dtime',\dfreq')\neq (\dtdf)}\Ex{}{\abs{\inner{\CHop\logon_{\dtime',\dfreq'}}{\slogon}}^{2}}
\es
\label{eq:first bound on e4} 
\ee
where (a)~follows because the~$\inp[\dtdf]$ are \iid and zero mean, and~(b) because~$\Ex{}{\abs{\inp[\dtdf]}^{2}}\leq 1$.
We now provide an expression for $\Ex{}{\abs{\inner{\CHop\logon_{\dtime',\dfreq'}}{\slogon}}^{2}}$ that is explicit in the channel's scattering function:
\be
\label{eq:each term ISI-ICI}
\bs
	\Ex{}{\abs{\inner{\CHop\logon_{\dtime',\dfreq'}}{\slogon}}^{2}}
	&\stackrel{(a)}{=}\Ex{}{\abs{\inner{\spf}{\af_{\slogon,\logon_{\dtime',\dfreq'} } } }^2 }\\
	&\stackrel{(b)}{=}\spreadint{\scafunp\abs{\af_{\slogon,\logon_{\dtime',\dfreq'}}(\doppler,\delay)}^{2}}\\
	&\stackrel{(c)}{=}\spreadint{\scafunp\abs{\af_{\logon}\lefto(\doppler+(\dfreq'-\dfreq)\fstep,\delay+(\dtime'-\dtime)\tstep\right)}^{2} }\\
	&\stackrel{}{=}\spreadint{\scafun\lefto(\doppler-(\dfreq'-\dfreq)\fstep,\delay-(\dtime'-\dtime)\tstep\right)\abs{\afp}^{2} }.
\es
\ee
Here, (a) follows from~\fref{prop:channel operator}, (b) from the WSSUS property, and (c) from Property~3. We finally substitute~\eqref{eq:each term ISI-ICI} in~\eqref{eq:first bound on e4} and obtain
\be
\bs
\aerr_{4} \leq& \mathop{\sum_{\dtime'=-\infty}^{\infty}\sum_{\dfreq'=-\infty}^{\infty}}_{(\dtime',\dfreq')\neq (\dtdf)}
\spreadint{\scafun\lefto(\doppler-(\dfreq'-\dfreq)\fstep,\delay-(\dtime'-\dtime)\tstep\right)\abs{\afp}^{2} }\\
=&\mathop{\sum_{\dtime=-\infty}^{\infty}\sum_{\dfreq=-\infty}^{\infty}}_{(\dtdf)\neq (0,0)} \spreadint{\scafun(\doppler-\dfreq\fstep,\delay-\dtime\tstep)\abs{\afp}^{2}}.
\es
\label{eq:isi-ici-error}
\ee
This error is small if the ambiguity surface~$\abs{\afp}^{2}$ of~\logonp takes on
small values on the periodically repeated rectangles~$[-\maxDoppler+\dfreq
\fstep,\maxDoppler+\dfreq\fstep]\times[-\maxDelay+\dtime\tstep,\maxDelay+\dtime
\tstep]$, except for the dashed rectangle centered at the origin (see \fref{fig:matchingrule}).
This condition can be satisfied if the channel is highly underspread and if the grid parameters~\tstep and~\fstep are chosen such that the solid rectangle centered at the origin in \fref{fig:matchingrule} has large enough area to allow~$\abs{\afp}^{2}$ to decay. 
If $\logonp$ has effective duration~\eftime and effective bandwidth~\efband,  the latter condition holds if~$\tstep\geq\maxDelay+\eftime$, and~$\fstep\geq\maxDoppler+\efband$. Given a constraint on the product~$\tstep\fstep$, good localization of~\logonp, both in time and frequency,  is necessary for the two inequalities above to hold.

The minimization of~$\aerr_{4}$ in~\fref{eq:isi-ici-error} over all
orthonormal Weyl-Heisenberg sets~$\{\slogon(\time)\}$ is a difficult task; numerical
methods to minimize~$\aerr_{4}$ are described in~\cite{matz07-05a}.
The simple rule on how to choose the grid parameters~\tstep
and~\fstep provided in~\eqref{eq:grid-matching} is derived from the following observation:
for known~\maxDelay and~\maxDoppler, and for a fixed product~$\tstep\fstep$, the
area~$4(\tstep-\maxDelay)(\fstep-\maxDoppler)$ of the solid rectangle centered at the origin in
\fref{fig:matchingrule} is maximized if~\cite{kozek97a,kozek98-10a,matz07-05a}
\be
	\frac{\tstep}{\fstep} = \frac{\maxDelay}{\maxDoppler}.
\een
%

\section{}
\label{app:mmse-mi}
 
\begin{lem}
\label{lem:logdetinf-1d}
Let~$\{\rndproc[\dtime]\}$ be a stationary random
process with correlation function
\bas
\rndproccorr[\dtime]\define
\Ex{}{\rndproc[\dtime'+\dtime]\conj{\rndproc}[\dtime']}
\eas
 and spectral density
\bas
	\rndprocspec(\specparam)\define\sum_{\Ddtime=-\infty}^{\infty}
		\rndproccorr[\Ddtime]e^{-\iu 2 \pi \Ddtime\specparam},
		\quad\abs{\specparam}\le 1/2.
\eas
Furthermore, let~$\rndvec\define\tp{\btm \rndproc[0]\, \rndproc[1]\, \dots\,
\rndproc[\tslots-1]\etm}$, and denote the $\tslots\times\tslots$ covariance
matrix of~\rndvec by~$\rndveccovmat\define\Ex{}{\rndvec\herm{\rndvec}}$. This
covariance matrix is Hermitian Toeplitz with 
entries~$\left[\rndveccovmat\right]_{i,j}=\rndproccorr[i-j]$. 
Then, for any deterministic  $\tslots$-dimensional vector~$\vecx$ with binary entries~$\{0,1\}$ and
for any~$\rho>0$, the following inequality holds:
	\ba \label{eq:logdetinf-1d}
		\inf_{\vecx}\frac{1}{\vecnorm{\vecx}^{2}}\logdet{\imat_{\tslots}+\rho
			(\vecx\herm{\vecx})\had\rndveccovmat}\ge
			\int_{-1/2}^{1/2}\log(1+\rho\rndprocspec(\specparam))d\specparam.
	\ea
Furthermore, in the limit~$\tslots\to\infty$, the above inequality is
satisfied with equality if the entries of~\vecx are all equal to~$1$.
\end{lem}

\begin{rem}
The second statement in \fref{lem:logdetinf-1d}---that the infimum can be
achieved by an all-$1$ vector in the limit for~$\tslots\to\infty$---was already
proved in~\cite[Sec.~VI.B]{sethuraman05-09a}. The proof in~\cite{sethuraman05-09a}
relies on rather  technical
set-theoretic arguments, so that it is not easy to see how the
structure of the problem---the stationarity of the 
process~$\{\rndproc[\dtime]\}$---comes into play.
Therefore, it is cumbersome to extend the proof in~\cite{sethuraman05-09a} to accommodate two-dimensional
stationary processes as used in this paper. Here, we provide an alternative proof
that is significantly shorter, explicitly uses the stationarity property, can be directly generalized
to two-dimensional stationary processes (see Corollary~\ref{cor:logdetinf-2d} below), and yields the 
new lower bound~\eqref{eq:logdetinf-1d} as an important additional result. 
\end{rem}

Our proof is based on the relation between mutual information MMSE
discovered recently by Guo {\it et al.}~\cite{guo05-04a}. In the following lemma,
we restate, for convenience, the mutual information-MMSE relation for JPG random
vectors\footnote{For a proof of~\fref{lem:mmse-mi}, see~\cite[Sec.~V.D]{guo05-04a}.}
\begin{lem}
\label{lem:mmse-mi}
Let~\rndvec be a $\tslots$-dimensional random vector that satisfies~$\Ex{}{\vecnorm{\rndvec}^{2}}<
\infty$, and let~\wgnvec be a zero-mean JPG vector,~$\wgnvec
\distas\jpg(\veczero,\imat_{\tslots})$, that is independent of~\rndvec. Then, for any
deterministic $\tslots$-dimensional vector~\vecx,
\ba
\label{eq:mmse-mi}
	\frac{d}{d\gamma}\mi(\sqrt{\gamma}\vecx \had \rndvec+\wgnvec;\rndvec)=
		\Ex{}{\vecnorm{\vecx \had\rndvec-\vecx \had\Ex{}{\rndvec\given\sqrt{\gamma}\vecx
		\had\rndvec+\wgnvec}}^{2}}.
\ea
\end{lem}
The expression on the RHS in~\fref{eq:mmse-mi} is the MMSE obtained when~$\vecx \had \rndvec$ is
estimated from the noisy observation~$\sqrt{\gamma}\vecx \had\rndvec+\wgnvec$.

\begin{IEEEproof}[Proof of \fref{lem:logdetinf-1d}]
We first derive the lower bound~\fref{eq:logdetinf-1d} and then show
achievability in the limit $\tslots\to\infty$ in a second step.
To apply \fref{lem:mmse-mi}, we rewrite the LHS of~\fref{eq:logdetinf-1d}
as
\ba
\label{eq: first rewriting MI-MMSE appendix}
\frac{1}{\vecnorm{\vecx}^{2}}\logdet{\imat_{\tslots}+\rho(\vecx\herm{\vecx})\had
		\rndveccovmat}
		=\frac{1}{\vecnorm{\vecx}^{2}}\mi(\sqrt{\rho}\vecx\had
			\rndvec+	\wgnvec;\rndvec)
\ea
where~$\wgnvec\distas\jpg(\veczero,\imat_{\tslots})$ is a JPG vector.
Without loss of generality, we assume that the vector~\vecx has exactly~\numnz
nonzero entries, with corresponding indices in the set~\iset. Then,
\be
\bs
	\frac{1}{\vecnorm{\vecx}^{2}}\mi(&\sqrt{\rho}\vecx\had\rndvec
		+\wgnvec;\rndvec)=\\
	&\stackrel{(a)}{=} \frac{1}{\vecnorm{\vecx}^{2}}\int_{0}^{\rho}
		\Ex{}{\vecnorm{\vecx\had\rndvec - \vecx\had\Ex{}{\rndvec\given
		\sqrt{\gamma}\vecx\had\rndvec+\wgnvec}}^{2}}d\gamma\\
	&\stackrel{(b)}{=} \frac{1}{\numnz}\int_{0}^{\rho}\sum_{m \in\iset}
		\Ex{}{\abs{\rndproc[m] - \Ex{}{\rndproc[m]\given\lefto\{\sqrt{
		\gamma}\rndproc[\dtime] + \wgn[\dtime]\right\}_{\dtime\in\iset}}}^{2}}
		d\gamma\\
	& \stackrel{(c)}{\geq}\frac{1}{\numnz}\int_{0}^{\rho}\sum_{m\in\iset}
		\Ex{}{\abs{\rndproc[m] - \Ex{}{\rndproc[m]\given\lefto\{\sqrt{
		\gamma}\rndproc[\dtime] + \wgn[\dtime]\right\}_{\dtime=-\infty}^
		{\infty}}}^{2}}d\gamma\\
	& \stackrel{(d)}{=}\int_{0}^{\rho}\Ex{}{\abs{\rndproc[0] - \Ex{}{\rndproc[0]
		\given\lefto\{\sqrt{\gamma}\rndproc[\dtime] + \wgn[\dtime]\right
		\}_{\dtime=-\infty}^{\infty}}}^{2}}d\gamma.
\es	\label{eq:logdet-lb}
\ee
Here, (a) follows from the relation between mutual information and MMSE in
\fref{lem:mmse-mi} in the form given in~\cite[Eq.~(47)]{guo05-04a}. Equality~(b) holds because~\vecx has exactly~\numnz
nonzero entries with corresponding indices in~\iset, and because the components of the
observation that contain only noise do not influence the estimation error. 
The argument underlying inequality~(c) is that the MMSE can only decrease if each~$\rndproc[m]$ is estimated not just from a
finite set of noisy observations of the random process~$\{\rndproc[\dtime]\}$, but also from noisy observations of the process' infinite past and
future. This is the so-called {\em infinite-horizon noncausal} MMSE.
Finally, we obtain~(d) because the process~$\{\rndproc[\dtime]\}$ is stationary
and its infinite horizon noncausal~MMSE is, therefore, the same for all
indices~$m\in\iset$~\cite[Sec.~V.D.1]{poor94a}.

The infinite-horizon noncausal~MMSE can be expressed in terms of the spectral density of the process~$\{\rndproc[\dtime]\}$~\cite[Eq.~(V.D.28)]{poor94a}:
\be
\label{eq:noncausal-mmse}
	\Ex{}{\abs{\rndproc[0] - \Ex{}{\rndproc[0]\given\lefto\{\sqrt{\gamma}
		\rndproc[\dtime] + \wgn[\dtime]\right\}_{\dtime=-\infty}^{\infty}}}^{2}}
		= \int_{-1/2}^{1/2}\frac{\rndprocspec(\specparam)}{1+\gamma
		\rndprocspec(\specparam)}d\specparam.
\ee
To obtain the desired inequality~\fref{eq:logdetinf-1d}, we
substitute~\fref{eq:noncausal-mmse} in~\fref{eq:logdet-lb}, and~\fref{eq:logdet-lb} in~\fref{eq: first rewriting MI-MMSE appendix}, and note that the resulting
lower bound does not depend on~\vecx. We have therefore established a lower bound on the LHS of~\fref{eq:logdetinf-1d} as well. We finally integrate
over~$\gamma$ and get
\be
\bs
	\inf_{\vecx}\frac{1}{\vecnorm{\vecx}^{2}}\logdet{\imat_{\tslots}+\rho
		(\vecx\herm{\vecx})\had\rndveccovmat}  
	&\geq \int_{-1/2}^{1/2}\int_{0}^{\rho}\frac{\rndprocspec(\specparam)}
		{1+\gamma\rndprocspec(\specparam)}d\gamma d\specparam\\
	& = \int_{-1/2}^{1/2}\log\bigl(1 + \rho\rndprocspec(\specparam)\bigr)
		d\specparam.
\es
\een

To prove the second statement in \fref{lem:logdetinf-1d}, we choose~\vecx in~\eqref{eq: first rewriting MI-MMSE appendix} to be
the all-$1$ vector for any dimension~\tslots, and evaluate the limit~$\tslots\to\infty$ of the LHS 
of~\eqref{eq: first rewriting MI-MMSE appendix}  by means of Szeg\"o's theorem on the asymptotic eigenvalue distribution of a Toeplitz matrix~\cite{grenander84a,gray05a}:
\ba
\label{eq:szegoe}
	\limintime\frac{1}{\tslots}\logdet{\imat_{\tslots}+\rho\rndveccovmat} =
		\int_{-1/2}^{1/2}\log\bigl(1 +\rho\rndprocspec(\specparam)\bigr)
		d\specparam.
\ea
This shows that the lower bound in~\fref{eq:logdetinf-1d} can indeed be achieved in the limit~$\tslots\to\infty$
when $\vecx$ is the all-$1$ vector.
\end{IEEEproof}

Our proof allows for a simple generalization of
\fref{lem:logdetinf-1d} to two-dimensional stationary processes, which are relevant
to the problem considered in this paper.  The generalization is stated in the following corollary.
%
\begin{cor}\label{cor:logdetinf-2d}
Let~$\{\rndproc[\dtdf]\}$ be a
random process that is stationary in~\dtime and~\dfreq with two-dimensional
correlation function~$\rndproccorr[\dtime,\dfreq]\define\Ex{}{
\rndproc[\dtime+\dtime',\dfreq+\dfreq']\conj{\rndproc}[\dtime',\dfreq']}$ and
two-dimensional spectral density
\be
	\rndprocspec(\specparam,\altspecparam)\define\sum_{\Ddtime=-\infty}^{\infty}
		\sum_{\Ddfreq=-\infty}^{\infty}\rndproccorr[\Ddtime,\Ddfreq]e^{-\iu 2\pi
		(\Ddtime\specparam-\Ddfreq\altspecparam)},
		\quad\abs{\specparam},\abs{\altspecparam}\le 1/2.
\een
Furthermore, let~$\rndvec[\dtime]\define\tp{\btm \rndproc[\dtime,0] &
\rndproc[\dtime,1] & \cdots & \rndproc[\dtime,\fslots-1]\etm}$, 
let the $\tslots\fslots$-dimensional stacked vector~$\rndvec\define\tp{\btm \tp{\rndvec}[0]\,
\tp{\rndvec}[1]\, \dots\, \tp{\rndvec}[\tslots-1]\etm}$, and denote the $\tslots
\fslots\times\tslots\fslots$~covariance matrix of~\rndvec by~$\rndveccovmat
\define\Ex{}{\rndvec\herm{\rndvec}}$. This covariance matrix is a two-level
Toeplitz matrix. Then, for any $\tslots\fslots$-dimensional vector~$\vecx$ with binary entries~$\{0,1\}$ and
for any~$\rho>0$, the following inequality holds:
\ba \label{eq:logdetinf-2d}
	\inf_{\vecx}\frac{1}{\vecnorm{\vecx}^{2}}\logdet{\imat_{\tslots\fslots}+
	\rho(\vecx\herm{\vecx})\had\rndveccovmat}\ge
		\int_{-1/2}^{1/2}\int_{-1/2}^{1/2}\log(1+\rho
		\rndprocspec(\specparam,\altspecparam))d\specparam d\altspecparam.
\ea
Furthermore, in the limit~$\tslots,\fslots\to\infty$, the above inequality
is satisfied with equality if the entries of~\vecx are all equal to~$1$.
\end{cor}

\begin{IEEEproof}
Without loss of generality, we assume that the vector~\vecx has exactly~\numnz
nonzero elements, with corresponding indices in the set~\iset. The arguments used
in the proof of \fref{lem:logdetinf-1d} directly apply, and we obtain
\begin{multline*}
	\frac{1}{\vecnorm{\vecx}^{2}}\logdet{\imat_{\tslots\fslots} + \rho
		(\vecx\herm{\vecx})\had\rndveccovmat}\geq\\
	\int_{0}^{\rho}\Ex{}{\abs{\rndproc[0,0] - \Ex{}{\rndproc[0,0]\given
		\lefto\{\sqrt{\gamma}\rndproc[\dtdf] + \wgn[\dtdf]\right\}_{\dtdf
		=-\infty}^{\infty}}}^{2}}d\gamma.
\end{multline*}
To complete the proof, we use the two-dimensional counterpart
of~\fref{eq:noncausal-mmse}---the closed-form expression for the
two-dimensional noncausal MMSE~\cite[Eq.~(2.6)]{helstrom67-03a}---and 
we compute the two-dimensional equivalent of~\fref{eq:szegoe} by means of the
extension of Szeg\"o's theorem to two-level Toeplitz
matrices provided, e.g., in~\cite{voois96a}.
\end{IEEEproof}

\section{}
\label{app:avpopt-proof}

In this appendix, we show that a sufficient condition for
\ba
	\avpoptp=\min\lefto\{1,\, \frac{\bandwidth}{\tfstep}\left(\frac{1}
		{\ubpenaltyp}-\frac{1}{\Pave}\right)\right\} = 1,
\label{eq:avpopt-max}
\ea
with~\ubpenaltyp defined in~\eqref{eq:ubTFpeak-penalty},  is that
\bas
&0\leq \frac{\Pave}{\bandwidth}\leq \frac{1}{\tfstep},\quad \text{and}\quad  \spread\leq \frac{\papr}{3\tfstep}\\
\intertext{or that}
&\frac{1}{\tfstep}< \frac{\Pave}{\bandwidth}< \frac{\spread}{\papr}\left[\exp\lefto(
		\frac{\papr}{2\tfstep\spread}\right)-1\right].
 \eas
For notational convenience, we set~$\SNR\define\Pave/\bandwidth$.
The necessary and sufficient condition under
which~\fref{eq:avpopt-max} holds can be restated as
\bas
\frac{\bandwidth}{\ubpenaltyp}\geq \frac{1}{\SNR} + \tfstep 
\eas
or, equivalently, as
\ba
\label{eq:avpopt-suffc}
	\frac{1}{\papr} \spreadint{\log\lefto(1+
	 	\SNR\papr\scafunp\right)} \leq \left( \frac{1}{\SNR} + \tfstep \right)^{-1}.
\ea
We now use Jensen's inequality as in~\eqref{eq:ubTFpeak-penalty-ub} to upper-bound the 
LHS of~\eqref{eq:avpopt-suffc} and get the following sufficient condition for~$\avpoptp=1$:
\ba
\label{eq: upper-bound first term inequality}
	\frac{\spread}{\papr}\log\left(1 +\frac{\papr\SNR}{\spread}\right)\leq \left( \frac{1}{\SNR} + \tfstep \right)^{-1}.
\ea
We next distinguish between two cases: $\SNR> 1/(\tfstep)$ and $\SNR \leq 1/(\tfstep)$.
\subsubsection*{Case $\SNR> 1/(\tfstep)$}
We use the inequality
\bas
	\left(\frac{1}{\SNR} + \tfstep \right) \leq 2\tfstep
\eas
to lower-bound the RHS of~\eqref{eq: upper-bound first term inequality} and obtain the following sufficient condition for~$\avpoptp=1$:
\bas
	\frac{\spread}{\papr}\log\left(1 +\frac{\papr\SNR}{\spread}\right)\le
		\frac{1}{2\tfstep}.
\eas
This condition can be expressed in terms of~\SNR as
\ba
\label{eq:avpotp-cond-upper}
	\SNR < \frac{\spread}{\papr}\left[\exp\lefto(
		\frac{\papr}{2\tfstep\spread}\right)-1\right].
\ea

\subsubsection*{Case $\SNR \leq (1/\tfstep)$}
We further upper-bound the LHS of~\eqref{eq: upper-bound first term inequality} by means of the inequality
\bas
	\frac{1}{x}\log(1+x)\le\frac{1}{\sqrt{1+x}},\quad\text{for all }x\ge0
\eas
and get the following sufficient condition for~$\avpoptp=1$:
\bas
	\frac{\SNR}{\sqrt{1+ \papr\SNR /\spread}} \leq \left( \frac{1}{\SNR} +
		\tfstep \right)^{-1}.
\eas
This condition is satisfied for all~$\SNR\in[0,1/(\tfstep)]$ as long as
\ba
\label{eq:avpotp-cond-lower}
	\spread\le\papr/(3\tfstep).
\ea
If we combine~\fref{eq:avpotp-cond-upper}
and~\fref{eq:avpotp-cond-lower}, the sufficient
condition~\fref{eq:avpopt-conditions} follows.

\section{Proof of \fref{lem:lbTFpeak-penalty-bound}}
\label{app:penalty term}

\subsubsection{Upper bound}
We restate the penalty term in~\fref{eq:lbTFpeak} in the more convenient form\footnote{For simplicity and without loss of generality,
we set~$\param=1$.}
\ba
\label{eq:lbTFpeak-penalty}
	\frac{1}{\tstep}\!\int_{-1/2}^{1/2}\logdet{\imat_{\fslots}
		+\frac{\Pave\tstep}{\fslots}\mvchspecfunp}d\specparam.
\ea
We seek an upper bound on~\eqref{eq:lbTFpeak-penalty} that 
can be evaluated efficiently, even for large~\fslots, and
that is tight in the limit~$\fslots\to\infty$. To obtain such a bound, we need to solve two
problems: first, the eigenvalues of the $\fslots \times \fslots$ Toeplitz matrix~\mvchspecfunp are
difficult to compute; second, the determinant expression in~\eqref{eq:lbTFpeak-penalty}
needs to be evaluated for all~$\specparam \in[-1/2,1/2]$. 
To upper-bound~\eqref{eq:lbTFpeak-penalty}, we will replace~\mvchspecfunp with a 
suitable circulant matrix that is asymptotically equivalent~\cite{gray05a} to~\mvchspecfunp.
Asymptotic equivalence guarantees tightness of the resulting bound in the limit~$\fslots\to\infty$. As the
eigenvalues of a circulant matrix can be computed efficiently via the
 discrete Fourier transform~(DFT), the first problem
is solved. To solve the second problem, we use Jensen's inequality.

We shall need the following result on the asymptotic equivalence between
Toeplitz and circulant matrices.
\begin{lem}[see~\cite{pearl73-03a}]
\label{lem:asymptotic-equivalence}
Let~\matT be an~$\fslots\times \fslots$ Hermitian Toeplitz matrix. Furthermore, let~\dftmat be the
DFT~matrix, i.e., the matrix~$\dftmat=[\dftcol_{0}\, \dftcol_{1}\,\cdots\,
\dftcol_{\fslots-1}]$ whose columns~$\dftcol_{\dfreq}=\tp{[\beta^{0\dfreq}\, \beta^{1\dfreq}\,
\cdots\, \beta^{(\fslots-1)\dfreq}]}/\sqrt{\fslots}$ contain powers of the $\fslots$th root of
unity,~$\beta=\cex{/\fslots}$. Construct from the matrix~$\herm{\dftmat} \matT\dftmat$
the diagonal matrix~\matD so that the entries on the main diagonal of~\matD and
on the main diagonal of~$\herm{\dftmat}\matT\dftmat$ are equal. Then,~\matT
and the circulant matrix~$\dftmat\matD\herm{\dftmat}$ are asymptotically
equivalent, i.e., the Frobenius norm~\cite[Sec.~5.6]{horn85a} of the matrix $\bigl(\matT-\dftmat\matD\herm{\dftmat}\bigr)/\sqrt{\fslots}$ converges to zero as~$\fslots\to\infty$.
\end{lem}

Our goal is to upper-bound a function of the form~$\logdet{\imat_{\fslots}+\matT/\fslots}$. Because~\dftmat is unitary, and by Hadamard's inequality,
\be
\label{eq:ae-bound-example}
\bs
	\logdet{\imat_{\fslots}+\frac{1}{\fslots}\matT} &=\logdet{\imat_{\fslots}+\frac{1}{\fslots}
		\herm{\dftmat}\matT\dftmat}\\
	&\leq \logdet{\imat_{\fslots} + \frac{1}{\fslots}\matD} \\
	&=\logdet{\imat_{\fslots} +\frac{1}{\fslots}\dftmat\matD\herm{\dftmat}}.
\es
\ee
Since~\matT and~$\dftmat\matD\herm{\dftmat}$ are asymptotically equivalent, we expect 
the difference between the LHS and the RHS of the inequality~\eqref{eq:ae-bound-example} to  vanish
as~$\fslots$ grows large. We formalize this result in the following lemma, which follows directly from Szeg\"o's theorem on the asymptotic eigenvalue distribution of Toeplitz
matrices.
%
\begin{lem}
\label{lem:taylor expansion is the same}
	Let~$\{\toep_{\dfreq}\}$ be a sequence that satisfies~$\toep_{-\dfreq}=\conj{\toep}_{\dfreq}$ for all $\dfreq$,
	and has Fourier transform
	\bas
		\specp=\sum_{\dfreq=-\infty}^{\infty}\toep_{\dfreq}\cexn{\dfreq\altspecparam}, \quad \abs{\altspecparam}\leq 1/2.
	\eas
Let~\matT be the~$\fslots\times\fslots$ Hermitian Toeplitz matrix constructed as
\ba
\label{eq:toeplitz matrix}
\matT=\mat
		\toep_{0} &\toep_{-1}	& \hdots & \toep_{-(\fslots-1)} \\
		\toep_{1} & \toep_{0}& \hdots &\toep_{-(\fslots-2)}\\
		\vdots & \vdots & \ddots & \vdots \\
		\toep_{\fslots-1} &\toep_{\fslots-2} & \hdots& \toep_{0}
	\emat.
\ea
Then, the function $\logdet{\imat_{\fslots}+\matT/\fslots}$ admits the following $\taylexp$th-order Taylor series expansion around the point~$1/\fslots=0$:
\ba
	\logdet{\imat_{\fslots}+\frac{1}{\fslots}\matT}=\sum_{\taylexpindex=0}^{\taylexp} \frac{(-1)^{\taylexpindex}}{(\taylexpindex+1)\fslots^{\taylexpindex}}\int_{-1/2}^{1/2}[\specp]^{\taylexpindex+1}d\altspecparam+\osmall{\frac{1}{\fslots^{\taylexp}}}.
\label{eq:Taylor expansion logdet expression}
\ea
Furthermore, let~\dftmat and~\matD be as in~\fref{lem:asymptotic-equivalence}. Then,~$\logdet{\imat_{\fslots} +\dftmat\matD\herm{\dftmat}/\fslots}$ has the same~$\taylexp$th-order Taylor series expansion around~$1/\fslots=0$
as~$\logdet{\imat_{\fslots}+\matT/\fslots}$.
\end{lem}
\begin{IEEEproof}
Let~\esssup be the essential supremum of~\specp, i.e.,~\esssup is the smallest number that satisfies~$\specp\leq \esssup$ for all~$\altspecparam$, except on a set of measure zero. Then for any~\fslots, the eigenvalues~$\{\lambda_{\dfreq}\}_{\dfreq=0}^{\fslots-1}$ of the matrix~\matT satisfy~$\lambda_{\dfreq}\leq \esssup$~\cite[Lemma~6]{gray05a}. We now use the expansion in power series
\bas
	\log(1+x)=\sum_{\taylexpindex=1}^{\infty}\frac{(-1)^{\taylexpindex+1}}{\taylexpindex}x^{\taylexpindex}, \quad\text{for } \abs{x}<1
\eas
to rewrite~$\genfunp=\logdet{\imat_{\fslots}+\matT/\fslots}$  as
\ba
\genfunp=\sum_{\dfreq=0}^{\fslots-1}\log\lefto(1 + \frac{\lambda_{\dfreq}}{\fslots}\right)&=
\sum_{\dfreq=0}^{\fslots-1}\sum_{\taylexpindex=1}^{\infty}\frac{(-1)^{\taylexpindex+1}}{\taylexpindex}\left(\frac{\lambda_{\dfreq}}{\fslots}\right)^{\taylexpindex}
\nonumber\\&=
\sum_{\taylexpindex=1}^{\infty}\frac{(-1)^{\taylexpindex+1}}{\taylexpindex}\frac{1}{\fslots^{\taylexpindex-1}}
\left[\frac{1}{\fslots}\sum_{\dfreq=0}^{\fslots-1}\lambda^{\taylexpindex}_{\dfreq}\right], \quad\text{for } \fslots\geq\esssup\label{eq:power series}.
\ea
To compute the Taylor series expansion of~\genfunp around~$1/\fslots=0$
we need to evaluate~\genfunp and its derivatives for~$\fslots\to\infty$. We observe that Szeg\"o's theorem on the asymptotic eigenvalue distribution of Toeplitz matrices implies that~\cite[Th.~9]{gray05a}
\be
\label{eq:szego's theorem 1D}
\liminfreq\frac{1}{\fslots}\sum_{\dfreq=0}^{\fslots-1}\lambda^{\taylexpindex}_{\dfreq}=\int_{-1/2}^{1/2}[\specp]^{\taylexpindex}d\altspecparam.
\ee
Consequently, it follows from~\eqref{eq:power series} that
\bas
	\genfun(0)&=\liminfreq\genfunp=\int_{-1/2}^{1/2}\specp d\altspecparam, \\
	\genfun'(0)&=\liminfreq \fslots[\genfunp-\genfun(0)]=-\frac{1}{2}\int_{-1/2}^{1/2}[\specp]^{2}d\altspecparam, 
\eas
and, for the $\taylexpindex$th derivative,
\bas
	\genfun^{(\taylexpindex)}(0)&=\liminfreq\taylexpindex !\,\fslots^{\taylexpindex} \left[\genfunp-\genfun(0)
	-\sum_{i=1}^{\taylexpindex-1} i!\,\fslots^{i}\genfun^{(i)}(0) \right]\\
	&=l!\, \frac{(-1)^{\taylexpindex}}{\taylexpindex+1}\int_{-1/2}^{1/2}[\specp]^{\taylexpindex+1}d\altspecparam.
\eas
The proof of the first statement in~\fref{lem:taylor expansion is the same} is therefore concluded.
The second statement follows directly from the asymptotic equivalence 
between~\matT and~$\dftmat\matD\herm{\dftmat}$ (see~\fref{lem:asymptotic-equivalence}) and from~\cite[Th.~2]{gray05a}.

\end{IEEEproof}

To apply the bound~\fref{eq:ae-bound-example} to our problem of upper-bounding the  penalty term~\fref{eq:lbTFpeak-penalty}, we need to compute the diagonal entries of~$\herm{\dftmat}\mvchspecfunp\dftmat$. Similarly to~\eqref{eq:toeplitz matrix},  we denote the entries of the power spectral density Toeplitz matrix~\mvchspecfunp as~$\{\mvchspecfunentryp\}_{n=-(\fslots-1)}^
{\fslots-1}$. As a consequence of~\eqref{eq:mv-covmat} and~\eqref{eq:mvspec},~\mvchspecfunp is Hermitian, i.e.,
~$\mvchspecfunentry_{-\dfreq}(\specparam)=\conj{\mvchspecfunentry}_{\dfreq}(\specparam)$.
Furthermore, again by~\eqref{eq:mv-covmat} and~\eqref{eq:mvspec}, each entry~\mvchspecfunentryp is related to the
discrete-time discrete-frequency correlation function~$\chcorr[\dtime,\dfreq]$
according to
\be
\bs
	\mvchspecfunentryp &=\sum_{\dtime=-\infty}^{\infty}\chcorr[\dtime,\dfreq]
		\cexn{\dtime\specparam} \\
	&\stackrel{(a)}{=} \frac{1}{\tstep} \sum_{\dtime=-\infty}^{\infty} \int_{\delay}
	\scafun\lefto(\frac{\specparam-\dtime}{\tstep},\delay\right)
		\cexn{ \dfreq \fstep\delay}d\delay\\
	&\stackrel{(b)}{=}\frac{1}{\tstep} \sum_{\dtime=-\infty}^{\infty} \int_{-\maxDelay}^
		{\maxDelay}\scafun\lefto(\frac{\specparam-\dtime}{\tstep},\delay\right)
		\cexn{ \dfreq \fstep\delay}d\delay
\es
\label{eq: c_n of theta}
\ee
where~(a) follows from the Fourier transform relation~\eqref{eq:scafun-chcorr}, and the Poisson
summation formula as in~\eqref{eq:specfun-scafun}, and in~(b) we used that~\scafunp is zero
outside~$[-\maxDelay,\maxDelay]$. 
Consequently, the~$\indvar$th element on the main diagonal of~$\herm{\dftmat}
\mvchspecfunp\dftmat$, which we denote as~\diagelcircp, can be expressed as
a function of the entries of~\mvchspecfunp as follows
\be
\bs
	\diagelcircp&=\frac{1}{\fslots} \sumz{p}{\fslots}
		\sumz{q}{\fslots}\beta^{-\indvar q}\mvchspecfunentry_{q-p}(\specparam)
		\beta^{\indvar p} \\
	&=\frac{1}{\fslots} \sumz{p}{\fslots}\sumz{q}{\fslots}
		\mvchspecfunentry_{q-p}(\specparam)\beta^{-\indvar(q- p)} \\
	&= \frac{1}{\fslots}\sum_{\dfreq=-(\fslots-1)}^{\fslots-1}
		(\fslots-\abs{\dfreq})\mvchspecfunentry_{\dfreq}(\specparam)
		\cexn{\frac{\indvar\dfreq}{\fslots}}\\
	&=\Re\lefto\{\frac{2}{\fslots}\sumz{\dfreq}{\fslots}(\fslots-\dfreq)
		\mvchspecfunentryp\cexn{\frac{\indvar\dfreq}{\fslots}}\right\}
		-\mvchspecfunentry_{0}(\specparam)
\es
\label{eq:diag-element}
\ee
where we set~$\dfreq=q-p$ and used~$\mvchspecfunentry_{-\dfreq}(\specparam)=\conj{\mvchspecfunentry}_{\dfreq}(\specparam)$.
We can now establish an upper bound on the penalty term~\fref{eq:lbTFpeak-penalty} in terms of the~$\{\diagelcircp\}$ on the basis of~\fref{eq:ae-bound-example}: 
%
%
\be
\label{eq:first upper bound penalty term}
\bs
	\frac{1}{\tstep}\int_{-1/2}^{1/2}\logdet{\imat_{\fslots}
		+\frac{\Pave\tstep}{\fslots}\mvchspecfunp}d\specparam 
	&=\frac{1}{\tstep}\int_{-1/2}^{1/2}\logdet{\imat_{\fslots}
		+\frac{\Pave\tstep}{\fslots}\herm{\dftmat}\mvchspecfunp\dftmat}
		d\specparam \\ 
	&\leq \frac{1}{\tstep}\int_{-1/2}^{1/2} \sumz{\indvar}{\fslots} \log\lefto(1
		+\frac{\Pave\tstep}{\fslots}\diagelcircp\right)d\specparam \\
	&\stackrel{(a)}{=}\int_{-1/(2\tstep)}^{1/(2\tstep)} \sumz{\indvar}{\fslots} \log\lefto(1 +
		\frac{\Pave\tstep}{\fslots}\diagelcirc_{i}(\doppler\tstep) \right)
		d\doppler\\  
	&\stackrel{(b)}{=}\int_{-\maxDoppler}^{\maxDoppler} \sumz{\indvar}{\fslots} \log\lefto(1 +
		\frac{\Pave\tstep}{\fslots}\diagelcirc_{i}(\doppler\tstep) \right)
		d\doppler
\es
\ee
where~(a) follows from the change of variables~$\doppler=\specparam/\tstep$ and~(b) holds because~$\scafunp$
is zero for~\doppler outside~$[-\maxDoppler,\maxDoppler]$, and because, by assumption~$\tstep\leq 1/(2\maxDoppler)$, so that~$\scafun(\doppler-\dtime/\tstep,\delay)$ is zero whenever~$\dtime \neq 0$; hence, by~\eqref{eq: c_n of theta} and~\eqref{eq:diag-element}, also~$\mvchspecfunentry_{\dfreq}(\doppler\tstep)$ and~$\diagelcirci(\doppler\tstep)$ are zero for~\doppler outside~$[-\maxDoppler,\maxDoppler]$.

We proceed to remove the dependence on~\doppler. To this end, we further upper-bound~\fref{eq:first upper bound penalty term} by means of Jensen's inequality
and obtain the desired upper bound in~\fref{eq:lbTFpeak-penalty-bound};
\be
\bs
	\int_{-\maxDoppler}^{\maxDoppler} \sumz{\indvar}{\fslots}\log\lefto(1 +
		\frac{\Pave\tstep}{\fslots}\diagelcirc_{\indvar}(\doppler\tstep)
		\right)d\doppler &\leq
	2\maxDoppler \sumz{\indvar}{\fslots}\log\lefto(1 +\frac{\Pave\tstep}
		{2\maxDoppler\fslots}\int_{-\maxDoppler}^{\maxDoppler}
		\diagelcirc_{\indvar}(\doppler\tstep) d\doppler \right) \\
	&= 2\maxDoppler\sumz{\indvar}{\fslots}\log\lefto(1 +
		\frac{\Pave}{2\maxDoppler\fslots} \diagelcirc_{\indvar}
		\right)
\es
\label{eq:lbTFpeak-penalty-ub}
\ee
where we set~$\diagelcirc_{\indvar} = \tstep\int_{-\maxDoppler}^
{\maxDoppler}\diagelcirc_{\indvar}(\doppler\tstep) d\doppler$. As we have by~\eqref{eq: c_n of theta} that
\bas
	\tstep\int_{-\maxDoppler}^{\maxDoppler} \mvchspecfunentry_{\dfreq}(\doppler
		\tstep) d\doppler
	&=\sum_{\dtime=-\infty}^{\infty}\
		\int_{-\maxDoppler}^{\maxDoppler}\int_{-\maxDelay}^{\maxDelay}\scafun\lefto(\doppler-\frac{\dtime}
		{\tstep},\delay\right)\cexn{\dfreq\fstep\delay} d\delay d\doppler \\
	&=\int_{-\maxDoppler}^{\maxDoppler}\int_{-\maxDelay}^{\maxDelay}\scafunp
		\cexn{\dfreq\fstep\delay}d\delay d\doppler\\
	&=\chcorr[0,\dfreq],
\eas
it follows from~\fref{eq:diag-element} that
\bas
	\diagelcirc_{\indvar} = \Re\lefto\{\frac{2}{\fslots}\sumz{\dfreq}
		{\fslots}(\fslots-\dfreq)\chcorr[0,\dfreq]\cexn{\frac{\indvar\dfreq}
		{\fslots}}\right\}-1
\eas
as defined in~\fref{eq:lb-diag-elem}. 

As a consequence of~\fref{lem:taylor expansion is the same}, the penalty term~\eqref{eq:lbTFpeak-penalty} and its upper bound in~\eqref{eq:first upper bound penalty term} have the same Taylor series expansion around the point~$1/\fslots=0$, while  the upper bound on the penalty term given on the RHS of~\eqref{eq:lbTFpeak-penalty-ub} has the same Taylor series expansion around the point~$1/\fslots=0$ as~\eqref{eq:lbTFpeak-penalty} only when the Jensen penalty in~\eqref{eq:lbTFpeak-penalty-ub} is zero. 
This happens for scattering functions that are flat in the Doppler
domain, or, equivalently, that satisfy~\fref{eq:doppler-flat-scafun}.

We next provide an explicit expression for the Taylor series expansion of the penalty term~\eqref{eq:lbTFpeak-penalty} around~$1/\fslots=0$; this expression will be needed in the next section, as well as in \fref{app:lbTFpeak-taylor-proof}. As the Fourier transform $\sum_{\dfreq=-\infty}^{\infty}\mvchspecfunentryp\cex{\dfreq\altspecparam }$ of the sequence~$\{\mvchspecfunentryp\}$ is the two-dimensional power spectral density~\chspecfunp defined in~\eqref{eq:chspecfun}, we have by~\fref{lem:taylor expansion is the same} that
\be
\bs
	\frac{1}{\tstep}\!\int_{-1/2}^{1/2}\logdet{\imat_{\fslots}
		+\frac{\Pave\tstep}{\fslots}\mvchspecfunp}d\specparam
		&=\frac{1}{\tstep} \sum_{\taylexpindex=0}^{\taylexp} \frac{(-1)^{\taylexpindex}}{(\taylexpindex+1)\fslots^{\taylexpindex}}\int_{-1/2}^{1/2}\int_{-1/2}^{1/2}[\Pave\tstep\chspecfunp]^{\taylexpindex+1}d\altspecparam d\specparam+\osmall{\frac{1}{\fslots^{\taylexp}}}\\
		&= \Pave\sum_{\taylexpindex=0}^{\taylexp} \frac{(-1)^{\taylexpindex}}{\taylexpindex+1}\left(\frac{\Pave}{\fslots\fstep}\right)^{\taylexpindex}\spreadint{[\scafunp]^{\taylexpindex+1}}+\osmall{\frac{1}{\fslots^{\taylexp}}}
\es
\label{eq:taylor expansion penalty term}
\ee
where in the last step we first used~\eqref{eq:specfun-scafun} and then proceeded as in~\eqref{eq:pathloss}.
\subsubsection{Lower bound}
To lower-bound the penalty term~\fref{eq:lbTFpeak-penalty}, we use
\fref{lem:logdetinf-1d} in \fref{app:mmse-mi} for the case when~\vecx is an~$\fslots$-dimensional
vector with all-$1$ entries and obtain
\be
\label{eq: lower bound penalty term appendix}
\bs	
	\frac{1}{\tstep}\int_{-1/2}^{1/2}\logdet{\imat_{\fslots}
		+\frac{\Pave\tstep}{\fslots}\mvchspecfunp}d\specparam
	&\geq\frac{\fslots}{\tstep}\int_{-1/2}^{1/2}\int_{-1/2}^{1/2}
	\log\lefto(1+\frac{\Pave\tstep}{\fslots}\chspecfunp\right) d\altspecparam d\specparam\\
	&=\fslots\fstep\!\spreadintlog{ 1 + \frac{\Pave\tstep}{\fslots}\scafunp}
\es
\ee
where in the last step we again first used~\eqref{eq:specfun-scafun} and then proceeded as in~\eqref{eq:pathloss}.
We next show that the penalty term~\fref{eq:lbTFpeak-penalty} and its lower bound~\eqref{eq: lower bound penalty term appendix} have the same Taylor series expansion [given in~\eqref{eq:taylor expansion penalty term}].  For any fixed~$(\doppler,\delay)$ the function~$\fslots\fstep\log\lefto(1 + \Pave\tstep\scafunp\right/\fslots)$ is nonnegative, and monotonically increasing in~\fslots. Hence, by the monotone convergence theorem~\cite[Th.~11.28]{rudin76a}, 
we can expand the logarithm inside the integral on the RHS of~\eqref{eq: lower bound penalty term appendix} into a Taylor series. The resulting Taylor series expansion coincides with the Taylor series expansion of~\fref{eq:lbTFpeak-penalty} stated in~\eqref{eq:taylor expansion penalty term}.

\section{Proof of Lemma~\ref{lem:ubTFpeak-taylor}}
\label{app:ubTFpeak-taylor-proof}

To prove \fref{lem:ubTFpeak-taylor}, we need to evaluate~$\liminbw
\bandwidth\ubTFpeakp$, where~\ubTFpeakp is the upper bound
in~\fref{eq:ubTFpeak}. Our analysis is similar to the asymptotic analysis of an
upper bound on capacity in~\cite[Prop.~2.1]{sethuraman08a}, with the main
difference that we deal with a time- and frequency-selective channel whereas the
channel analyzed in~\cite{sethuraman08a} is frequency flat.
We start by computing the first-order Taylor series expansion of~\ubpenaltyp in~\eqref{eq:ubTFpeak-penalty}
around~$1/\bandwidth=0$.
This first-order Taylor series expansion follows directly from~\fref{app:penalty term}, and is given by:
\be
\label{eq:ubpenalty-taylor}
\bs
\ubpenaltyp&=\frac{\bandwidth}{\papr}\spreadintlog{1+\frac{\papr\Pave}
		{\bandwidth}\scafunp}\\
		&=\Pave-\frac{\papr\Pave^{2}}{2\bandwidth}\underbrace{\spreadint{
	\scafunpsq}}_{\peakiness}+\osmall{\frac{1}{\bandwidth}}.
\es
\ee
We now use~\eqref{eq:ubpenalty-taylor} to evaluate the minimum in~\fref{eq:ubTFpeak-avpopt}.
\be
\bs
	\liminbw\frac{\bandwidth}{\tfstep}\left(\frac{1}{\ubpenaltyp}-\frac{1}
		{\Pave}\right) 
	&=\liminbw\frac{\bandwidth}{\tfstep}\left(\frac{1}{\Pave - \papr
		\peakiness\Pave^{2}/(2\bandwidth) +\osmall{1/\bandwidth} } -\frac{1}
		{\Pave} \right) \\
	&= \liminbw\frac{\bandwidth}{\tfstep\Pave}\left(\frac{1}{1 - \papr
		\Pave\peakiness/(2\bandwidth) +\osmall{1/\bandwidth} } - 1
		\right)\\
	&\stackrel{(a)}{=} \liminbw\frac{\bandwidth}{\tfstep\Pave}
		\left(\frac{\papr \Pave\peakiness}{2\bandwidth} + \osmall{
		\frac{1}{\bandwidth}}\right)	= \frac{\papr\peakiness}{2\tfstep}
\es
\label{eq:ubTFpeak-avp-lim}
\ee
where we used the Taylor series expansion~$1/(1-x)=1+x+\osmall{x}$ for~$x\to0$ to
obtain equality~(a). Because~\avpoptp is defined in~\fref{eq:ubTFpeak-avpopt}
as the minimum
\be
	\avpoptp\define\min\lefto\{1,\, \frac{\bandwidth}{\tfstep}\left(\frac{1}
		{\ubpenaltyp}-\frac{1}{\Pave}\right)\right\}
\een
we need to distinguish two cases. 
\begin{itemize}
\item If~$\papr>2\tfstep/\peakiness$, we get~$\liminbw
	\avpoptp=1$, so that, for
	sufficiently large bandwidth, the upper bound~\fref{eq:ubTFpeak-core} can
	be expressed as
	%
	%
	\be
	\bs
		\ubTFpeakp &= \frac{\bandwidth}{\tfstep}\log\lefto(1 + \Pave
			\frac{\tfstep}{\bandwidth}\right) -\ubpenaltyp\\
		&= \Pave - \frac{1}{2}\Pave^{2}
			\frac{\tfstep}{\bandwidth} - \Pave + \frac{\papr\Pave^{2}}
			{2\bandwidth}\peakiness + \osmall{\frac{1}{\bandwidth}}\\
		&= \frac{\Pave^{2}}{2\bandwidth}\left(\papr\peakiness-\tfstep
			\right)+\osmall{\frac{1}{\bandwidth}}.
	\es\label{eq:ubTFpeak-taylor-paprlarge}
	\ee
	Consequently, we obtain the first-order Taylor series coefficient
	\be
		\taylorone=\liminbw\bandwidth\ubTFpeakp = \frac{\Pave^{2}}{2}
			\left(\papr\peakiness - \tfstep\right).
	 \een
\item If~$\papr\leq 2\tfstep/\peakiness$, we
	get
	\be
		\liminbw\avpoptp=\liminbw\frac{\bandwidth}{\tfstep}\left(\frac{1}{\ubpenaltyp}
		- \frac{1}{\Pave}\right)
	\een
	so that for sufficiently large bandwidth
	\be
	\bs
		\ubTFpeakp &= \frac{\bandwidth}{\tfstep}\log\lefto(\frac{\Pave}
			{\ubpenaltyp}\right) + \frac{\bandwidth}{\tfstep}\left(
			\frac{\ubpenaltyp}{\Pave} - 1\right)\\
		&= \frac{\bandwidth}{\tfstep}\left(\frac{\ubpenaltyp}{\Pave}
			- 1 -\log\lefto(1 + \frac{\ubpenaltyp}{\Pave} -1 \right)
			\right).
	\es\label{eq:ubTFpeak-paprsmall}
	\ee
	We now use the Taylor series~$x-\log(1+x)=x^{2}/2
	+\osmall{x^{2}}$	for~$x\to0$ on the RHS of~\fref{eq:ubTFpeak-paprsmall} to get
		\be
	\bs
		\ubTFpeakp &= \frac{\bandwidth}{2\tfstep}\left(\frac{\ubpenaltyp}
			{\Pave}-1\right)^{2} + \osmall{\frac{1}{\bandwidth}}\\
		&\stackrel{(a)}{=}\frac{\bandwidth}{2\tfstep}\left(\frac{\papr\Pave
			\peakiness}{2\bandwidth} + \osmall{\frac{1}{\bandwidth}}
			\right)^{2}+\osmall{\frac{1}{\bandwidth}}\\
		&= \frac{(\papr\Pave\peakiness)^{2}}{8\tfstep\bandwidth}
			+ \osmall{\frac{1}{\bandwidth}}
	\es\label{eq:ubTFpeak-taylor-paprsmall}
	\ee
	where~(a) follows from the Taylor series expansion of~\ubpenaltyp
	in~\fref{eq:ubpenalty-taylor}. Hence, the first-order Taylor series
	coefficient of the upper bound~\ubTFpeakp is given by
	\be
		\taylorone=\liminbw\bandwidth\ubTFpeakp = \frac{(\papr\Pave
			\peakiness)^{2}}{8\tfstep}.
	\een	
\end{itemize}
Both cases taken together yield~\fref{eq:TFpeak-taylor}.

\section{Proof of \fref{lem:lbTFpeak-taylor}}
\label{app:lbTFpeak-taylor-proof}

To prove \fref{lem:lbTFpeak-taylor}, we need to evaluate 
$\liminbw \bandwidth\lbTFpeakp$, where~\lbTFpeakp is the lower bound~\fref{eq:lbTFpeak}.
The first term in~\fref{eq:lbTFpeak} is the coherent mutual information of a scalar
Rayleigh-fading channel with zero-mean constant-modulus input.
This mutual information has the following first-order Taylor series expansion around $1/\bandwidth=0$~\cite[Th.~14]{verdu02-06a}:
\ba
	\frac{\bandwidth}{\param\tfstep} \mi(\outp;\inp\given\ch) = \Pave -\frac{\param P^{2} \tfstep}{\bandwidth} +\landauo\lefto(\frac{1}{\bandwidth}\right).
	\label{eq: taylor expansion first part lower bound}
\ea

We now analyze the second term in~\fref{eq:lbTFpeak}; its Taylor series expansion around $1/\bandwidth=0$ (for the case~$\param=1$) is given in~\eqref{eq:taylor expansion penalty term}. If we truncate this expansion to first order and take into account the factor~\param, we obtain
\ba
\frac{1}{\param\tstep}
		\int_{-1/2}^{1/2}&\logdet{\imat_{\fslots}+\frac{
		\param\Pave\tfstep}{\bandwidth}\mvchspecfunp}
		d\specparam=\Pave-\frac{\param\Pave^{2}}{2\bandwidth}\peakiness+\osmall{\frac{1}{\bandwidth}}
\label{eq:taylor expansion second part lower bound}
\ea
where~\peakiness is defined in~\eqref{eq:peakiness}.
We then combine~\fref{eq: taylor expansion first part lower bound} and~\fref{eq:taylor expansion second part lower bound} to get  the desired result
\bas
	\liminbw \bandwidth\lbTFpeakp &= \liminbw  \max_{1\le\param\le\papr}
		\bandwidth\Biggl[ \Pave -\frac{\param\Pave^{2}\tfstep}{\bandwidth}
		-\Pave + \frac{\param \Pave^{2}\peakiness}{2\bandwidth} +\landauo
		\lefto(\frac{1}{\bandwidth} \right) \Biggr]\\
	&=\papr\Pave^{2}(\peakiness/2 - \tfstep).
\eas
%

\section{Proof of \fref{thm:TFpeak-taylor}}
\label{app:altlbTFpeak-taylor}

To prove \fref{thm:TFpeak-taylor}, we need to find  a lower bound
on~\capacityp whose first-order Taylor series expansion matches that of the upper
bound~\ubTFpeakp given in~\fref{eq:TFpeak-taylor}. To obtain such a
lower bound, we compute the mutual information for a specific input distribution
that (slightly) generalizes the input distribution used in~\cite{sethuraman08a}. 
For a given time duration~$\tslots\tstep$ and bandwidth~$\fslots\fstep$,
we shall first specify the distribution of the input symbols that belong to a generic
$\tslotsb \times \fslotsb$ rectangular block in the time-frequency plane, where~\tslotsb and~\fslotsb are fixed and
$\tslotsb \leq \tslots$, $\fslotsb \leq \fslots$, and then describe
the joint  distribution of all input symbols in the overall~$\tslots \times \fslots$ rectangle; transmission over the~$\tslots \times \fslots$ rectangle is denoted  as a {\em channel use}. 
Within a $\tslotsb \times \fslotsb$  block, we use~\iid zero-mean constant-modulus signals.
We arrange these signals in a $\tslotsb\fslotsb$-dimensional
vector~\mvscminpb in the same way as in~\fref{eq:stacked-input}, i.e., we
stack first in frequency and then in time. Finally, we let 
the input vector for the $\tslotsb \times \fslotsb$  block be~$\mvsinpb=\ampfactor\,\mvscminpb$, where~\ampfactor
is a binary RV with distribution
\be
	\ampfactor=\begin{cases}
		\sqrt{{\papr\Pave\tstep}/{\fslots}}, &\text{with 
			probability~\dutycycle,}\\
		0, &\text{with probability~$1-\dutycycle$.}
	\end{cases}
\een
This means that the~\iid constant-modulus vector~\mvscminpb undergoes on-off
modulation with {\em duty cycle}~\dutycycle. The above signaling scheme
satisfies the peak constraint~\fref{eq:peak-per-tfslot} by construction. The
covariance matrix of the input vector~\mvsinpb is given by
\be
	\Ex{}{\mvsinpb\herm{\mvsinpb}}=\Ex{\ampfactor}{\Ex{\mvsinpb}{\mvsinpb
		\herm{\mvsinpb} \given\ampfactor}}=\dutycycle\frac{\papr\Pave\tstep}
		{\fslots}\imat_{\tfslotsb}
\een
so that for~$\dutycycle\le1/\papr$ the signaling scheme also satisfies the 
power constraint~$\Ex{}{\vecnorm{\mvsinpb}^{2}}\leq\tslotsb\fslotsb\Pave\tstep/\fslots$. 
In the remainder of 
this appendix we will assume that~$\dutycycle\le1/\papr$. The
input-output relation for the transmission of the~$\tslotsb\times\fslotsb$ block can now be written as
\bas    
	\mvsoutpb=\mvsinpb\had\mvschb + \mvswgnb
\eas
where the \tfslotsb-dimensional stacked output vector~\mvsoutpb, the
corresponding stacked channel vector~\mvschb, and the stacked noise
vector~\mvswgnb are defined in the same way as the stacked input
vector~\mvsinpb.  Finally, we define the correlation matrix of the channel vector~\mvschb 
as~$\mvchcovmatb=\Ex{}{\mvschb\herm{\mvschb}}$.

Let now~$\modtime=\floor{\tslots/\tslotsb}$ and~$\modfreq=\floor{\fslots/\fslotsb}$. In a channel use,
we let the $\tfslots$-dimensional input vector~\altinpvec with entries~$\{\altinp[\dtdf]\}$ be constructed as follows: we use~$\modtime\tslotsb\cdot\modfreq\fslotsb$ out of the~$\tslots\fslots$ entries of~\altinpvec to form~$\modtime\modfreq$ subvectors, each of dimension~$\tslotsb\fslotsb$, and we leave the remaining~$\tslots\fslots-\modtime\tslotsb\cdot\modfreq\fslotsb$ entries unused. For $\allz{\modtimeindex}{\modtime}$ and $\allz{\modfreqindex}{\modfreq}$, the~$(\modtimeindex,\modfreqindex)$th subvector is constructed from the entries of~\altinpvec in the set~$\{\altinp[\dtdf] \sothat
\dtime=\modtimeindex\tslotsb,\modtimeindex\tslotsb+1,\ldots,(\modtimeindex+1)\tslotsb-1;\,
\dfreq=\modfreqindex\fslotsb,\modfreqindex\fslotsb+1,\ldots,(\modfreqindex+1)\fslotsb-1\}$.
Finally, we assume that the~$\modtime\modfreq$ subvectors are independent and are distributed as~\mvsinpb, so that
\bas
	\Ex{}{\vecnorm{\altinpvec}^{2}}=\modtime\modfreq\Ex{}{\vecnorm{\mvsinpb}^{2}}\leq\modtime
	\modfreq\tslotsb\fslotsb\Pave\tstep/\fslots \leq \tslots\Pave\tstep.
\eas
Hence, the vector~\altinpvec satisfies both the average power constraint and the peak constraint~\eqref{eq:peak-per-tfslot}  in~\fref{sec:power constraints}.
Finally, we have 
\be
\label{eq:lbTFpeak-blockfading}
\bs
\capacityp=\limintime\frac{1}{\tslots\tstep}\sup_{\dsetpapr}
		\mi(\mvsoutp;\mvsinp)
		&\ge\limintime\frac{1}{\tslots\tstep}\mi(\mvsoutp;\altinpvec)\\
		&\stackrel{(a)}{\ge}	\limintime \frac{\modtime\modfreq}{\tslots\tstep}\mi(\mvsoutpb;\mvsinpb)\\
		&\stackrel{(b)}{=}\frac{\modfreq}{\tslotsb\tstep}\mi(\mvsoutpb;\mvsinpb)
\es
\ee
where~(a) follows from the chain rule of mutual information (the intermediate steps are
detailed in~\cite[App.~A]{sethuraman08a}), and in~(b) we used
\bas
	\limintime \frac{\modtime}{\tslots}=\limintime \frac{\floor{\tslots/\tslotsb}}{\tslots}=\frac{1}{\tslotsb}.
\eas

Because we are only interested in the asymptotic behavior of the lower bound~\eqref{eq:lbTFpeak-blockfading}, it
suffices to analyze the second-order Taylor series expansion of $\mi(\mvsoutpb;\mvsinpb)$
around~$1/\fslots=0$. As the entries of~\mvsinpb are peak-constrained, and~\mvschb is a proper complex vector, we can use the expansion derived 
in~\cite[Cor.~1]{prelov04-08a} to obtain\footnote{Differently from~\cite[Cor.~1]{prelov04-08a}, the Taylor series expansion is for $\fslots\to\infty$; furthermore, we have~$N_{0}=1$, and the SNR is given by~$\tslotsb\fslotsb\Pave\tstep/\fslots$.}
\begin{multline}
	\mi(\mvsoutpb;\mvsinpb)=\frac{1}{2}\tr\lefto\{\Ex{\mvsinpb}{\Bigl(
		\Ex{\mvschb}{(\mvschb\had\mvsinpb)\herm{\bigl(\mvschb\had \mvsinpb
		\bigr)}}\Bigr)^{2}} \right\}\\
	-\frac{1}{2}\tr\lefto\{\left(\Ex{\mvschb,\mvsinpb}{(\mvschb\had\mvsinpb)
		\herm{\bigl(\mvschb\had \mvsinpb\bigr)}}\right)^{2}\right\}
		+\osmall{\frac{1}{\fslots^{2}}}.
\label{eq:lbTFpeak-block-mi-prelov}
\end{multline}
In the following, we analyze the two trace terms separately.

The first term is:
\be
\bs
	\tr\Biggl\{&\Ex{\mvsinpb}{\Bigl(\Ex{\mvschb}{(\mvschb\had\mvsinpb)
		\herm{\bigl(	\mvschb\had \mvsinpb\bigr)}}\Bigr)^{2}}\Biggr\}\\
	&\stackrel{(a)}{=}\tr\lefto\{\Ex{\mvsinpb}{\left(\mvchcovmatb\had
		\bigl(\mvsinpb\herm{\mvsinpb}\bigr)\right)^{2}}\right\}\\
	&\stackrel{(b)}{=}\tr\lefto\{\Ex{\mvsinpb}{\herm{\Bigl(\mvchcovmatb
		\had	\bigl(\mvsinpb\herm{\mvsinpb}\bigr)\Bigr)}\Bigl(\mvchcovmatb\had
		\bigl(\mvsinpb\herm{\mvsinpb}\bigr)\Bigr)}\right\}\\
	&\stackrel{(c)}{=}\tr\lefto\{\Ex{\mvsinpb}{\herm{\mvchcovmatb}\Bigl(\bigl(
		\conj{\mvsinpb}\tp{\mvsinpb}\bigr) \had \mvchcovmatb \had \bigl(\mvsinpb
		\herm{\mvsinpb}\bigr) \Bigr) }   \right\} \\
	&\stackrel{(d)}{=} \dutycycle\tr\lefto\{\herm{\mvchcovmatb}\left(
		\mvchcovmatb \had \Ex{\mvsinpb}{\bigl(\conj{\mvsinpb}\tp{\mvsinpb}\bigr)
		\had\bigl(\mvsinpb\herm{\mvsinpb}\bigr)\Bigg|\ampfactor=\sqrt{\frac{
		\papr\Pave\tstep}{\fslots}}} \right)\right\}\\
	&\stackrel{(e)}{=} \dutycycle \left(\frac{\papr\Pave\tstep}{\fslots}
		\right)^{2} \tr\lefto\{\herm{\mvchcovmatb}\mvchcovmatb\right\}.
\es
\label{eq:lbTFpeak-trace}
\ee
Here,~(a) follows from~\fref{eq:had-gramian}, 
(b)~follows because $\mvchcovmatb$ and $\mvsinpb\herm{\mvsinpb}$ are Hermitian and~(c)
follows from the identity~\cite[p.~42]{luetkepohl96a}
\bas
	\tr\lefto\{\herm{\bigl(\matA \had \matB\bigr)}\matC \right\}
		=\tr\lefto\{ \herm{\matA} (\conj{\matB} \had \matC) \right\}.
\eas
We obtain~(d) as the Hadamard product is commutative and~(e) holds because the
entries of the matrix~$\bigl(\conj{\mvsinpb}\tp{\mvsinpb}\bigr)\had\bigl(
\mvsinpb\herm{\mvsinpb}\bigr)$ are all equal to~$(\papr\Pave\tstep)^{2}/
\fslots^{2}$ \wpone given that~$\ampfactor=\sqrt{\papr\Pave\tstep/\fslots}$.

To evaluate the second trace term in~\fref{eq:lbTFpeak-block-mi-prelov}, we
once more use the identity~\fref{eq:had-gramian}:
\be
\bs
	\tr\lefto\{\left(\Ex{\mvschb,\mvsinpb}{(\mvschb\had\mvsinpb)\herm{\bigl(
		\mvschb\had \mvsinpb\bigr)}}\right)^{2}\right\}
	&=\tr\lefto\{\left(\mvchcovmatb\had\frac{\dutycycle\papr\Pave\tstep}
		{\fslots}\imat_{\tfslotsb}\right)^{2}\right\}\\
	&=\tfslotsb\left(\frac{\dutycycle\papr\Pave\tstep}{\fslots}
		\right)^{2}
\es
\label{eq:lbTFpeak-asymp-trace2}
\ee
where the last equality follows because we normalized~$\chcorr[0,0]= \pathloss=1$ (see \fref{sec:dtdf-io}).

Next, we substitute the trace terms~\fref{eq:lbTFpeak-trace}
and~\fref{eq:lbTFpeak-asymp-trace2} into the second-order expansion of mutual
information in~\fref{eq:lbTFpeak-block-mi-prelov}, which, together with the
lower bound in~\fref{eq:lbTFpeak-blockfading}, results in the following lower
bound on~$\liminbw\bandwidth\capacityp$, valid for any
fixed~\tslotsb and~\fslotsb:
\be
\bs
	\liminbw\bandwidth\capacityp 
	&\ge\liminfreq\frac{\modfreq\fslots\fstep}{\tslotsb\tstep}
		\mi(\mvsoutpb;\mvsinpb)\\
	&=\liminfreq\frac{\modfreq\fslots\fstep}{2\tslotsb\tstep}\Biggl[\dutycycle
		\left(\frac{\papr\Pave\tstep}{\fslots}\right)^{2}\tr\lefto\{\herm{
		\mvchcovmatb}\mvchcovmatb\right\}\\
	&\hphantom{=}\qquad-\tfslotsb\left(\frac{\dutycycle\papr\Pave\tstep}
		{\fslots}	\right)^{2} +\osmall{\frac{1}{\fslots^{2}}}\Biggr]\\
	&=\left(\liminfreq\frac{\modfreq}{\fslots}\right)\frac{(\dutycycle\papr\Pave)^{2}}{2}\Biggl[
		\frac{\tfstep}{\dutycycle\tslotsb}\tr\lefto\{\herm{
		\mvchcovmatb}\mvchcovmatb\right\}-\fslotsb\tfstep\Biggr] \\
	&=\frac{(\dutycycle\papr\Pave)^{2}}{2}\Biggl[
		\frac{\tfstep}{\dutycycle\tfslotsb}\tr\lefto\{\herm{
		\mvchcovmatb}\mvchcovmatb\right\}-\tfstep\Biggr]
\es
\label{eq:lbasymTFpeak}
\ee
where in the last step we used~$\liminfreq\modfreq/\fslots=\liminfreq\floor{\fslots/\fslotsb}/\fslots=1/\fslotsb$.

If we now take~\tslotsb and \fslotsb sufficiently large, the RHS of~\eqref{eq:lbasymTFpeak} can be made arbitrarily close to its limit for~$\tslotsb \to \infty$ and~$\fslotsb \to \infty$. This limit admits a closed-form expression in~\scafunp. In fact,
\be
\label{eq: intermediate term}
\bs
	\lim_{\tslotsb,\fslotsb\to\infty}\frac{1}{\tfslotsb}\tr\lefto\{\herm{
		\mvchcovmatb}\mvchcovmatb\right\} 
		&\stackrel{(a)}{=}\lim_{\tslotsb,\fslotsb\to\infty}\frac{1}{\tfslotsb}\sumo{\dtime}{\tslotsb}\sumo{\dfreq}{\fslotsb}
		\lambda^{2}_{\dtime,\dfreq}(\mvchcovmatb)\\
		&\stackrel{(b)}{=} \int_{-1/2}^{1/2}\int_{-1/2}^{1/2}\bigl[\chspecfunp\bigr]^{2}d\specparam
		d\altspecparam\\
	       &\stackrel{(c)}{=} \frac{1}{\tfstep}\underbrace{\spreadint{\bigl[\scafunp
		\bigr]^{2}}}_{\peakiness}.
\es
\ee
Here,~(a) follows because~$\mvchcovmatb$ is Hermitian and its~\tfslotsb eigenvalues~$\{\lambda_{\dtime,\dfreq}\}$
are real. The matrix~\mvchcovmatb is two-level Toeplitz and its entries belong to the sequence~$\{\chcorr[\Ddtime,\Ddfreq] \}$ with two-dimensional power spectral density~\chspecfunp defined in~\eqref{eq:chspecfun}; then, (b) follows from the extension of~\eqref{eq:szego's theorem 1D} to two-level Toeplitz matrices provided in~\cite{voois96a}. Finally, to obtain~(c) we proceed as in~\eqref{eq:pathloss}. 
If we now replace~\eqref{eq: intermediate term} in~\eqref{eq:lbasymTFpeak} for~$\tslotsb \to \infty$ and~$\fslotsb \to \infty$ we obtain,
\ba
\label{eq:limit lower bound}
	\lim_{\tslotsb,\fslotsb\to\infty}\liminbw\bandwidth\capacityp=\frac{(\dutycycle\papr\Pave)^{2}}{2}\left(\frac{\peakiness}
		{\dutycycle}-\tfstep \right).
\ea
If we choose~$\dutycycle=1/\papr$ whenever~$\papr>2\tfstep/\peakiness$, and~$\dutycycle=\peakiness/(2\tfstep)$ otherwise, the limit~\eqref{eq:limit lower bound} equals the first-order Taylor series coefficient~\taylorone of the upper
bound~\ubTFpeakp in~\fref{eq:TFpeak-taylorone-coeff}. Hence, the first-order
Taylor series expansion of the lower bound~\fref{eq:lbasymTFpeak} can be made to match
the first-order Taylor series expansion of the upper bound~\fref{eq:ubTFpeak} as
closely as desired.

\section{Proof of \fref{thm:viterbi}}
\label{app:viterbi-result-proof}

To obtain a lower bound on~\infcapacity, we compute the rate achievable in the
infinite-bandwidth limit for a specific signaling scheme. Similarly to the proof
of \fref{thm:TFpeak-taylor} in \fref{app:altlbTFpeak-taylor}, it suffices to specify only the distribution
of the input symbols that belong to a generic rectangular block in the time-frequency plane. Differently from \fref{app:altlbTFpeak-taylor}, we take the generic block to be of dimension~$\tslotsb\times\fslots$,
where~\tslotsb is fixed and~$\tslotsb\leq\tslots$.
We denote the input
symbols  in each time-frequency slot of the~$\tslotsb\times\fslots$ block as~$\inpb[\dtdf]$ and 
arrange them in a vector where---differently from
\fref{sec:dtdf-io}---we first stack along time and then along frequency.
The \tslotsb-dimensional vector that contains the input symbols in the $\dfreq$th
frequency slot is defined as
\bas
	\mvsinpb[\dfreq]\define\tp{\mat\inpb[0,\dfreq]\; \inpb[1,\dfreq]\; \cdots\;
		\inpb[\tslotsb-1,\dfreq]\emat}
\eas
and the \tfslotsbv-dimensional vector that contains all symbols in the block is
\ba
\label{eq:viterbi-vecstack}
	\mvsinpb\define\tp{\mat\tp{\mvsinpb}[0]\; \tp{\mvsinpb}[1]\; \cdots\;
		\tp{\mvsinpb}[\fslots-1]\emat}.
\ea
We define the stacked channel vector~\mvschb, the stacked noise vector~\mvswgnb, and the stacked output vector~\mvsoutpb in a similar way. The input-output
relation corresponding to the~$\tslotsb\times\fslots$ block is
\ba\label{eq:viterbi-block-io}
	\mvsoutpb=\mvsinpb\had\mvschb + \mvswgnb.
\ea
Finally, we denote the correlation matrix of the channel vector~\mvschb
by~\mvchcovmatb; this matrix is again two-level Toeplitz.
Within the~$\tslotsb\times\fslots$ block, we use a signaling scheme that is a generalization  of
the on-off FSK scheme proposed in~\cite{gursoy06a}, and can be viewed as FSK in the channel's eigenspace.
\begin{dfn}[On-off Weyl-Heisenberg keying---OO-WHK]
\label{def:whk}
Let~$\mvsinpb_{\subcindex}$ for~$\allz{i}{\fslots}$
denote a~\tfslotsbv-dimensional vector with entries~$\inpb_{\subcindex}[\dtdf]$
that satisfy~$\abs{\inpb_{\subcindex}[\dtdf]}^{2}=\papr \Pave \tstep \krond
[\subcindex-\dfreq]$. We transmit each~$\mvsinpb_{\subcindex}$ with probability
$p=1/(\fslots \papr)$, for~$\allz{i}{\fslots}$, and the all-zero~\tfslotsbv-dimensional
vector~$\veczero$ with probability~$1-1/(\fslots \papr)$.
\end{dfn}

\begin{figure}
\centering
	\includegraphics[width=\figwidth]{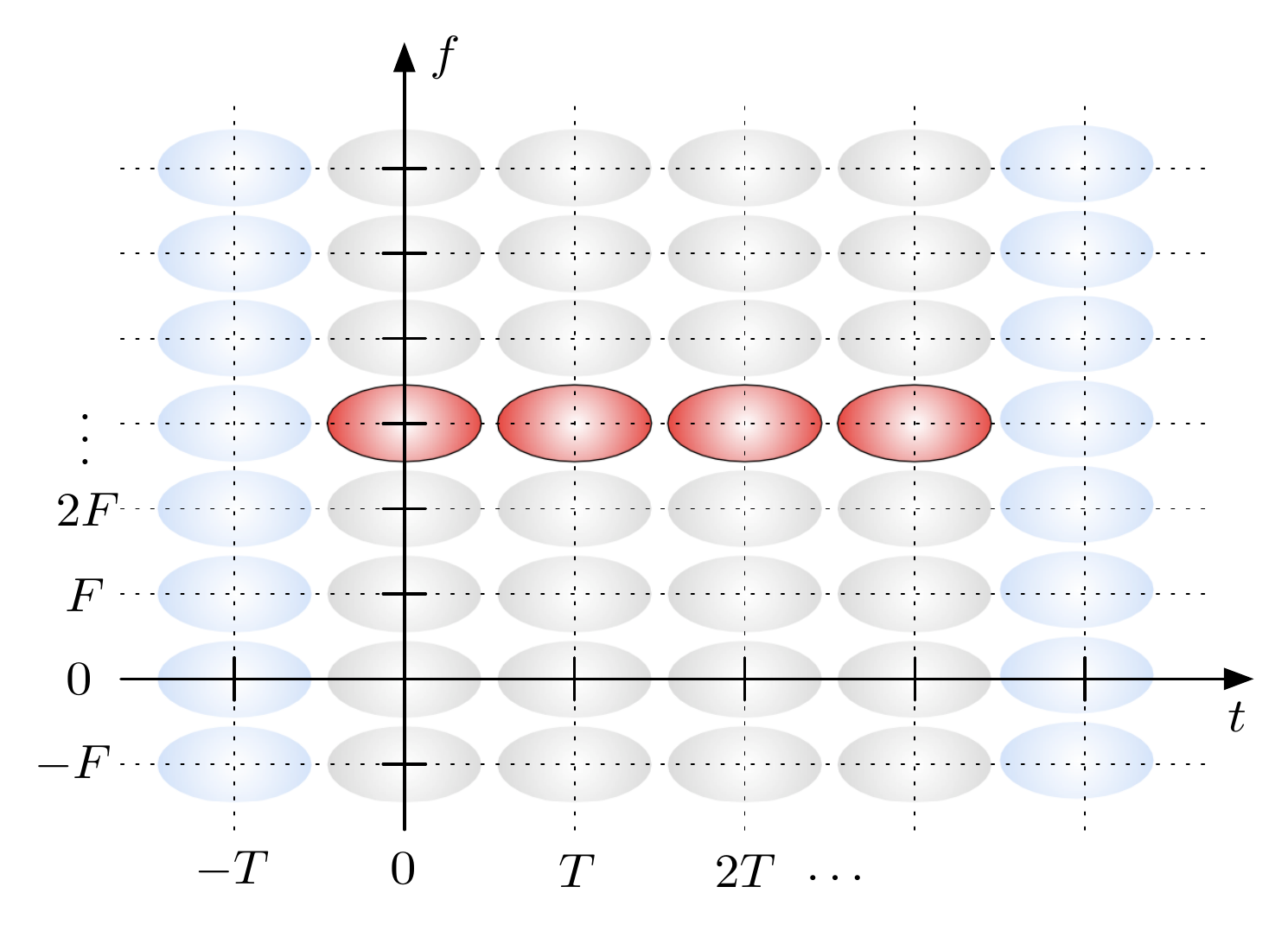}
	\caption{Slots in the time-frequency plane occupied by the symbol $\mvsinpb_{3}$
		for the case $K'=4$.}
\label{fig:oowhk}
\end{figure}

 \fref{fig:oowhk} shows the time-frequency slots occupied by
the symbol~$\mvsinpb_{3}$ for~$\tslotsb=4$.
Steps similar to the one detailed in~\fref{app:altlbTFpeak-taylor} [see~\eqref{eq:lbTFpeak-blockfading}]
yield the following lower bound on~\infcapacity:
\ba\label{eq: inf capacity first lower bound}
	\infcapacity =  \liminfreq \limintime \sup_{\dsetpeaktime} \frac{1}{\tslots
		\tstep} \mi(\mvsoutp;\mvsinp)
		\geq \liminfreq\frac{1}{\tslotsb\tstep} \mi(\mvsoutpb;\mvsinpb).
\ea
Since this lower bound holds for any finite~\tslotsb we can tighten it if
we take the supremum over~\tslotsb; this leads to
\ba
\label{eq:lb-inf}
	\infcapacity \geq \sup_{\tslotsb} \liminfreq \frac{1}{\tslotsb \tstep}
		\mi(\mvsoutpb;\mvsinpb).
\ea
We next decompose the mutual
information in~\fref{eq:lb-inf} as the difference of
KL~divergences~\cite[Eq.~(10)]{verdu90-09a} 
\ba
\label{eq:mi-decomposition-kl}
	\frac{1}{\tslotsb \tstep}\mi(\mvsoutpb;\mvsinpb)= \frac{1}{\tslotsb \tstep}
		\Ex{\mvsinpb}{\kulleib{\mvsoutpb \given \mvsinpb}{\mvsoutpb\given
		\mvsinpb=\veczero}} -\frac{1}{\tslotsb \tstep}\kulleib{\mvsoutpb}
		{\mvsoutpb \given \mvsinpb=\veczero}
\ea
and evaluate the two terms
separately.
As $\cdf{\mvsoutpb \given \mvsinpb}=\jpg\lefto(\veczero,\imat_{\tfslotsbv}
+ \left(\mvsinpb\herm{\mvsinpb}\right)\had \mvchcovmatb  \right)$, we can use
the closed-form expression for the KL divergence of two JPG
random vectors~$\veca\distas\jpg\lefto(\veczero,\covmat{\veca}\right)$ and~$\vecb
\distas \jpg\lefto(\veczero,\imat\right)$~\cite[Eq.~(59)]{verdu02-06a}
\be
\label{eq:kl-jpg}
	\kulleibprob{\jpg\lefto(\veczero,\covmat{\veca}\right)}{\jpg\lefto(\veczero,
		\imat\right)}
		= \tr\lefto(\covmat{\veca} -\imat\right) -\logdet{\covmat{\veca}}.
\ee
Thus, the expected divergence in~\fref{eq:mi-decomposition-kl} can be expressed
as
\be
\bs
	\frac{1}{\tslotsb\tstep}\Ex{\mvsinpb}{\kulleib{\mvsoutpb\given \mvsinpb}
		{\mvsoutpb\given\mvsinpb=\veczero}}
	&= \frac{1}{\tslotsb\tstep}\Ex{\mvsinpb}{\tr\lefto\{\left(\mvsinpb
		\herm{\mvsinpb}\right)\had \mvchcovmatb  \right\}}\\
	&\phantom{=}\quad-\frac{1}{\tslotsb\tstep}\Ex{\mvsinpb}{\logdet{\imat_{\tslotsb\fslots}
		+\left(\mvsinpb\herm{\mvsinpb}\right)\had \mvchcovmatb  } }\\
	&\stackrel{}{=}\Pave - \frac{1}{\tslotsb \tstep \fslots \papr}
		\sumz{\subcindex}{\fslots}\logdet{\imat_{\tfslotsbv}+	\left(
		\mvsinpb^{(\subcindex)}\herm{\left(\mvsinpb^{(\subcindex)}\right)}
		\right)\had \mvchcovmatb}.
\es
\label{eq:mi-kl-first}
\ee
The last step follows because each nonzero vector is transmitted with
probability~$1/(\fslots\papr)$ in the OO-WHK signaling scheme of
\fref{def:whk}, and because the diagonal entries of~\mvchcovmatb are
normalized to~$1$. We next exploit the structure of the signaling scheme,
and the fact that the correlation matrix~\mvchcovmatb is two-level Toeplitz, to
simplify the determinant in the second term on the RHS of~\fref{eq:mi-kl-first}
as
\ba
\label{eq:det-property}
	\det\lefto(\imat_{\tslotsb\fslots} + \left(\mvsinpb^{(\subcindex)}
		\herm{\left(\mvsinpb^{(\subcindex)}\right)}\right)\had \mvchcovmatb
		\right)
	= \det\lefto( \imat_{\tslotsb} + \papr \Pave \tstep \chcovmatb \right)
\ea
for all~\subcindex, and where $\mvschb[0]\define\tp{\btm \ch[0,0]\; \ch[1,0]\;
\cdots\; \ch[\tslotsb-1,0]\etm}$ and $\chcovmatb=\Exop[\mvschb[0]
\herm{\mvschb}[0]]$. We next substitute our intermediate
results~\fref{eq:mi-decomposition-kl}, \fref{eq:mi-kl-first},
and~\fref{eq:det-property} into the lower bound~\fref{eq:lb-inf} to obtain
\be
\label{eq:lb-inf-expanded}
	\infcapacity \geq \Pave - \inf_{\tslotsb} \Biggl\{ \frac{1}{\papr \tslotsb
		\tstep} \logdet{\imat_{\tslotsb} + \papr \Pave \tstep \chcovmatb}
	+ \liminfreq \frac{1}{\tslotsb \tstep} \kulleib{\mvsoutpb}{\mvsoutpb
		\given \mvsinpb=\veczero}\Biggr\}.
\ee
In \fref{app:martingale-proof} it is shown that
\bas
	\liminfreq \frac{1}{\tslotsb \tstep} \kulleib{\mvsoutpb}{\mvsoutpb\given
		\mvsinpb=\veczero}=0.
\eas
To conclude, we simplify the second term on the RHS of~\fref{eq:lb-inf-expanded} as
\bas
	\inf_{\tslotsb} \frac{1}{\papr\tslotsb  \tstep} \logdet{\imat_{\tslotsb}
		+ \papr \Pave \tstep \chcovmatb}
	&\stackrel{(a)}{=} \frac{1}{\papr \tstep} \int_{-1/2}^{1/2} \log\lefto(1
		+ \papr \Pave \sum_{\dtime=-\infty}^{\infty}\pDop\lefto(\frac{\specparam
		+\dtime}{\tstep}\right) \right) d \specparam  \\
	&\stackrel{(b)}{=} \frac{1}{\papr} \dopplerintlog{ 1 + \papr \Pave \pDopp}.
\eas
Here, in~(a) we used \fref{lem:logdetinf-1d} in
\fref{app:mmse-mi} for the case when~\vecx is a \tslotsb-dimensional
vector with all-$1$ entries, as well as 
\bas
	\chspecfunoned(\specparam)&=\sum_{\dtime=-\infty}^{\infty} \chcorr[\dtime,0]
		\cexn{\dtime \specparam} \\
		&=\spreadint{\scafunp\sum_{\dtime=-\infty}^{\infty}\cex{\Ddtime\tstep
		\left(\doppler-\frac{\specparam}{\tstep}\right)}}\\
		&=\frac{1}{\tstep}\sum_{\dtime=-\infty}^{\infty}
		\pDop\lefto(\frac{\specparam-\dtime}{\tstep}\right).
\eas
Finally, (b) holds because~\pDopp is
compactly supported on~$[-\maxDoppler,\maxDoppler]$, and~$\tstep\leq 1/(2\maxDoppler)$. 
A change of variables~$\doppler=\specparam /\tstep$ yields the final result.

\section{}
\label{app:martingale-proof}

\begin{lem}
\label{lem:martingales}
Consider a channel with input-output relation\footnote{To keep the notation compact, in this appendix  we drop the tilde notation~[cf.~\eqref{eq:viterbi-block-io}].}\fref{eq:viterbi-block-io}
\bas
	\mvsoutp=\mvsinp\had\mvsch + \mvswgn
\eas
where the~$\tfslotsbv$-dimensional vectors~\mvsoutp, \mvsinp, \mvsch,
and~\mvswgn are defined as in~\fref{eq:viterbi-vecstack}, i.e., stacking is
first along time and then along frequency. Then,
\begin{equation}\label{eq: App I tesi}
    \liminfreq \frac{1}{\tslotsb}
    \kulleib{\mvsoutp}{\mvsoutp\given\mvsinp=\veczero}=0
\end{equation}
for the OO-WHK scheme in \fref{def:whk} of
\fref{app:viterbi-result-proof}.
\end{lem}

\begin{IEEEproof}
Let~$\pdf{\mvsoutp}$ and~$\pdf{\mvsoutp \given \mvsinp}$ be the probability
density functions~(PDFs) associated with the probability
distributions~$\cdf{\mvsoutp}$ and~$\cdf{\mvsoutp \given \mvsinp}$,
respectively. By definition of the KL~divergence,
\ba
\label{eq:kl-divergence}
	\kulleib{\mvsoutp}{\mvsoutp\given\mvsinp=\veczero} = \Ex{\mvsoutp}{
		\log\lefto( \frac{\pdf{\mvsoutp}(\mvsoutp)}{\pdf{\mvsoutp \given
		\mvsinp=\veczero} (\mvsoutp) }\right)}.
\ea
For the OO-WHK scheme in \fref{def:whk}, the PDF~$\pdf{\mvsoutp}$ of the output
vector can be written as
\ba
\label{eq: total prob theorem}
	\pdf{\mvsoutp} = \left(1-\frac{1}{\papr}\right) \pdf{\mvsoutp \given \mvsinp
		=\veczero} + \frac{1}{\fslots\papr}\sum_{\subcindex=0}^{\fslots -1}
		\pdf{\mvsoutp \given \mvsinp=\mvsinp_{\subcindex}}.
\ea
The output random vector~\mvsoutp has the same distribution as the noise
vector~$\mvswgn \distas \jpg(\veczero,\imat_{\tslotsb\fslots})$ when~$\mvsinp =
\veczero$. Hence,  $\pdf{\mvsoutp \given \mvsinp =\veczero} = \pdf{\mvswgn}$. To
express~\fref{eq:kl-divergence} in a more convenient form, we define the
following RV:
\bas
	\Sfunp =  \sum_{\subcindex=0}^{\fslots -1} \underbrace{\left[ \left(1
		-\frac{1}{\papr}\right) + \frac{1}{\papr}\frac{ \pdf{\mvsoutp \given
		\mvsinp=\mvsinp_{\subcindex}}(\mvswgn)}{\pdf{\mvswgn}(\mvswgn)}
		\right]}_{\sfunp}.
\eas
We can express the KL divergence~\fref{eq:kl-divergence} as a function of the
RV~\Sfunp as follows:
\bas
	\Ex{\mvsoutp}{\log\lefto( \frac{\pdf{\mvsoutp}(\mvsoutp)}{\pdf{\mvsoutp
		\given \mvsinp=\veczero} (\mvsoutp) }\right)}
	&=\int_{\mvsoutp}  \log\lefto( \frac{\pdf{\mvsoutp}(\mvsoutp)}{
		\pdf{\mvsoutp \given \mvsinp =\veczero}(\mvsoutp)} \right)
		\pdf{\mvsoutp}(\mvsoutp)d \mvsoutp \\
	&= \int_{\mvsoutp} \log\lefto( \left(1-\frac{1}{\papr}\right)
		+\frac{1}{\fslots\papr} \sum_{\subcindex=0}^{\fslots -1}
		\frac{\pdf{\mvsoutp \given \mvsinp=\mvsinp_{\subcindex}} (\mvsoutp)}
		{\pdf{\mvsoutp \given \mvsinp =\veczero}(\mvsoutp)}  \right)  \\
	&\phantom{=}\quad\times\left[ \left(1-\frac{1}{\papr}\right) \pdf{\mvsoutp \given \mvsinp
		=\veczero}(\mvsoutp) + \frac{1}{\fslots\papr}\sumz{\subcindex}{\fslots}
		\pdf{\mvsoutp \given \mvsinp=\mvsinp_{\subcindex}} (\mvsoutp) \right]
		d \mvsoutp  \\
	&=\int_{\mvsoutp} \frac{\Sfun(\mvsoutp)}{\fslots} \log\lefto(\frac{
		\Sfun(\mvsoutp)}{\fslots}\right) \underbrace{\pdf{\mvsoutp \given
		\mvsinp =\veczero}}_{\pdf{\mvswgn}}(\mvsoutp)  d\mvsoutp \\
	&=\Ex{\mvswgn}{\frac{\Sfunp}{\fslots}\log\lefto(\frac{\Sfunp}{\fslots}
		\right)}.
\eas

To prove \fref{lem:martingales}, it
suffices to show that the sequence of RVs~$\{\Vfunp\}$ where
\bas
	\Vfunp=\frac{\Sfunp}{\fslots}\log\lefto(\frac{\Sfunp}{\fslots}\right)
\eas
converges to~$0$ in mean as~$\fslots \to \infty$. To prove this result, we first
show that~$\{\Vfunp\}$ converges to~$0$~\wpone. Then we argue that the sequence forms a
backward submartingale~\cite[p.~474 and p.~499]{grimmett01} so that it converges
to~$0$ also in mean by the submartingale convergence
theorem~\cite[Sec. 32.IV]{loeve77a}.

\subsection{Convergence \wpone}

The RVs~\sfunp are~\iid for~$\allz{\subcindex}{\fslots}$. As this
result is rather tedious to prove, we postpone its proof to
\fref{app:rv-independence}. It is instead straightforward to prove 
that these RVs have mean~$1$. In fact,
\be
	\Ex{\mvswgn}{\sfunp}=\int_{\mvswgn} \left[\left(1
		-\frac{1}{\papr}\right) + \frac{1}{\papr}\frac{ \pdf{\mvsoutp \given
		\mvsinp=\mvsinp_{\subcindex}}(\mvswgn)}{\pdf{\mvswgn}(\mvswgn)}\right]\pdf{\mvswgn}(\mvswgn)d\mvswgn=1.
\een
It then follows  from
the strong law of large numbers that
\bas
	\liminfreq \frac{\Sfunp}{\fslots} =\Ex{\mvswgn}{\sfunzerop} =1 \quad \wpone
\eas
and, as the function  $\fun{x}=x\log x$ is continuous, we have by~\cite[Th.~4.6]{rudin76a} that
\bas
	\liminfreq \Vfunp= \liminfreq\fun{\frac{\Sfunp}{\fslots}} =
		\fun{\liminfreq \frac{\Sfunp}{\fslots}} = 0\quad \wpone.
\eas
\subsection{Convergence in Mean}

As the RVs~$\{\sfunp\}$ are~\iid, the sequence~$\{ \Vfunp \}$
and the decreasing sequence of $\sigma$-fields~$\left\{ \sigmaSet \right\}$,
where~\sigmaSet is the smallest $\sigma$-field with respect to which the random
variables~$\{\Sfunp,\Sfunplus(\mvswgn),\cdots\}$ are measurable, form a backward (or
reverse) submartingale~\cite[p.~474 and p.~499]{grimmett01}. This result  
follows  because the pair~$\left(\left\{ \Sfunp/ \fslots
\right\},\{\sigmaSet\}\right)$  is a backward martingale~\cite[p.~499]{grimmett01}, and because the
function~$\fun{x}=x\log x$ is convex.

Since~$\{ \Vfunp\}$ is a backward submartingale and~$\{ \Vfunp\}$ converges
to~$0$~\wpone as~$\fslots \to \infty$, ~$\{ \Vfunp\}$ converges to~$0$
as~$\fslots \to \infty$ also in mean. This result follows by the backward
submartingale convergence theorem below:
\begin{thm}[see~{\cite[Sec.~32.IV]{loeve77a}}]
\label{thm:submartingale-convergence}
Let~$\{X_{N}\}$ be a backward submartingale with respect to a decreasing
sequence of $\sigma$-fields~$\left\{ \sigmaSet \right\}$. Then~$\{X_{N}\}$
converges~\wpone and in mean to~$X<\infty$ if and only
if~$\Ex{}{\abs{X_{1}}}<\infty$ and~$\liminfreq \Ex{}{X_{N}}> -\infty$.
\end{thm}

To conclude the proof, we need to show that the technical conditions in
\fref{thm:submartingale-convergence} hold, i.e., that the sequence~$\{ \Vfunp \}$
satisfies
\ba
	&\liminfreq \Ex{\mvswgn}{\Vfunp} > -\infty
\label{eq: first property for conv. theorem}
\intertext{and} 
	&\Ex{\mvswgn}{\abs{\Vfunonep}}=\Ex{\mvswgn}{\abs{\sfunzerop \log \sfunzerop}} < \infty.
\label{eq: second property for conv. theorem}
\ea
The first inequality follows from Jensen's inequality and because the~\sfunp have
mean~$1$:
\bas
	\Ex{\mvswgn}{\Vfunp}=\Ex{\mvswgn}{\fun{\frac{\Sfunp}{\fslots}}} \geq \fun{\Ex{\mvswgn}{
		\frac{\Sfunp}{\fslots}}} =0 \quad \forall \fslots.
\eas
The second inequality is proven in \fref{app:boundedness-proof}.
\end{IEEEproof}

\subsection{The Random Variables~\sfunp are~\iid}
\label{app:rv-independence}

To show that the RVs
\bas
	\sfunp =\left[ \left(1-\frac{1}{\papr}\right) + \frac{1}{\papr}
		\frac{ \pdf{\mvsoutp \given \mvsinp=\mvsinp_{\subcindex}}(\mvswgn)}
		{\pdf{\mvswgn}(\mvswgn) } \right]
\eas
are~\iid, we first simplify~$\pdf{\mvsoutp \given \mvsinp=\mvsinp_{\subcindex}}$
as
\be
\label{eq: conditional probability computed}
\bs	
	\pdf{\mvsoutp \given \mvsinp=\mvsinp_{\subcindex}}(\mvswgn)
	&=\frac{\exp\lefto[ -\herm{\mvswgn}\left( \imat_{\tslotsb\fslots} +
		\left(\mvsinp_{\subcindex}\herm{\mvsinp_{\subcindex}}
		\right)\had \mvchcovmat \right)^{-1} \mvswgn \right]}{\pi^{\tslotsb
		\fslots}\det\lefto(\imat_{\tslotsb\fslots} + \left(\mvsinp_{\subcindex}
		\herm{\mvsinp_{\subcindex}}\right)\had \mvchcovmat
		\right)} \\
	&=\dfrac{ \exp\biggl(\displaystyle -\mathop{\sum_{\dfreq = 0}^{\fslots -1}}_{\dfreq \neq
		\subcindex} \vecnorm{\mvswgn[\dfreq]}^{2} - \herm{\mvswgn}[\subcindex]
		\covmatalt^{-1} \mvswgn[\subcindex]\biggr)}{\pi^{\tslotsb\fslots}
		\det(\covmatalt)}
\es
\ee
where we set
\ba
\label{eq:matA}
	\covmatalt=\imat_{\tslotsb} + \papr \Pave \tstep \chcovmat
\ea
and where, as usual,~$\mvswgn\define\tp{\btm\tp{\mvswgn}[0]\; \tp{\mvswgn}[1]\;
\cdots\;	\tp{\mvswgn}[\fslots-1]\etm}.$ To
obtain~\fref{eq: conditional probability computed} we apply the determinant
equality~\fref{eq:det-property} to simplify the denominator. For the numerator,
we used that, for the OO-WHK in \fref{def:whk}, the matrix $\imat_{\tslotsb\fslots} + \left(\mvsinp_{\subcindex}\herm{\mvsinp_{\subcindex}}\right)\had \mvchcovmat$ is block diagonal,
with~$\fslots-1$ blocks equal to~$\imat_{\tslotsb}$ and one block equal to
$\covmatalt=\imat_{\tslotsb} + \papr \Pave \tstep \chcovmat$. Hence, its inverse
is also block diagonal, with~$\fslots-1$ blocks equal to~$\imat_{\tslotsb}$ and
one block equal to~$\inv{\covmatalt}$. Next, we
use~\fref{eq: conditional probability computed} to express the
ratio~$\pdf{\mvsoutp \given \mvsinp=\mvsinp_{\subcindex}} / \pdf{\mvswgn}$ as
\ba
\label{eq: prob. ratio}
	 \frac{\pdf{\mvsoutp \given \mvsinp=\mvsinp_{\subcindex}}(\mvswgn)}{
	 	\pdf{\mvswgn}(\mvswgn) }= \frac{1}{\det\lefto( \covmatalt \right)}
		\exp\lefto[\vecnorm{\mvswgn[\subcindex]}^{2} -\herm{\mvswgn}[\subcindex] 
		\covmatalt^{-1} \mvswgn[\subcindex] \right].
\ea
This last result implies that each~$\sfunp$ depends only on the random noise
vector~$\mvswgn[\subcindex]$. As the noise is white, the random
vectors~$\mvswgn[\subcindex]$ are~\iid for all~\subcindex. Hence, the
RVs~$\sfunp$ are~\iid as well.

\subsection{Proof of Inequality~\fref{eq: second property for conv. theorem}}
\label{app:boundedness-proof}
As $x\log x \geq -e^{-1}$ for all~$x>0$, we have that $\abs{x\log x}\leq x\log x +2e^{-1}$; hence, 
\bas
	\Ex{\mvswgn}{\abs{\sfunzerop \log \sfunzerop}}\leq  \Ex{\mvswgn}{\sfunzerop \log \sfunzerop} +2e^{-1}.
\eas
We next use the convexity of~$x\log x$ and that~$\papr\geq 1$ to upper-bound
$\sfunzerop \log \sfunzerop$ as
\be
\bs
	\sfunzerop \log \sfunzerop &= \left[ \left(1-\frac{1}{\papr}\right)
		+ \frac{1}{\papr}\frac{\pdf{\mvsoutp\given\mvsinp=\mvsinp_{0}}(\mvswgn)}{
		\pdf{\mvswgn}(\mvswgn)}\right]\log\lefto[ \left(1-\frac{1}{\papr}\right)
		+ \frac{1}{\papr}\frac{\pdf{\mvsoutp \given \mvsinp=\mvsinp_{0}}(\mvswgn)}{
		\pdf{\mvswgn}(\mvswgn)}\right]\\
	&\stackrel{(a)}{\leq} \frac{1}{\papr} \left[\frac{ \pdf{\mvsoutp \given \mvsinp
		=\mvsinp_{0}}(\mvswgn)}{\pdf{\mvswgn}(\mvswgn)}\right]\log\lefto[\frac{\pdf{\mvsoutp
		\given\mvsinp=\mvsinp_{0}}(\mvswgn)}{\pdf{\mvswgn}(\mvswgn) } \right] \\
	&\stackrel{(b)}{\leq} \left[\frac{\pdf{\mvsoutp \given \mvsinp=\mvsinp_{0}}(\mvswgn)}{\pdf{\mvswgn}(\mvswgn)}
		\right]\log\lefto[ \frac{ \pdf{\mvsoutp \given \mvsinp=\mvsinp_{0}}(\mvswgn)}
		{\pdf{\mvswgn}(\mvswgn)}\right]
\es
\label{eq: upper bound on r(s)}
\ee
where~(a) follows from the definition of convexity, and in~(b) we used that~$\papr\geq1$.
If we take the expectation on both sides
of~\fref{eq: upper bound on r(s)}, we get
\bas
	\Ex{}{\sfunzerop\log\sfunzerop} &\leq\int_{\mvswgn}\left[\frac{
		\pdf{\mvsoutp \given \mvsinp=\mvsinp_{0}}(\mvswgn)}{\pdf{\mvswgn}
		(\mvswgn)}\right]\log\lefto[ \frac{ \pdf{\mvsoutp\given\mvsinp
		=\mvsinp_{0}}(\mvswgn)}{\pdf{\mvswgn}(\mvswgn)}\right]\pdf{\mvswgn}
		(\mvswgn)d\mvswgn \\
	&\stackrel{(a)}{\leq} \int_{\mvswgn}\pdf{\mvsoutp \given \mvsinp=\mvsinp_{0}}
		(\mvswgn) \abs{\log\lefto[ \frac{ \pdf{\mvsoutp \given \mvsinp
		=\mvsinp_{0}}(\mvswgn)}{\pdf{\mvswgn}(\mvswgn)  }\right]} d\mvswgn \\
	&\stackrel{(b)}{=} \int_{\mvswgn}\frac{ \exp\lefto( -\sum_{\dfreq = 1}^{\fslots -1} \vecnorm{\mvswgn[\dfreq]}^{2} - \herm{\mvswgn}[0]\covmatalt^{-1} \mvswgn[0]\right)}{\pi^{\tslotsb\fslots}\det(\covmatalt)} \\
	&\hphantom{=\int_{\mvswgn}}\times\abs{\log\lefto(	\frac{\exp\lefto(\vecnorm{\mvswgn[0]}^{2}-\herm{\mvswgn[0]}
		\covmatalt^{-1} \mvswgn[0]\right)}{\det( \covmatalt)}
		\right)} d\mvswgn \\
	&\stackrel{(c)}{\leq}\int_{\mvswgn[0]}\frac{\exp\lefto(-\herm{\mvswgn}[0]
		\covmatalt^{-1} \mvswgn[0]\right)}{\pi^{\tslotsb}\det( \covmatalt)}\\
		&\hphantom{=\int_{\mvswgn}}\times
		\left[\vecnorm{\mvswgn[0]}^{2}+\herm{\mvswgn[0]} \covmatalt^{-1} \mvswgn[0]
		+ \log\lefto(\det\lefto(\covmatalt\right)\right)
		\right] d\mvswgn[0] \\
	&< \infty.
\eas
where (a) follows because $\pdf{\mvsoutp \given \mvsinp=\mvsinp_{0}}
		(\mvswgn)>0$ for all~\mvswgn; in (b) we used~\eqref{eq: conditional probability computed}
		and~\eqref{eq: prob. ratio}, while  to obtain~(c) we first integrated over~$\{\mvswgn[\dfreq]\}_{\dfreq=1}^{\fslots-1}$
		and then we used the triangle inequality and that~\matA is positive definite with eigenvalues larger or equal to~$1$ [see~\eqref{eq:matA}]. The last inequality holds because~$\covmatalt$ satisfies the trace constraint~$\tr(\matA)=\tslotsb(1+\papr\Pave\tstep)$, which implies that its eigenvalues are bounded.

\section{Proof of \fref{thm:inf-bw-ub}}
\label{app:viterbi-ub-proof}

\begin{figure}
\centering
	\includegraphics[width=\figwidth]{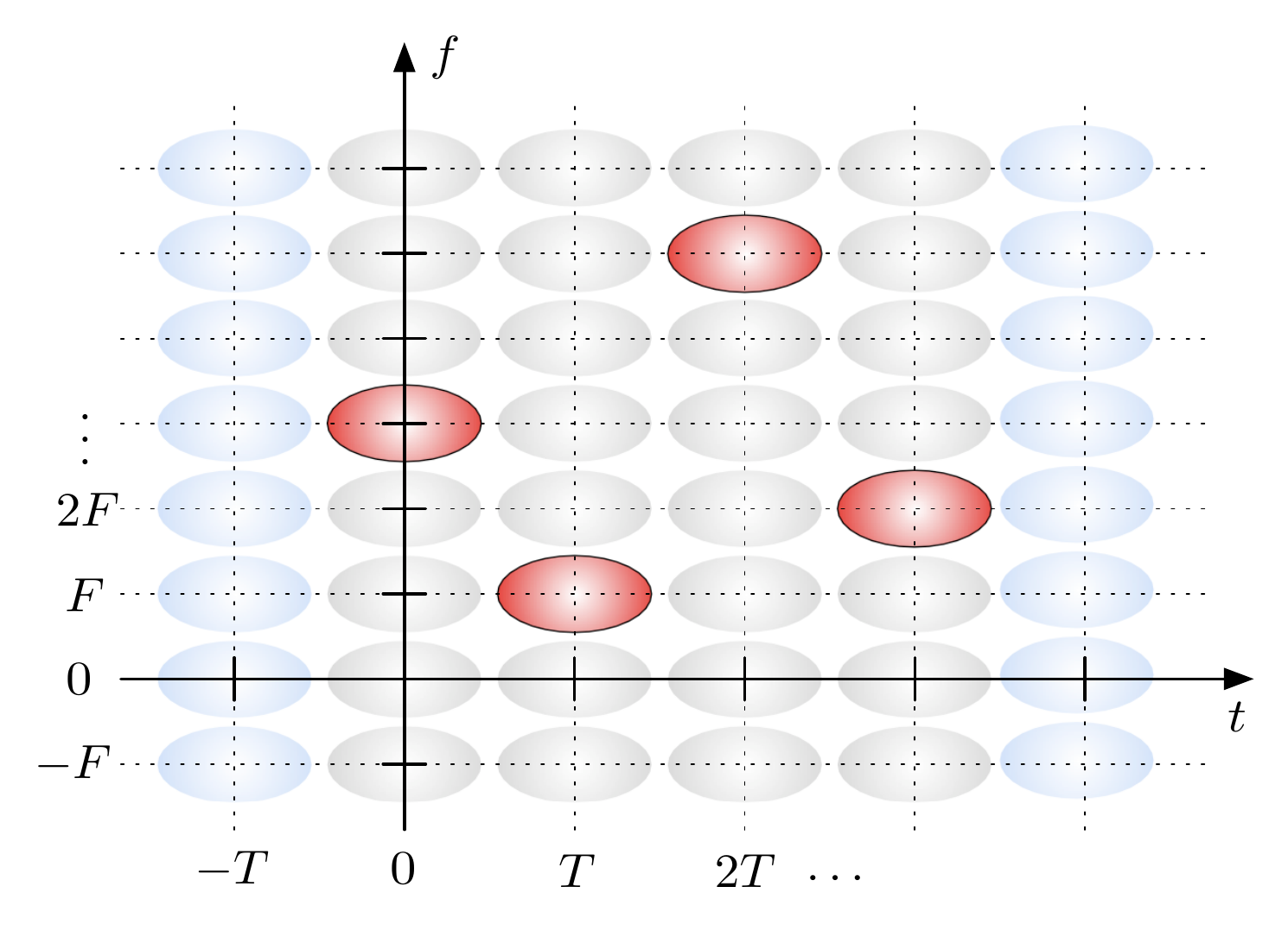}
	\caption{The entries in the time-frequency plane of a vector~\mvsinp 
	that satisfies the
		necessary condition to minimize $\logdet{\imat_{\tslots\fslots}+\left(\mvsinp\herm{\mvsinp}\right)\had \mvchcovmat} / \vecnorm{\mvsinp}^{2}$ in~\eqref{eq:setTpeak} for the case~$\tslots=4$.}
\label{fig:inf-whk-sig}
\end{figure}

We use the decomposition of mutual information as a difference of
KL~divergences~\fref{eq:mi-decomposition-kl}, and
upper-bound~$\sup_{\dsetpeaktime} \mi(\mvsoutp;\mvsinp)$ in~\eqref{eq:infinite-bandwidth capacity} because the KL~divergence
is nonnegative:
\ba
\label{eq: first decomposition}	
	\sup_{\dsetpeaktime} \mi(\mvsoutp;\mvsinp) &=  \sup_{\dsetpeaktime}
		\left\{\Ex{\mvsinp}{\kulleib{\mvsoutp \given \mvsinp}{\mvsoutp\given
		\mvsinp=\veczero}} - \kulleib{\mvsoutp}{\mvsoutp \given
		\mvsinp=\veczero}\right\} \\
	& \leq \sup_{\dsetpeaktime}  \Ex{\mvsinp}{\kulleib{\mvsoutp \given \mvsinp}
		{\mvsoutp\given\mvsinp=\veczero}}.
\ea
As in the proof of \fref{thm:ubTFpeak}, we rewrite the supremum over the
distributions in the set~\dsetpeaktime as a double supremum over~$\pparam\in[0,1]$
and over the restricted set of input distributions~\dsetpeaktimeres that satisfy the
average power constraint~\avPeq and the peak
constraint~\fref{eq:peak-per-tslot}. Then, we use the closed-form expression
for the KL divergence of two multivariate Gaussian vectors~\fref{eq:kl-jpg} and
we follow the same arguments as in the proof of \fref{thm:ubTFpeak}:
\ba
	\frac{1}{\tslots \tstep}\sup_{\dsetpeaktime}& \Ex{\mvsinp}{\kulleib{\mvsoutp
		\given \mvsinp}{\mvsoutp\given\mvsinp=\veczero}} \notag\\
	&=\supavp \sup_{\dsetpeaktimeres}\left\{\alpha \Pave - \frac{1}{\tslots \tstep}
		\Ex{}{\logdet{\imat_{\tslots\fslots} + \left(\mvsinp\herm{\mvsinp}
		\right)\had \mvchcovmat }}\right\} \notag\\
	&=\supavp \left\{\alpha \Pave - \inf_{\dsetpeaktimeres} \frac{1}{\tslots \tstep}
		\Ex{}{\logdet{\imat_{\tslots\fslots} +\left(\mvsinp\herm{\mvsinp}\right)
		\had \mvchcovmat}} \right\} \notag\\
	&\leq \supavp\left\{ \alpha \Pave  -  \alpha\Pave  \inf_{\mvsinp}
		\frac{\logdet{\imat_{\tslots\fslots} +\left(\mvsinp\herm{\mvsinp}\right)
		\had \mvchcovmat}}{\vecnorm{\mvsinp}^{2}} \right\}  \notag\\
	&= \Pave  - \Pave \inf_{\mvsinp} \frac{\logdet{\imat_{\tslots\fslots}
		+\left(\mvsinp\herm{\mvsinp}\right)\had \mvchcovmat}}{\vecnorm{
		\mvsinp}^{2}}.
\label{eq: second decomposition}
\ea
The infimum in~\fref{eq: second decomposition} has the same structure as the
infimum~\fref{eq:ubTFpeak-term2-inf} in the proof of \fref{thm:ubTFpeak}.
Hence, as~\mvchcovmat is positive semidefinite, we can conclude that 
the infimum~\fref{eq: second decomposition} is achieved on the boundary of
the admissible set. Differently from the proof of \fref{thm:ubTFpeak}, however, the input
signal is subject to a peak constraint in time so that the admissible set is
defined by the two conditions
\be
\label{eq:setTpeak}
\bs
	\abs{\inp[\dtdf]}^2 &\in \{0,\papr\Pave\tstep\} \\
	\sum_{\dfreq=0}^{\fslots-1}\abs{\inp[\dtdf]}^2 &\leq \papr\Pave\tstep, \quad \wpone.
\es
\ee
Hence, a necessary condition for a vector~\mvsinp to minimize  
$\logdet{\imat_{\tslots\fslots}+\left(\mvsinp\herm{\mvsinp}\right)\had \mvchcovmat} / \vecnorm{\mvsinp}^{2}$
is the following: for any fixed~\dtime,~$\inp[\dtdf]$ may be different from~$0$ only for at most one discrete frequency~\dfreq. 
An example of such a vector
is shown in \fref{fig:inf-whk-sig}.
Even if the structure of the vector minimizing the second term
on the RHS of~\fref{eq: second decomposition} is known, the
infimum~\fref{eq: second decomposition} does not seem to admit a closed-form
expression. We can obtain, however, the following closed-form lower bound on the
infimum if we replace the  constraint   $\sum_{\dfreq=0}^{\fslots -1}
\abs{\inp[\dtdf]}^{2} \leq \papr \Pave \tstep$ \wpone in~\fref{eq:setTpeak} with the less stringent constraint
$\abs{\inp[\dtdf]}^{2} \leq \papr \Pave \tstep$ \wpone for all~\dtime and~\dfreq. The infimum of $\logdet{\imat_{\tslots\fslots}+\left(\mvsinp\herm{\mvsinp}\right)\had \mvchcovmat} / \vecnorm{\mvsinp}^{2}$
over the vectors~\mvsinp that belong to the new admissible set can be bounded as 
in~\fref{eq:ubTFpeak-term2-immse-lb}, after replacing~$\papr\Pave\tstep/\fslots$ by~$\papr\Pave\tstep$
and proceeding as in~\eqref{eq:pathloss}:
\be
\label{eq: lower bound to infimum}
\bs
	\inf_{\mvsinp} \frac{1}{\vecnorm{\mvsinp}^{2}}  \logdet{\imat_{\tslots
		\fslots} + \left(\mvsinp\herm{\mvsinp}\right)\had \mvchcovmat} &\geq
		\frac{1}{\papr\Pave\tstep}\int_{-1/2}^{1/2}\int_{-1/2}^{1/2} \log\lefto(1 + \Ppeak\tstep
		 \chspecfunp\right)d\specparam 
		d\altspecparam \\
		&=
		\frac{\fstep}{\papr\Pave}\spreadintlog{1+\frac{\papr\Pave}{\fstep}
		\scafunp}.
\es
\ee
To conclude the proof, we substitute~\fref{eq: lower bound to infimum} in~\fref{eq: second decomposition} and obtain
the desired upper bound~\fref{eq:ubTpeak}.
%

\bibliographystyle{IEEEtran}
\bibliography{IEEEabrv,publishers,confs-jrnls,ulibib,
}
\end{document}